\documentclass[aps,prx,twocolumn,longbibliography,superscriptaddress,showpacs]{revtex4-1}
\usepackage{graphicx}
\usepackage{latexsym}
\usepackage{amssymb}
\usepackage{amsmath}
\usepackage{soul}
\usepackage{amsfonts}
\usepackage{mathtools}
\usepackage{dsfont}
\usepackage{bm}
\usepackage{multirow}
\usepackage{color}
\usepackage{caption}
\usepackage{subcaption}
\usepackage{array}
\usepackage{tikz}
\usepackage[mathscr]{eucal}
\usepackage{mathrsfs}
\usetikzlibrary{decorations.pathreplacing,calligraphy,shapes}
\usepackage{comment}
\usepackage{xcolor}
\newcommand{\ket}[1]{\vert#1\rangle}

\newcommand{\bra}[1]{\langle#1\vert}
\usepackage[colorlinks=true, citecolor={blue!80!black}, urlcolor={blue!50!black}, linkcolor = {blue!80!black}]{hyperref}
\usepackage[percent]{overpic}
 \usepackage[utf8]{inputenc}

\newcommand{\Tau}{\mathcal{T}}

\DeclareMathAlphabet{\mathbbold}{U}{bbold}{m}{n}

\def\i{\textbf{i}}

\definecolor{forestgreen}{rgb}{0.43, 0.45, 0.53}
\definecolor{gr}{rgb}{0,0.82,0.18}
\definecolor{DarkerRed}{rgb}{0.9,0.05,0.25}
\DeclareMathOperator\arctanh{arctanh}

\newcommand\groupequation[2][17pt]{
  \setbox0=\hbox{$\displaystyle#2$}
  \stackengine{0pt}{\copy0}{
    \makebox[\linewidth]{\hfill$\left.\rule{0pt}{\ht0}\right\}$\kern#1}}
    {O}{c}{F}{T}{L}
}
\begin{document}
\title{
Highly complex novel critical behavior from {the}
intrinsic randomness of quantum mechanical measurements on 
critical ground state{s}
- a controlled renormalization group
{analysis}}

\author{Rushikesh A. Patil}
\author{Andreas W. W. Ludwig}
\affiliation{Department of Physics, University of California, Santa Barbara, CA 93106, USA}

\date{\today}

\begin{abstract}

We consider the effects of weak measurements on the quantum critical ground state
of  the  one-dimensional (a)  tricritical and (b) critical quantum Ising model, by measuring in (a) the local energy and in (b) the local spin operator in a lattice formulation. By employing a controlled renormalization group (RG) analysis
we find that 
each
problem exhibits highly complex novel scaling behavior, arising from the intrinsically indeterministic (`random') nature of quantum mechanical measurements,  which is  governed by a measurement-dominated RG fixed point that we study within an $\epsilon$ expansion. In the tricritical  Ising case (a) we find (i): multifractal scaling behavior of energy and spin correlations in the measured groundstate, corresponding to an infinite hierarchy of independent critical exponents and,  equivalently, to a continuum of universal scaling exponents for each of these correlations; (ii): the presence of logarithmic factors multiplying powerlaws in correlation functions, a hallmark of `logarithmic conformal field theories' (CFT); (iii): universal `effective central charges' $c^{({\rm eff})}_n$ for the prefactors of the logarithm of subsystem size of the $n$th R\'enyi entropies, which are independent of each other for different $n$, in contrast to the unmeasured critical ground state, and (iv): a universal (``Affleck-Ludwig'') `effective boundary entropy' $S_{\rm{eff}}$ which we show, quite generally, to be  related to the system-size independent part of the Shannon entropy of the measurement record, computed explicitly here to 1-loop order.
 --  
A subset of these results have so-far also been obtained within the $\epsilon$ expansion for the measurement-dominated critical point in
the critical Ising case (b).
\end{abstract}

\maketitle

\tableofcontents
\section{\label{sec:intro}Introduction}

Effects of measurements have recently attracted substantial attention especially in the context of measurement-induced quantum phase transitions in deep quantum circuits, and related problems, which exhibit novel universality classes of phase transitions in such non-equilibrium quantum systems
\cite{li2018quantum,li2023cross,MajidyAgrawalGopalakrishnanPotterVasseurHalpern2023,li2019measurement,skinner2019measurement,chan2019unitary,Choi_2020,potter2022entanglement,fisher2023random,cao2019entanglement,lavasani2021measurement,NahumRoySkinnerRuhman,JianShapourianBauerLudwig,fava2023nonlinear,gullans2020dynamical,gullans2020scalable,LiFisher2021,PhysRevX.12.041002,PhysRevLett.129.200602,PhysRevLett.129.120604,BaoChoiAltman2019,JianYouVasseurLudwig2019,li2023entanglement,LiVasseurFisherLudwig,vasseur2019entanglement,zhou2019emergent,NahumWiese,LiChenLudwigFisher,ZabaloGullansWilsonVasseurLudwigGopalakrishnanHusePixley,KumarKemalChakrabortyLudwigGopalakrishnanPixleyVasseur}.
Another class of quantum systems subjected to measurements was recently introduced in Ref.~[\onlinecite{GarrattWeinsteinAltman2022}], and subsequent 
works~\cite{WeinsteinSajithAltmanGaratt,YangMaoJian,MurcianoSalaYueMongAlicea}, considering the effects of measurements on one-dimensional quantum critical ground states. Ref.~[\onlinecite{GarrattWeinsteinAltman2022}] considered a Luttinger liquid, and provided a field theory formulation with measurements acting on the one-dimensional zero-time slice in space-time, exhibiting a version of the Kosterlitz-Thouless transitions, while Ref.~[\onlinecite{WeinsteinSajithAltmanGaratt,YangMaoJian,MurcianoSalaYueMongAlicea}]  similarly considered several types of measurements, with and without postselection, performed on the ground state of the critical one-dimensional quantum Ising model.\footnote{Compare also Refs. \onlinecite{LeeJianXu,ZouSangHsieh,Myerson-JainXuHughes,AshidaFurukawaOshikawa}
which consider different but related problems of effects of decoherence on 1d critical ground states, which we do not study in the present paper}
The aim of the present paper is to exhibit novel universality classes of critical behavior with highly complex and novel scaling behavior that can emerge when (weak)  measurements are performed (without postselection) on quantum critical ground states. In the examples we discuss such critical behavior originates from a measurement-dominated fixed point occurring at a finite measurement strength, which we treat using a controlled
 renormalization group (RG) analysis, i.e. an $\epsilon$ expansion. Physically, the complexity of the scaling behavior originates from the intrinsic indeterministic (``random'')  nature of quantum mechanical measurements. While so-far analytically-based tools for understanding measurement-induced transitions in deep quantum circuits have been largely elusive, problems involving measurements performed on one-dimensional quantum critical ground states are typically simpler  technically, and thus more  susceptible to a controlled RG analysis, as we demonstrate in the examples we study. Yet, they exhibit similar complex scaling behavior as that in the deep circuits.

 The first problem we study, ``problem (a)'',  consists of measurements with the local energy operator on the ground state of the tricritical quantum Ising model, within a lattice formulation.
 After introducing replicas this can be written as the field theory of the $(1+1)$-d tricritical Ising model in space-time with a perturbation acting solely on the $\tau=0$  equal time-slice describing the measurements. This perturbation is
 relevant in the  RG sense, and it flows to an infrared fixed point at finite measurement strength which can be controlled within an $\epsilon$ expansion. This is analogous to the Wilson-Fisher $\epsilon=4-d$ expansion, except  that here the dimension of space-time is always two, and a small parameter $\epsilon$ is obtained by generalizing the tricritical Ising model to the
tricritical $q$-state Potts model, and expanding about $q=4$ where the perturbation becomes marginal [which is in essence, an expansion in the small parameter $(4-q)$]. This allows for a systematic calculation of all universal scaling properties at the infrared fixed point.

One reflection of the complexity of the finite measurement strength fixed point appears in correlation functions of the spin and energy operator taken in the measured ground state with measurement outcomes ${\vec m}$. When raised to the $N$th power and averaged over measurement outcomes with the Born probability these $N$th moments decay, for spin and energy correlations, with independent exponents, one for each moment order $N$. Thus, associated with each of the two observables (spin and energy) there is an infinite hierarchy of scaling exponents, in contrast to standard critical behavior.  This is referred to as multifractal scaling. Since this scaling behavior has its origin~\cite{LUDWIG1990infinitehierarchy,ZabaloGullansWilsonVasseurLudwigGopalakrishnanHusePixley}
in a universal
scaling form of the entire probability distribution for the correlation function,
also the non-integer $N$ moments will scale giving rise to a continuous spectrum of scaling dimensions for  each, the spin and the energy correlations.

Another exotic feature of the measurement-dominated fixed point that we find is that of a so-called {``logarithmic conformal field 
theory"}~\cite{Gurarie1993,GurarieLudwig2005,VasseurJacobsenSaleur,CardyLogarithm1999}.
While at RG fixed points associated with conventional critical points 
correlation functions decay with powerlaws, here we observe in certain averaged correlation functions a powerlaw multiplied by a logarithm. This arises because at the finite measurement fixed point a rescaling of distances does not act diagonally on all observables, but may  act in the form of a two-dimensional~\footnote{in simplest manifestation} non-diagonalizable ``Jordan-form" matrix. When translated into the behavior of averaged correlation function this amounts to the presence of the multiplicative logarithm.

The entanglement entropies exhibit further complex universal behavior at the finite measurement strength critical point. While the universal coefficients of the logarithm of subsystem size in the $n$th R\'enyi entropy of the unmeasured ground state,
${1\over 3} c_n$,
are all related to the central charge $c$,  i.e. $c_n= {c \over 2}[1+{1\over n}]$,
at the finite measurement critical point the universal coefficients ${1\over 3} c^{({\rm eff})}_n$ are all unrelated to each other and have a more complicated $n$ dependence already to first order in $\epsilon$ which we calculate, and which  is further modified in higher order in $\epsilon$. This represents another hierarchy of independent universal quantities, similar to those encountered in spin and energy correlations functions discussed above. (Specifically, these are correlation functions of the ``$n$-twist field'', Sect.~\ref{sec:EntanglementEntropy}.) We furthermore show that the universal quantities $c^{({\rm eff})}_n$ also appear in the coefficient 
of the linear temperature dependence of  the {\it extensive} measurement averaged R\'enyi entropies of  the full mixed thermal Gibbs state of the system at finite temperature.

The problem of performing measurements on a quantum critical ground state can be viewed as a problem of the unmeasured critical (here conformally invariant) field theory in space-time with a 
defect at the zero-time slice, and the finite measurement strength fixed point represents a scale-invariant, in fact conformally invariant defect. After folding along the slice it becomes a boundary  condition on the (doubled)  unmeasured conformal field theory (CFT). In general, to any 
boundary of a CFT is associated a universal constant, the ``Affleck-Ludwig''  boundary entropy~\cite{AffleckLudwig1991}. Here we establish quite generally that for problems of measurements on 1d quantum critical ground states, the corresponding universal ``effective boundary entropy'' $S_{{\rm eff}}$ is the constant, system size independent piece of the Shannon entropy of the measurement record. (We have computed it here explicitly to lowest order in the $\epsilon$ expansion.) 
It may be viewed as a boundary analog of the
``effective central charge''
at measurement-induced transitions in deep quantum circuits, which arises from the universal finite-size
dependence
of the Shannon entropy of the measurement record of the (bulk) space-time of the circuit \cite{ZabaloGullansWilsonVasseurLudwigGopalakrishnanHusePixley}.

 Lastly, we address the problem of weak  measurements
performed on the ground state of the {\it critical}  Ising model
with the Pauli ${\hat \sigma}^z_i$ operator at lattice sites $i$,  via an extension of the controlled RG analysis and $\epsilon$ expansion developed for the {\it tri}critical Ising case. We find that these measurements lead to similar complex scaling behavior 
governed by another measurement-dominated RG fixed point, occurring at a finite measurement strength. In particular, we obtain (to two-loop order) an infinite hierarchy of independent multifractal critical exponents for the set of measurement averaged moments of the 
${\hat \sigma}^z_i$ correlation function, leading again to a continuous spectrum of critical exponents and an independent scaling exponent of the typical connected correlation function. 
Moreover, we find, in analogy to the tricritical Ising case{,} independent coefficients ${1\over 3} c^{{\rm (eff)}}_n$ of the logarithm of subsystem size for the measurement averaged $n$-th R\'enyi entropies for different values of $n$, which we compute to leading order in the $\epsilon$ expansion. The presence of multiplicative logarithms in measurement averaged correlation function{s} (logarithmic CFT features) is currently being studied as well in the Ising case.
\par
The remaining {parts of the paper are}
structured in the following manner: In Section \ref{Sec:TheModel}, we introduce the O'Brien-Fendley model and discuss its zero temperature phase diagram which has a critical point in the universality class of the tricritical Ising point. 
For this quantum tricritical ground state, in Section \ref{section:Measurements}, we describe a measurement protocol with explicit Kraus  operators
corresponding to weak measurements. 
In Section \ref{sec:FieldTheoryandFixedPoint}, we develop a replica field theory to analyze the problem of described weak measurements on the tricritical ground state.
We analyze 
the
infrared behavior of the obtained replica field theory using {a controlled} perturbative RG  expansion and 
determine 
the new `non-trivial' fixed point 
in an $\epsilon$-expansion.
In Section \ref{sec:CorrelationFunctions},  we determine the long distance behavior of measurement averaged moments of correlation functions for the spin $\hat{\sigma}^{z}$ and the energy $\hat{\boldsymbol{E}}$ operator (defined in Section \ref{section:Measurements}) and 
 demonstrate
the logarithmic CFT features of the measurement-dominated fixed point.
In Section \ref{sec:EntanglementEntropy}, we calculate the measurement averaged $n^{\text{th}}$ R\'enyi entanglement entropies and the von Neumann entanglement entropy.
In Section \ref{Sec:EffectiveGroundStateDegeneracyGeff}, 
we discuss the Shannon entropy of the measurement record and the relationship of its constant universal part with the `effective boundary entropy'.
{In Section \ref{Sec:IsingModel}, we discuss the case of Ising critical point under measurements with the $\hat{\sigma}^{z}$ spin operator.}
Section \ref{Sec:Conclusion}
{is}
reserved for conclusions and discussion of results.

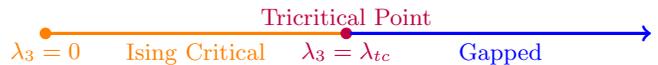
\begin{figure}
\centering
\begin{tikzpicture}
\filldraw [orange] (-6,0) circle (2pt) node[below]{$\lambda_{3}=0$};
\draw[orange, very thick] (-6,0) -- node[below]{Ising Critical} ++ (4,0);
\filldraw [purple] (-2,0) circle (2pt) node[below]{$\lambda_{3}=\lambda_{tc}$} node[above]{Tricritical Point};
\draw[blue, very thick,->] (-1.95,0) --node[below]{Gapped} ++ (4,0);
\end{tikzpicture}
\caption{Phase Diagram of the O'Brien-Fendley Chain}\label{fig:PhaseDiagramO'Brien-Fendley}
\end{figure}

\vskip .8cm
 {\section{The O'Brien-Fendley Model\label{Sec:TheModel}}}
{A 
{variety}
of 
quantum mechanical systems
in one-dimensional space with different microscopic appearance are known to exhibit
a quantum critical point in the universality class of the tricritical Ising model,}
{see e.g.}
Ref. \cite{AlcarezDrugowichFelicioKoberleStilck,RahmaniZhuAffleckFranz,O'BrienFendley,Slagleetal,GroverShengVishwanath}.
In 
{the present} paper, we will consider
{one such microscopic realization convenient for our purposes,}
the O'Brien-Fendley chain introduced in Ref. \onlinecite{O'BrienFendley}.
The O'Brien-Fendley chain is a 1d quantum chain with spin-$\frac{1}{2}$ 
{(qubit)}
degrees of freedom at each site and is described by the Hamiltonian {$H$} 
\begin{subequations}\label{eq:HamiltoniansTricriticalIsing}
\begin{eqnarray}
       &&H=H_{I}+\lambda_{3}H_{3} \label{eqn:O'BrienFendleyHamiltonian}\\
       &&H_{I}=-\sum_{j}(\hat{\sigma}^{z}_{j}\hat{\sigma}^{z}_{j+1}+\hat{\sigma}^{x}_{j})\label{EqIsingCriticalHamiltonian}\\
       &&H_{3}=\sum_{j} (\hat{\sigma}^{x}_{j}\hat{\sigma}^{z}_{j+1}\hat{\sigma}^{z}_{j+2}+\hat{\sigma}^{z}_{j}\hat{\sigma}^{z}_{j+1}\hat{\sigma}^{x}_{j+2})\label{Eq:Isingthreesiteinteraction}
\end{eqnarray}
\end{subequations}
where 
{$\hat{\sigma}^a$ ($a = x,y,z$) are}
{the standard} Pauli matrices. 
Note that at $\lambda_{3}=0${,} the Hamiltonian $H$ reduces to the {Hamiltonian $H_{I}$ of 
the \textit{critical} 1d quantum Ising chain.} 
{As seen by inspection, the}
term $H_{3}$ in the Hamiltonian is invariant under the Kramers-Wannier (K-W) transformation given by
\begin{equation}\label{Eqn:K-WTransformation}
    \begin{split}
\hat{\sigma}^{z}_{j}\hat{\sigma}^{z}_{j+1}=\tau^{x}_{j+\frac{1}{2}}\\
\hat{\sigma}^{x}_{j}=\tau^{z}_{j-\frac{1}{2}}\tau^{z}_{j+\frac{1}{2}}.
\end{split}
\end{equation}
{Since} there are no RG relevant K-W self-dual operators
{at the Ising critical point,}
{for sufficiently} small $\lambda_{3}\neq 0$ the chain in Eq. \ref{eqn:O'BrienFendleyHamiltonian} 
{is described by a K-W invariant line of 
second order transitions parametrized by $\lambda_3$, all in the Ising universality class.}
However{,} as discussed in Ref. \onlinecite{O'BrienFendley}, for large enough $\lambda_3$ the spectrum becomes gapped{,} and since the Hamiltonian is self-dual under the K-W transformation{,} 
{there is}
a line of first order phase transitions
{on the phase boundary between the ferromagnetic and the paramagnetic phase}. 
The phase diagram of the chain  {along the K-W line}
is shown in 
{Fig. \ref{fig:PhaseDiagramO'Brien-Fendley} 
    (from Ref.~\onlinecite{O'BrienFendley}).}
The {renormalization group (RG)} unstable critical point at $\lambda_{3}=\lambda_{tc}\approx 0.856$, which separates the Ising second order phase transition line from the first order phase transition line, lies in the universality class of {the} tricritical Ising model.
{In the present paper{,} we will consider performing measurements on the ground state  of the
{\it tricritical} quantum Ising Hamiltonain $H$ of Eq.(\ref{eqn:O'BrienFendleyHamiltonian}) at $\lambda=\lambda_{tc}$.}
\vskip .8cm
{\section{Measurement Protocol
and Replica Trick\label{section:Measurements}}}
{Coming first back to ordinary Ising case $\lambda_3=0$, the ground state of the  critical quantum Ising chain $H_I$ subject to} measurements with operators 
$\hat{\sigma}^{x}_{j}$
{or}  $\hat{\sigma}^{z}_{j}\hat{\sigma}^{z}_{j+1}$ has been investigated in Ref. \onlinecite{WeinsteinSajithAltmanGaratt}, \onlinecite{YangMaoJian} and \onlinecite{MurcianoSalaYueMongAlicea}.
{Measurements with $\hat{\sigma}^{z}_{j}\hat{\sigma}^{z}_{j+1}$ and $\hat{\sigma}^{x}_{j}$
on the Ising critical ground state
result in the same universal effects~\cite{WeinsteinSajithAltmanGaratt}{,}
 because both operators represent (after subtraction of their expectation values) the energy operator
{$\mathfrak{e}$}
 of the Ising critical point 
{with scaling dimension $X_{\mathfrak{e}}=1$,} 
 up to corrections from subleading operators which are RG-irrelevant. In other words, their connected equal time correlation functions (denoted by a subscript ${}_c$)
 in the Ising critical ground state, describing the correlations of the subtracted operators, decay asymptotically with the same exponent,}
 \begin{equation}
\begin{split}
\langle\hat{\sigma}^{z}_{i}\hat{\sigma}^{z}_{i+1} \hat{\sigma}^{z}_{j}\hat{\sigma}^{z}_{j+1} \rangle_{c} \sim \frac{1}{|i-j|^{2X_{\mathfrak{e}}
}}
\text{and }
\langle\hat{\sigma}^{x}_{i}\hat{\sigma}^{x}_{j}\rangle_{c} \sim \frac{1}{|i-j|^{
2X_{\mathfrak{e}}
}}
\end{split}
\end{equation}
as  $|i-j|>>1$.
\vspace{0.4cm}
\par
{We  now move  on to the O'Brien-Fendley chain which, as recalled above, has the {\it same} underlying
lattice spin-$\frac{1}{2}$ (qubit)  degrees of freedom as the ordinary Ising chain. Therefore, it is natural to study effects of measurements of similar operators for the O'Brien-Fendley chain.}
Interestingly, {equal-time correlation functions of the {\it same} operators
 $\hat{\sigma}^{z}_{j}\hat{\sigma}^{z}_{j+1}$ and $\hat{\sigma}^{x}_{j}$ in the ground state of}
the tricritical point of the O'Brien-Fendley chain
{decay asymptotically 
(after subtraction of the expectation values) 
with the critical exponent $X_{\mathcal{E}}=
\frac{1}{5}$ of the energy scaling operator $\mathcal{E}$ of the Ising {\it tricritical} point}~\cite{ZouVidal},
{
\begin{equation}
\begin{split}
\langle\hat{\sigma}^{z}_{i}\hat{\sigma}^{z}_{i+1} \hat{\sigma}^{z}_{j}\hat{\sigma}^{z}_{j+1} \rangle_c \sim \frac{1}{|i-j|^{2{X}_{\mathcal{E}}}} \text{and }\langle\hat{\sigma}^{x}_{i}\hat{\sigma}^{x}_{j}\rangle_c \sim \frac{1}{|i-j|^{2{X}_{\mathcal{E}}}}
\end{split}
\end{equation}}
{as $|i-j| \gg 1$.}
{Unlike the ordinary}  {transverse field}
quantum Ising model, the subdominant contributions to both $\hat{\sigma}^{z}_{j}\hat{\sigma}^{z}_{j+1}$ and $\hat{\sigma}^{x}_{j}$ are now \textit{not} RG irrelevant.
{Specifically, these two lattice operators 
are represented~\cite{ZouVidal} by}
{
\begin{equation}\label{eqn:ContinuumLimitofTwoLatticeOperators}
\begin{split}
&\hat{\sigma}^{z}_{j}\hat{\sigma}^{z}_{j+1} \sim 
\chi(x)\equiv 
A I +B\mathcal{E}(x)+  
C\mathcal{E}'(x) + D \mathcal{E}^{''} +
...,\\
&\hat{\sigma}^{x}_{j}
\sim \chi'(x)\equiv A I -B \mathcal{E}(x) + 
C\mathcal{E}'(x) - D \mathcal{E}^{''} +
..., 
\\
&
(A,  B, C, D =\text{non-universal constants}),
\end{split}
\end{equation}}
{{
where  $I$ is the identity field with  $A$ the tricritical expectation value of the corresponding lattice operator,  
while $\mathcal{E}$ ({with scaling dimension} $X_{\mathcal{E}}=\frac{1}{5}$),  $\mathcal{E}'$ ({with scaling dimension} $X_{\mathcal{E}'}=\frac{6}{5}$), and
$\mathcal{E}^{''}$ ({with scaling dimension} $X_{\mathcal{E}^{''}}=3$)
are the energy, the subdominant energy, and the further subleading energy}
scaling operators at the
Ising {\it tricritical} point, 
{the first two} of which are RG-relevant {as bulk operators}; the ellipses denote more subleading operators.}
Forming
{the difference}
of $\hat{\sigma}^{z}_{j}\hat{\sigma}^{z}_{j+1}$ and $\hat{\sigma}^{x}_{j}$, 
{we define} the operator $\hat{\boldsymbol{E}}_{j+\frac{1}{2}}$ given by
\begin{equation}\label{eqn:boldsymbol{E}}
    \hat{\boldsymbol{E}}_{j+\frac{1}{2}}\equiv\frac{1}{\sqrt{2}}(\hat{\sigma}^{z}_{j}\hat{\sigma}^{z}_{j+1}-\hat{\sigma}^{x}_{j}){,}
\end{equation}
{which changes sign under the K-W duality transformation, Eq.~\ref{Eqn:K-WTransformation}. Owing to Eq.~\ref{eqn:ContinuumLimitofTwoLatticeOperators}{,}
this operator  
is a lattice representation of}
the energy scaling operator  $\mathcal{E}$ of the 
 tricritical Ising 
 {critical point which is consequently also odd under K-W duality, with corrections from solely RG {\it irrelevant} (K-W odd) operators.
 For the same reason, the linear combination 
 \begin{equation}
 \label{eqn:boldsymbol{E}Prime}
 \hat{\boldsymbol{E}}'_{j+
 \frac{1}{2}}
 \equiv
 \frac{1}{\sqrt{2}}(\hat{\sigma}^{z}_{j}\hat{\sigma}^{z}_{j+1}+\hat{\sigma}^{x}_{j}){,}
 \end{equation}
 with the opposite sign than in
 Eq.~\ref{eqn:boldsymbol{E}}{,} is {\it even} under K-W-duality and is (after subtraction of its expectation value) a lattice representation of the subleading energy operator $\mathcal{E}'(x)$ which, consequently, is K-W even (together with  all occurring subleading operators)}~\footnote{{To re-iterate, these results}
can be understood by using Kramers-Wannier duality. The O'Brien-Fendley chain is invariant under Kramers-Wannier duality throughout {the line in} its phase diagram {depicted in Fig.~\ref{fig:PhaseDiagramO'Brien-Fendley}}. 
The energy scaling operator, {$\mathcal{E}(x)$, and the subleading energy scaling operator, $\mathcal{E}'(x)$ are, respectively, odd and even under the K-W transformation at the tricritical Ising point. 
Since $\hat{\sigma}^{z}_{j}\hat{\sigma}^{z}_{j+1}-\hat{\sigma}^{x}_{j}$ is odd under the K-W transformation (see Eq. \ref{Eqn:K-WTransformation}), it \textit{cannot} contain any contribution from the subleading energy field $\mathcal{E}'(x)$ in the continuum limit and is given by the energy field $\mathcal{E}(x)$ with corrections from solely RG irrelevant (K-W odd) operators.}}{.}
Note that $\hat{\boldsymbol{E}}_{j+\frac{1}{2}}$ lies on the link of the lattice connecting site at $j$ and $j+1${,} and
\begin{equation}
\nonumber
\begin{split}
&(\hat{\boldsymbol{E}}_{j+\frac{1}{2}})^2=
\\
& =\frac{1}{2}((\hat{\sigma}^{z}_{j}\hat{\sigma}^{z}_{j+1})^2+(\hat{\sigma}^{x}_{j})^2-\hat{\sigma}^{x}_{j}\hat{\sigma}^{z}_{j}\hat{\sigma}^{z}_{j+1}-\hat{\sigma}^{z}_{j}\hat{\sigma}^{z}_{j+1}\hat{\sigma}^{x}_{j}) = 1\\
\end{split}
\end{equation}
{implying that}
{
\begin{equation}
\ \hat{\boldsymbol{E}}_{j+\frac{1}{2}}\, \text{has eigenvalues } \pm 1. 
\end{equation}}
{We note in passing that, as verified by inspection,  the operator ${\hat{\boldsymbol{E}}'}_{i+
\frac{1}{2}}$ 
also squares to the identity and therefore also has eigenvalues $\pm 1$.~\footnote{{${\hat{\boldsymbol{E}}'}_{i+{1\over 2}}$ does not commute with the operator $\hat{\boldsymbol{E}}_{i+{1\over 2}}$.}}}\\
{In the following, we will describe a protocol for performing measurements on the ground state of the O'Brien-Fendley chain. }
\vskip .6cm
\subsection{
Measuring \texorpdfstring{$\hat{\boldsymbol{E}}_{i+\frac{1}{2}}$}{Lg} on Even Links \label{subsec:MeasProto1}}
Note that the operators  $\hat{\boldsymbol{E}}_{i+\frac{1}{2}}$ on neighbouring links do not commute with each other, 
{because} $\hat{\sigma}^{z}_i$ and $\hat{\sigma}^{x}_j$ on the same site do not commute. 
However, if we take $\hat{\boldsymbol{E}}_{i+\frac{1}{2}}$ operators on alternate links, say even links {(those where $i$ is even)}, all of them commute with each other 
(see Fig. \ref{fig:evenlinks}).
\begin{figure}
\centering
\begin{tikzpicture}
\filldraw[orange] (2,0) circle (2pt) circle (2pt) node[anchor=west]{$\hspace{0.2cm}\ldots\hspace{0.2cm}$ } node[anchor=north]{1};
\filldraw[blue] (0,0) circle (2pt) node[anchor=north]{0};
\filldraw[orange] (-2,0) circle (2pt) node[anchor=north]{-1};
\filldraw[blue] (-4,0) circle (2pt) node[anchor=east]{$\hspace{0.2cm}\ldots\hspace{0.2cm}$ } node[anchor=north]{-2};
\draw [decorate,
    decoration = {calligraphic brace,raise=5pt}] (0,0) --  (2,0) node[pos=0.5,above=10pt,black]{$ \hat{\boldsymbol{E}}_{\frac{1}{2}}=\frac{1}{\sqrt{2}}(\textcolor{blue}{\hat{\sigma}^{z}_{0}}\textcolor{orange}{\hat{\sigma}^{z}_{1}}-\textcolor{blue}{\hat{\sigma}^{x}_{0}})$};
\draw [decorate,
    decoration = {calligraphic brace,raise=5pt}] (-4,0) --  (-2,0) node[pos=0.5,above=10pt,black]{$ \hat{\boldsymbol{E}}_{-\frac{3}{2}}=\frac{1}{\sqrt{2}}(\textcolor{blue}{\hat{\sigma}^{z}_{-2}}\textcolor{orange}{\hat{\sigma}^{z}_{-1}}-\textcolor{blue}{\hat{\sigma}^{x}_{-2}})$};
\end{tikzpicture}
\caption{`Energy' Operator on Even Links}
\label{fig:evenlinks}
\end{figure}
{Our measurement protocol consists in measuring the operators $\hat{\boldsymbol{E}}_{i+\frac{1}{2}}$
on even links.~\footnote{{
We note that the operator $\hat{\boldsymbol{E}}_{i+\frac{1}{2}}$ (for an even link $i$) has support on the two lattice sites $i$ and $i+1$, and one can check that each of its eigenvalues $\pm 1$ is two-fold degenerate.
This means that there will be two (linearly independent) eigenstates of $\hat{\boldsymbol{E}}_{i+\frac{1}{2}}$, which can be chosen orthogonal, that will be associated 
with each of the measurement outcomes $+1$ and $-1$.
This does \textit{not} imply that the
post-measurement state
is ambiguous.
The state that results after observing
any set of measurement outcomes is uniquely obtained by acting with the projector (or more generally, in the case of weak measurements, a Kraus operator) corresponding to the eigenspace associated with the measurement outcomes on the `incoming'  state before measurement. In the case of interest to us this will be, as we will discuss below, the ground state of the {\it tricritical} O'Brien-Fendley chain.
See e.g. Eq. \ref{EqStateObtainedUponMeasurement} below. (Measurement operators with two eigenvalues which are not both non-degenerate, have also been discussed in a different context in Ref. 
\onlinecite{MajidyAgrawalGopalakrishnanPotterVasseurHalpern2023}
.)}}}~\footnote{{We note that performing weak measurements with the operator $\hat{\boldsymbol{E}}_{i+\frac{1}{2}}$ only on even links $i$ does \textit{not} imply that the
system will effectively collapse onto a trivial `staggered' state. 
Compare for example with
Refs. \onlinecite{WeinsteinSajithAltmanGaratt} and \onlinecite{YangMaoJian}, which consider performing measurements with the Pauli operator $\hat{\sigma}^{x}_{i}$ on  all sites $i$ of the critical quantum Ising chain (See Eq. \ref{EqIsingCriticalHamiltonian} for the Hamiltonian). 
In terms of the Majorana formulation of the quantum Ising chain, 
this measurement operator reads
$\hat{\sigma}^{x}_i=\hat{\gamma}_{2i}\hat{\gamma}_{2i+1}$, where $\hat{\gamma}_{2i}$ is a Majorana operator.
Thus measuring $\hat{\sigma}^{x}_i$ corresponds to performing measurements only on the even-links of the underlying Majorana chain, however a `staggered' state is not observed \cite{WeinsteinSajithAltmanGaratt,YangMaoJian}.
In fact, 
even though measurements are performed only on the even links of the Majorana chain, 
it has been observed that for Born-rule measurements \cite{WeinsteinSajithAltmanGaratt,YangMaoJian} the long-distance critical properties of the system are the same as that of the unmeasured state, i.e. the Ising critical ground state (which is not `staggered').
In the same spirit, in our measurement protocol of our system we also 
would not expect to see a trivial `staggered' state even though we are performing measurements only on the even-links of the chain (which are now the physical links of the spin-chain and not of the underlying Majorana chain).
In our case, the critical behaviour of the unmeasured state, however, does get modified dramatically due to the presence of measurements as we will
discuss 
in detail in the subsequent sections of the present paper. 
 -- A general proof of the impossibility of obtaining a trivial `stagggered' state, for our (and also the above) system  will follow from Eq.~\ref{eqn:bornruleavgN=1} with ${\hat {\cal O}}_1 :=$
$\hat{\boldsymbol{E}}_{i+\frac{1}{2}}$
$\hat{\boldsymbol{E}}_{j+\frac{1}{2}}$. This equation implies that the expectation value of ${\hat {\cal O}}_1$, a two point function, is unmodified by {Born-rule} measurements and thus exhibits the algebraic decay  with distance $|i-j|$ characteristic of the unmeasured (tri-)critical ground state, which would be in contradiction with an exponential decay in a {trivial} `staggered' state.}} 
Then on each even link{,} we can define the (weak-) measurement  {Kraus}
{operator,}
\begin{eqnarray}\label{LabelEqKrausOperator}
{\hat K}_{i+\frac{1}{2},\pm} :=
\frac{1 \pm \lambda \hat{\boldsymbol{E}}_{i+\frac{1}{2}}}{
\sqrt{2(1+\lambda^2)}},
\end{eqnarray}
 and the measurement operators {$\hat{\boldsymbol{E}}_{i+\frac{1}{2}}$} at different even links will commute with each other. For notational convenience, we will drop the 
 {lattice-position offsets}
 of $+\frac{1}{2}$ from now on and label both 
 operators {as $\hat{K}_i$ and $\hat{\boldsymbol{E}}_i$, i.e. by just $i$,} which denotes the site at the left end of the link.~\footnote{
 {I.e.
 ${\hat K}_{i+\frac{1}{2},m_i} \to {\hat K}_{i,m_i}$ where $m_i=\pm$, and correspondingly for $\hat{\boldsymbol{E}}$}}

When $\lambda=1$, the
{Kraus} operators 
{in Eq.~\ref{LabelEqKrausOperator}}
reduce to projection operators
{$\hat{K}_{i, m_i}=$}
$\frac{1}{2} ( 1 + m_{i} \hat{\boldsymbol{E}}_{i})$
onto the eigenstates 
of $\hat{\boldsymbol{E}}_i$ with eigenvalues $m_i=\pm 1$.
The parameter $ 0 \leq \lambda \leq 1$ controls the `strength' of the measurement. When $\lambda=0$ no measurements are performed at all.

When the eigenvalue $m_i$ is measured at site $i$, the measurement changes a (normalized) quantum state ${\ket \psi}$ to the following (normalized) state  {\it after} this measurement
\begin{eqnarray}
{\ket \psi} \to {
{\hat K}_{i,m_i} {\ket \psi}
\over
||{\hat K}_{i,m_i} {\ket \psi}||
}. 
\end{eqnarray}
Each measurement outcome $m_i= \pm  1$  at an even link 
$i$
occurs with `Born-rule' probability
\begin{eqnarray}
\label{LabelBornRuleProbability}
&&
p_B(m_i) = {\bra \psi} 
({\hat K}_{i,m_i})^\dagger
{\hat K}_{i,m_i}
{\ket \psi} =
\\ \nonumber
&&
=
{1\over 2 (1 + \lambda^2)}
\left (
1 + \lambda^2+ 2 m_i \lambda
{\bra \psi} \hat{\boldsymbol{E}}_i {\ket \psi}
\right ){,}
\end{eqnarray}
which depends on the {incoming}  state $|\psi\rangle$.
The measurement operators for each  {even} $i$ satisfy the condition
\begin{eqnarray}
\label{LableEq-POVM-Condition}
\sum_{m_i = \pm 1}
({\hat K}_{i,m_i})^\dagger
{\hat K}_{i,m_i} = {\bf 1}_i{,}
\end{eqnarray}
where the right hand side denotes the identity 
operator~\footnote{{We note, continuing a previous footnote, that degeneracies of the measurements operators $\hat{\boldsymbol{E}}_i$
and thus of the  Kraus operators does not affect the identity
Eq.~\ref{LableEq-POVM-Condition}. For a similar situation with degeneracies of the eigenvalues of the measurement operators, see the already previously mentioned Ref.  
\onlinecite{MajidyAgrawalGopalakrishnanPotterVasseurHalpern2023}
}}.
This ensures the normalization of the Born-rule probabilities $p_B(m_i)$ defined above.
 {Eq.~(\ref{LableEq-POVM-Condition}) is referred to as the 
POVM~\footnote{standing for``Positive Operator Valued Measure''} condition.}

\par
{
Let us take the quantum state on which we perform measurements to be the {\it ground state}
$\ket{0}$ of the  O'Brien-Fendley chain at the {\it tricritical} point.}
Since the measurement {operators $\hat{\boldsymbol{E}}_i$  on the even links $i$} commute with each other, the state obtained after measuring on all even links {with measurement outcomes ${\vec m} :=$ $\{m_i\}$} is
{
\begin{eqnarray}\label{EqStateObtainedUponMeasurement}
\ket{\Psi_{\vec m}}=
{\prod_{i\in\text{even}} {\hat K}_{i, m_i} {\ket 0}
\over
||
\prod_{i\in\text{even}} {\hat K}_{i, m_i} {\ket 0} ||
}
=
{{\hat {\bm K}}_{\vec m} {\ket 0}
\over
\sqrt{{\bra 0}
{\hat {\bm K}}^\dagger_{\vec m}
 {\hat {\bm K}}_{\vec m}
{\ket 0}
}
}{,}
\end{eqnarray}
}
where
\begin{eqnarray}\label{EqProductofLocalKraus}
{\hat {\bm K}}_{\vec m} := \prod_{i\in\text{even}}{\hat K}_{i, m_i},
\end{eqnarray}
{and} 
\begin{eqnarray}
\label{LabelDefBornRuleProb}
p_0({\vec m})= {\bra 0}
{\hat {\bm K}}^\dagger_{\vec m}
 {\hat {\bm K}}_{\vec m}
{\ket 0}
\end{eqnarray}
{is 
the} Born-rule probability to obtain measurement outcomes ${\vec m}=$ $ \{m_i\}$. 
{We will refer to the state obtained upon performing measurements and corresponding to a particular set of measurement outcomes as a `quantum trajectory'.}
{It will be convenient to write the RHS of Eq. \ref{LabelEqKrausOperator} as}
\begin{eqnarray}
\label{LabelEqKrausmiExonential}
{\hat K}_{i, \pm}={1\over {\cal N} }\exp\{\pm \tilde{\lambda} \hat{\boldsymbol{E}}_i\},
\end{eqnarray}
where $0\leq\tilde{\lambda} = \arctanh(\lambda)<\infty$, and ${\cal N}$ is 
{a suitable} 
normalization factor.
{Then}
we can write the product 
{in Eq. \ref{EqProductofLocalKraus}}
in the {following} form
{
\begin{eqnarray}
\label{LabelEqKrausOperatorProto1}
 {\hat {\bm K}}_{\vec m}
=
{1\over {\cal N}^{L/2}}
\exp
\{\tilde{\lambda} 
\sum_{i = \text{even}}
m_i\hat{\boldsymbol{E}}_{i}
\}{,}
\end{eqnarray}}
where $L$ denotes the number of sites.
{Let us define the variable $t_i$ s.t.
\begin{equation}
\label{LabelEqtilambdami}
    t_i=\tilde{\lambda}m_i.
\end{equation}
}
{Since the measurement outcome 
{is $m_{i}=\pm 1$, the variable $t_i$ takes on}
values $\pm \tilde{\lambda}$.}
{We can reformulate the measurements by ``softening'' the 
{variable $t_i=\pm \tilde{\lambda}$}
to
take on {\it continuous} values  $-\infty  < t_i < +\infty$ drawn from some 
distribution $P(t_i)$ which we take to be symmetric under $t_i\to - t_i$. Sometimes, it may be convenient  to choose  a Gaussian distribution whose variance is a measure of the  `strength' of measurements ${\tilde \lambda}$. 
}
{The formulation  given
in Eqs.~\ref{LabelEqKrausmiExonential},\ref{LabelEqKrausOperatorProto1}
above  in terms of discrete measurement outcomes  $m_i=\pm 1$ is simply 
a special case of this where the 
distribution $P(t_i)$ is  
the
(normalized)
sum of two delta functions peaked at $t_i=\pm {\tilde \lambda}$. 
It turns out that {\it only} the {\it cumulants} of  the random 
variable $t_i$ determined by the distribution $P(t_i)$~\footnote{{We consider only such distributions $P(t_i)$ for which all cumulants
exist, i.e. are finite.}} will enter our formulation below, and the essential physics will turn out to depend only on the second {cumulant and will thus be insensitive to other details of the 
distribution $P(t_i)$.}
This then also covers the case where, with some probability, no measurement is performed at a site, corresponding  to 
{the
symmetric distribution}  $P(t_i)$ which is a 
 (normalized) {weighted} sum of three delta functions, peaked at $t_i=0$ and at $t_i=\pm {\tilde \lambda}$.}

 {The  corresponding reformulated Kraus operators}
{
\begin{eqnarray}
\label{LabelEqKrausKti}
 {\hat  {\bm K}}_{\vec {\bf t}}\equiv
 {1\over ({{\cal N}'})^{L/2}}
 \exp
\{
\sum_{i\in\text{even}} t_{i}\hat{\boldsymbol{E}}_{i}
\},
\ \ {\vec t} \equiv \{t_i\}_{i \in {\rm even}},
\end{eqnarray}
}
{with a suitable choice of normalization factor ${{\cal N}'}$, 
satisfy again the  required POVM 
condition
}
{
\begin{eqnarray}
\label{LabelEqPOVMKt_i}
\left  [ 
\prod_{i\in {\rm even}}
\left (\int_{-\infty}^{+\infty} dt_i P(t_i)\right )
\right ]
\ 
 ({\hat  {\bm K}}_{\vec {\bf t}})^\dagger
 {\hat  {\bm K}}_{\vec {\bf t}}
 = {\bf 1}.
 \end{eqnarray}
 This 
 follows  from Eq.~\ref{LabelBornRuleProbability} for any  $P(t_i)$ symmetric under $t_i \to - t_i$~\footnote{{the role of $m_i\lambda$ being played by $\tanh(t_i)$}}.}
{
\subsection{Calculation of Observables and Replica Trick\label{SubSecReplicaTrick}
}}
Consider now  a general  measurement average (denoted by an `overbar') of the quantum mechanical expectation of $N$ (potentially different) operators 
${\hat {\cal O}}_1$, ${\hat {\cal O}}_2$, ..., ${\hat {\cal O}}_N$, {where each of these we 
consider here to be a local operator or a product of local operators}. We will compute this average using the Born-rule probability distribution 
{$p_0({\vec m})$, Eq.~\ref{LabelDefBornRuleProb}.}
{We will also assume, for now, that}
{each}
operator
$\hat{\mathcal{O}}_{i}$ 
{commutes}
with {the Kraus operator} ${\hat {\bm K}}_{\vec m}$,
but we will relax this assumption at the end of this section.
Then averaging over measurement outcomes we 
{obtain the measurement-averaged expectation values}
\begin{eqnarray}
\nonumber
 && \overline{[\langle{\hat {\cal O}}_1\rangle_{\vec m} ...
  \langle{\hat {\cal O}}_N\rangle_{\vec m}]}
 \\&&=
\sum_{\vec m} p_0({\vec m)
 {
{\bra 0}{\hat {\bm K}}^\dagger_{\vec m}
{\hat{\cal O}}_1 {\hat {\bm K}}_{\vec m}
{\ket 0} 
...
{\bra 0}{\hat {\bm K}}^\dagger_{\vec m}
{\hat{\cal O}}_N {\hat {\bm K}}_{\vec m}
{\ket 0} 
\over
p_0^N({\vec m})
}}\label{eqn:SetupForMeasurementAvg}
\\\nonumber
&&=
\lim_{R\to 1}
\sum_{\vec m}
\bigg({\bra 0}{\hat {\bm K}}^\dagger_{\vec m}
{\hat{\cal O}}_1 {\hat {\bm K}}_{\vec m}
{\ket 0} 
...
{\bra 0}{\hat {\bm K}}^\dagger_{\vec m}
{\hat{\cal O}}_N {\hat {\bm K}}_{\vec m}
{\ket 0} \times\nonumber\\
 && \hspace{3.5cm} \times[{\bra 0}
{\hat {\bm K}}^\dagger_{\vec m}
{\hat {\bm K}}_{\vec m}
{\ket 0}]^{R-N}\bigg)\nonumber
\end{eqnarray}
{
\begin{eqnarray}
&&=
\lim_{R\to 1}
\sum_{\vec m}
\bigg({\bra 0}{\hat{\cal O}}_1{\hat {\bm K}}^\dagger_{\vec m} {\hat {\bm K}}_{\vec m}
{\ket 0} 
...
{\bra 0} {\hat{\cal O}}_N{\hat {\bm K}}^\dagger_{\vec m} {\hat {\bm K}}_{\vec m}
{\ket 0} \times\nonumber\\
 && \hspace{3.5cm} \times[{\bra 0}
{\hat {\bm K}}^\dagger_{\vec m}
{\hat {\bm K}}_{\vec m}
{\ket 0}]^{R-N}\bigg).
\label{EqObservablesAreCommutedThroughKraus}
\end{eqnarray}}
Note that when $N=1$ 
the last factor in the above equation
disappears since $(R-N )\to 0$ in the required 
{$R \to 1$
limit.} Thus{,} the average 
{of a single expectation value of an operator or of a product of operators ${\hat {\cal O}}_1$ (such as e.g. those appearing in a 2-point function)}
is unaffected by measurements, 
\begin{equation}\label{eqn:bornruleavgN=1}
\begin{split}
\overline{[\langle{\hat {\cal O}}_1\rangle_{\vec m}]}=
\sum_{\vec m}
{\bra 0}&
{\hat{\cal O}}_1 
{\hat {\bm K}}^\dagger_{\vec m}
 {\hat {\bm K}}_{\vec m}
{\ket 0} 
=
{\bra 0}
{\hat{\cal O}}_1 
{\ket 0} {,}
\end{split}
\end{equation}
where the last equality follows from {the POVM condition,}
{
Eq.~\ref{LableEq-POVM-Condition}.}
Coming back to Eq. \ref{EqObservablesAreCommutedThroughKraus}, if we replicate the Hilbert space $R$ times, Eq. \ref{EqObservablesAreCommutedThroughKraus} can be written as,
\begin{eqnarray}
 && \overline{[\langle{\hat {\cal O}}_1\rangle_{\vec m} ...
  \langle{\hat {\cal O}}_N\rangle_{\vec m}]}
\nonumber \\ \nonumber
&&=\lim_{R\to 1}\sum_{\vec m} \prescript{R \otimes}{}{\bra{0}} \mathcal{O}_{1}^{(1)}\mathcal{O}_{2}^{(2)}\dots\mathcal{O}_{N}^{(N)} ({\hat {\bm K}}^\dagger_{\vec m}
 {\hat {\bm K}}_{\vec m})^{\otimes R}\ket{0}^{\otimes R}\nonumber\\
 &&
 =\lim_{R\rightarrow 1}
\text{Tr}(\hat{\mathcal{O}}_{1}^{(1)}\hat{\mathcal{O}}_{2}^{(2)}\dots\hat{\mathcal{O}}_{N}^{(N)}(\ket{0}\bra{0})^{\otimes R}\sum_{\vec m}({\hat {\bm K}}_{\vec m}^{\dagger}{\hat {\bm K}}_{\vec m})^{\otimes R}).\nonumber\\\label{eqn:Nthmomentsetup}
\end{eqnarray}
 {Here,}
 the trace `$\text{Tr}$' is now performed in the replicated Hilbert space{,} and superscripts on the operators indicate 
 {which Hilbert space factor, in the $R$-fold tensor product Hilbert space, they act on.}
 {For {the} measurement protocol
 discussed in subsection \ref{subsec:MeasProto1},}
{
after 
{``softening" the measurement outcomes to take on continuous values,} 
we can replace $\hat{\mathbf{K}}_{\vec{m}}$ in Eq. \ref{eqn:Nthmomentsetup} by $\hat{\mathbf{K}}_{\vec{t}}$ in Eq. \ref{LabelEqKrausKti}, and also replace the sum $\sum_{\vec m}$ by the 
integral {over $t_{i}$}
{as in}
Eq. \ref{LabelEqPOVMKt_i}}. {Thus{,} we will make the following substitution in Eq. \ref{eqn:Nthmomentsetup}
\begin{equation}\label{Eq:SofteningReplicaVersion}
\sum_{\vec{m}} ({\hat {\bm K}}_{\vec m}^{\dagger}{\hat {\bm K}}_{\vec m})^{\otimes R}\rightarrow \left  [ 
\prod_{i\in {\rm even}}
\left (\int_{-\infty}^{+\infty} dt_i P(t_i)\right )\right ] ({\hat {\bm K}}_{\vec t}^{\dagger}{\hat {\bm K}}_{\vec t})^{\otimes R}.
\end{equation}}
{Moreover, since $\hat{\boldsymbol{E}}_{i}$ is an hermitian operator,
{we have}
${{\bm K}}_{\vec t}^{\dagger}={{\bm K}}_{\vec t}$,
and using Eq. \ref{LabelEqKrausKti}, we can 
write}
{
\begin{eqnarray}\label{Eq:K_tDAGGERK_tReplica}
    ({\hat {\bm K}}_{\vec t}^{\dagger}{\hat {\bm K}}_{\vec t})^{\otimes R}=\frac{1}{(\mathcal{N'})^{RL}}\exp{\left\{2\sum_{i= \text{even}} t_{i}\left(\sum_{a=1}^{R}\hat{\boldsymbol{E}}_{i}^{(a)}\right)\right\}}.\nonumber\\
\end{eqnarray}}
{As discussed in Section \ref{subsec:MeasProto1}, $P(t_i)$ is a symmetric distribution under $t_{i}\rightarrow-t_{i}$. 
{Here} $\{t_{i}\}_{i\in\rm 
even
}$ are independent random variables with joint distribution $\tilde{P}(\{t_{k}\})=\prod_{i
\in {\rm 
even
}
}P(t_i)$, and the first and second moments of the distribution are given by}
\begin{eqnarray}
\overline{t_{i}} =0, 
\quad
\overline{t_{i} t_{j}}
= 2  \tilde{\Delta} \delta_{i,j}\label{eqn:tcumulants}.
\end{eqnarray}
{{Here,} $\tilde{\Delta}$
    quantifies the strength of the measurements, and we assume that higher 
{cumulants}
of $P(t_{i})$} {vanish; and
they will be shown to {\textit{not} change the physics at long distances} in App. \ref{Appendix:DetailsOfReplicaSummation}.}
{Then  using Eq. \ref{Eq:K_tDAGGERK_tReplica} and Eq. \ref{eqn:tcumulants}
{we obtain using the cumulant expansion}
\footnote{We note that the normalization of the distribution $P(t_i)$ is chosen such that it satisfies Eq. \ref{LabelEqPOVMKt_i} and hence it is \textit{not} normalized as a probability distribution. However, multiplying and dividing $P(t_i)$ by an appropriate overall trivial constant we can use the formula for cumulant expansion, which is valid for probability distributions.}
\begin{align}
    \int_{-\infty}^{\infty}&\big(\prod_{i
    =
    \text{even}}d t_{i}\,{P}(t_{i})\big)({\hat {\bm K}}_{\vec t}^{\dagger}{\hat {\bm K}}_{\vec t})^{\otimes R}
    \propto
    \nonumber\\
    &
    \propto
    \exp{\bigg\{\tilde{\Delta}\sum_{i=\text{even}}4\left(\sum_{a=1}^{R}\hat{\boldsymbol{E}}_{i}^{(a)}\right)^2\bigg\}}\label{EqIntermediateCumulant}\\
    &
    \propto
\exp{\bigg\{4\tilde{\Delta}\sum_{i=\text{even}}\sum_{\substack{a,b=1\\a\neq b}}^{R}\hat{\boldsymbol{E}}_{i}^{(a)}\hat{\boldsymbol{E}}_{i}^{(b)}\bigg\}} \label{Eq:MeasurementsSummedonLattice}
\end{align}
 where {i}n Eqs. \ref{EqIntermediateCumulant} and Eq. \ref{Eq:MeasurementsSummedonLattice}, we have dropped 
{unimportant overall multiplicative constants and consequently}
replaced the equality signs by proportionality signs, and in Eq. \ref{Eq:MeasurementsSummedonLattice} we have used $\hat{\boldsymbol{E}}_{i}^2=1$ (see Eq. \ref{eqn:boldsymbol{E}}).}
{Using Eqs. \ref{eqn:Nthmomentsetup}, \ref{Eq:SofteningReplicaVersion} and \ref{Eq:MeasurementsSummedonLattice}, we can 
{then} write the measurement averaged moments of expectation values as
\begin{equation}
\begin{split}
 \overline{[\langle{\hat {\cal O}}_1\rangle_{\vec m} ...
  \langle{\hat {\cal O}}_N\rangle_{\vec m}]}&\propto  \lim_{R\rightarrow 1}\text{Tr}\bigg(\hat{\mathcal{O}}_{1}^{(1)}\hat{\mathcal{O}}_{2}^{(2)}\dots\hat{\mathcal{O}}_{N}^{(N)}\times\\
 \times (\ket{0}&\bra{0})^{\otimes R} \exp{\bigg\{4\tilde{\Delta}\sum_{i=\text{even}}\sum_{\substack{a,b=1\\a\neq b}}^{R}\hat{\boldsymbol{E}}_{i}^{(a)}\hat{\boldsymbol{E}}_{i}^{(b)}\bigg\}}\bigg)\label{eqn:NthmomentsetupFINALProtocol1}
\end{split}
\end{equation}
}
\par
{In the derivation of Eq. \ref{eqn:NthmomentsetupFINALProtocol1}, we assumed that operators $\hat{\mathcal{O}}_{i}$ commute with the Kraus operator ${\hat {\bm K}}_{\vec m}$~\footnote{{For the O'Brien-Fendley chain, we will consider calculating measurement averaged moments of the correlation function for operators $\hat{\sigma}^{z}_i$ and $\hat{\boldsymbol{E}}_{j}$. As noted earlier, the operators $\hat{\boldsymbol{E}}_{j}$ at even sites $j$ commute with each other, and hence the operator $\hat{\boldsymbol{E}}_{j}$ at an even site $j$ also commutes with the Kraus operator $\hat{\boldsymbol{K}}_{\vec{m}}$ (see Eq. \ref{LabelEqKrausKti}). 
Moreover, if we choose the operator $\hat{\sigma}^{z}_i$ to lie on an odd sites $i$, it commutes with $\hat{\boldsymbol{E}}_{j}=\frac{1}{\sqrt{2}}(\hat{\sigma}^{z}_{j+1}\hat{\sigma}^{z}_{j}-\hat{\sigma}^{x}_j)$ for all even sites $j$ (see Fig. \ref{fig:evenlinks}) and thus, it also commutes with the Kraus operator $\hat{\mathbf{K}}_{\vec{m}}$.
Therefore, the positions of operators for which we study the correlation functions in this paper can always be slightly ``tuned" such that they commute with the Kraus operator $\hat{\mathbf{K}}_{\vec{m}}$}}.
We will close this section by discussing the case when this assumption is not satisfied.}
{See also the discussion at the end of App.~\ref{Appendix:DetailsOfReplicaSummation}.}
{Since
{each} operator $\hat{\mathcal{O}}_{i}$ is {either} a `local'
{operator}
or a product of `local' operators, it will commute with most 
{Kraus operators}
$\hat{K}_{j,m_j}$ in
{the product} ${\hat {\bm K}}_{\vec m}=\prod_{j}\hat{K}_{j,m_{j}}$, and it might \textit{not} commute with only a few $\hat{K}_{j,m_{j}}$ which have support on the same sites $j$ as the operator $\hat{\mathcal{O}}_i$. 
{We expect such local commutator terms of operator $\mathcal{\hat{O}}_{i}$ to generically be subleading in scaling dimension~\footnote{We note that in Ref. \cite{MurcianoSalaYueMongAlicea} they have made a similar observation in a related context.}, such that the leading order long distance behavior of the measurement averaged moments of {the ground state} expectation value {of 
operator
}$\hat{\mathcal{O}}_{i}$ is still given by Eq. \ref{eqn:NthmomentsetupFINALProtocol1}}.
}
\vskip .8cm
{
\section{Field Theory Representation and measurement RG Fixed Point\label{sec:FieldTheoryandFixedPoint}}
}
\subsection{Field Theory Representation\label{SubSection:FieldTheoryRepresentation}}
\begin{figure}
\centering
\includegraphics[width=0.45\textwidth]{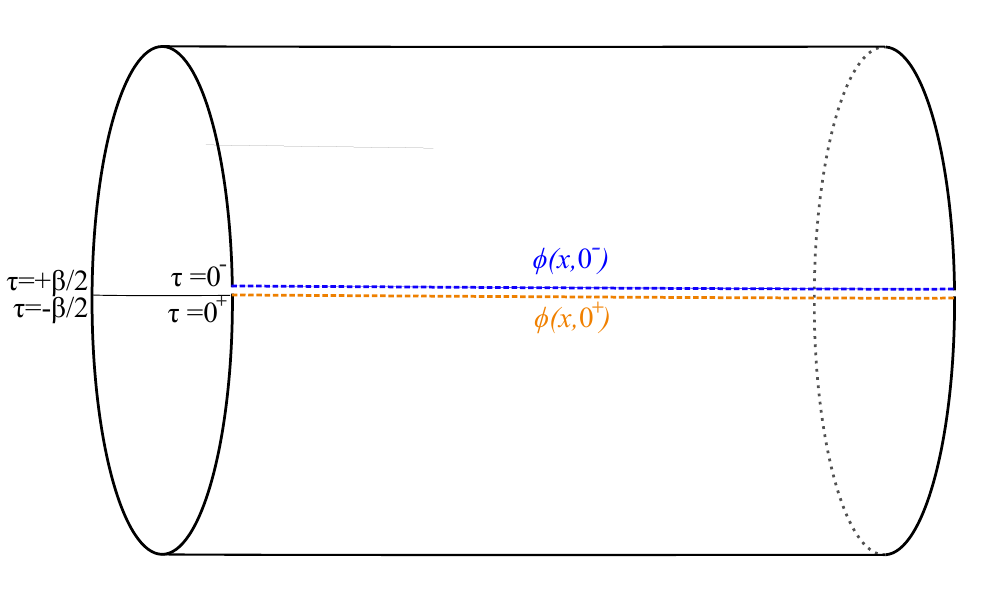}
\caption{The action in the path integral of Eq. \ref{eqn:vaccumpathint} is defined on the above
{cut cylinder} geometry. The direction along the axis and 
{along} the circumference of the cylinder are labeled by space coordinate $x$ and 
imaginary time $\tau$, respectively.
Note that the line $\tau=$~$+\beta/2$ is `glued' to $\tau=$~$-\beta/2$ line, and the field configurations on these two lines are identified 
with each other, i.e. $\phi(x,+\beta/2)=$ $\phi(x,-\beta/2)$ for all $x$.}
\label{fig:cylinder}
\end{figure}
In field theory language, the ground state density matrix of the O'Brien-Fendley chain at the
{{Ising}
tricritical point can be written
{as a path integral~\footnote{{Compare also analogous discussions for different systems in Refs. 
\cite{GarrattWeinsteinAltman2022}, \cite{LeeJianXu}
.}} over the cut cylinder shown in Fig.~\ref{fig:cylinder},}
{
\begin{eqnarray}
\ket{0}\bra{0}=&&\lim_{\beta \rightarrow \infty}\frac{e^{-\beta H}}{Z}=\frac{1}{Z}\int D \phi \,e^{-S_*}\ket{\phi(x,0^{-})}\bra{\phi(x,0^{+})}\nonumber\\
\label{eqn:vaccumpathint}\\
S_*=&&\int d \tau \int d x \bigg\{\frac{1}{2}(\partial_{x} \phi)^2+\frac{1}{2}(\partial_{\tau} \phi)^2+
{g^{*}_{\text{3}}\phi^6}
\bigg\},
\label{eqn:TricriticalIsingAction}\\
D\phi=&&\prod_{\tau=-\beta/2}^{+\beta/2}\prod_{x}d\phi(x,\tau),
\nonumber
\end{eqnarray}
}
where
{$\ket{0}$ is the ground state of the tricritical Ising Hamiltonian, and 
$S_*$ is the effective Landau-Ginzburg 
(-Zamolodchikov~\cite{Zamolodchikov1986})
fixed point action of the Ising tricritical point,}
defined on the {2d} {space-(imaginary)time}
geometry 
in Fig. \ref{fig:cylinder}.~\footnote{
{In the present case of two dimensions there is no meaning to a perturbative study of the Landau-Ginzburg action about the Gaussian theory, since the field $\phi$ is dimensionless by naive 
power-counting, and a non-perturbative tool is needed. This is provided in Ref.~\onlinecite{Zamolodchikov1986}
where it is shown that non-perturbative field identifications following from the exact equations of motion are exactly those obtained from the corresponding unitary minimal model CFT.}}
\par
{We can insert the path integral representation from Eq. \ref{eqn:vaccumpathint} into Eq. \ref{eqn:NthmomentsetupFINALProtocol1}, and replace} the \textit{local} operators
{(or products of local operators)}
{${\hat {\cal O}}_i$ with the corresponding continuum fields $\mathcal{O}^{(a_{i})}_{i}(x,0^{-})$
in the respective replica copy $``a_{i}"$.}}
{Following
Eqs.~\ref{eqn:ContinuumLimitofTwoLatticeOperators},
\ref{eqn:boldsymbol{E}}{,} 
we 
can also replace the
measurement operator $\hat{\boldsymbol{E}}^{(a)}_{i}$ in Eq.~\ref{eqn:NthmomentsetupFINALProtocol1} by the corresponding continuum energy scaling operator
$\mathcal{E}^{(a)}$ in replica copy $``a"$.}
{The field $\mathcal{E}$} is expressed in terms of the Landau-Ginzburg field $\phi$ by
\begin{equation}
\begin{split}\label{eqn:EnergyFieldInTermsofPhi}
\mathcal{E}(x,\tau)=\;:\phi^{2}&:(x,\tau){,}
\end{split}
\end{equation}
where `$:\;:$' indicates
{standard `normal ordering'~\footnote{{
subtraction of the singular terms in the operator product expansion [OPE]}}
of the field $\phi^2$~\cite{Zamolodchikov1986}}.
Thus{,} in
{continuum 
{language},
we obtain} {the following expression for {the} averages}
\begin{align}
& \overline{[\langle{\hat {\cal O}}_1\rangle_{\vec m} ...
  \langle{\hat {\cal O}}_N\rangle_{\vec m}]} \propto \nonumber\\ &\lim_{R\rightarrow 1} \int\limits_{\substack{\phi^{(a)}(x,0^{-})\\ =\phi^{(a)}(x,0^{+})}}\left[\prod_{a=1}^{R}D \phi^{(a)}\right]\; \mathcal{O}_{1}^{(1)}\mathcal{O}_{2}^{(2)}\dots\mathcal{O}_{N}^{(N)}
  e^{- \mathbb{S} }
  \label{LabelEqReplicaPathInt}\\&
\text{where, }-\mathbb{S}=
\sum_{a=1}^R
(-1)
{
S_*^{(a)}
}
+
\Delta \int_{-\infty}^{+\infty} dx \
\Phi(x)
\label{eqn:NthMomentReplicaFieldTheory}\\
&\Phi(x) :=
 \sum_{\substack{a,b=1 \\ a\neq b}}^{R}
\mathcal{E}^{(a)}(x,0) \mathcal{E}^{(b)}(x,0)\,\text{and }\,\Delta=(\text{constant})\times \tilde\Delta.\nonumber\\
\label{eqn:Phi(x)}
\end{align}
{
{Due} to the trace in Eq. \ref{eqn:NthmomentsetupFINALProtocol1}, the $\tau=0^-$ and $\tau=0^+$ boundaries of the cut cylinder in Fig. \ref{fig:cylinder} will be glued{,} and the field configurations $\phi^{(a)}(x,0^{-})$ and $\phi^{(a)}(x,0^{+})$ are identified for all replica indices ``$a$" as shown in Eq. \ref{LabelEqReplicaPathInt}~\footnote{Important differences in the gluing of boundary field configurations could occur if the fields $\mathcal{O}_{i}^{(a)}$ cannot be expressed locally in terms of the Landau-Ginzburg field $\phi$ and its normal ordered higher powers. This issue {is} addressed in App. \ref{LabelAppB3LocalityOfFields}}.}\par
{A different perspective to verify the form of defect interaction appearing in Eq. \ref{eqn:NthMomentReplicaFieldTheory} and \ref{eqn:Phi(x)} is to consider symmetries of the system, and in particular 
Kramers-Wannier duality.}
{Note that although our ground state $\ket{0}$ is 
{invariant} under Kramers-Wannier duality,
{an individual quantum trajectory will generally not be invariant under it.}
}
{However, since we average over all measurement {outcomes,}
we expect this symmetry to be restored in an average sense.}
{Thus{,} the K-W symmetry will appear as an 
{average (``weak") symmetry}
of the ensemble of quantum trajectories.} 
{This 
{implies{,} in particular{,}}
that although the total replica action {(in IR)} will be not invariant if we take $\mathcal{E}^{(a)}\rightarrow-\mathcal{E}^{(a)}$ in a single replica, the action will be invariant if we perform the transformation $\mathcal{E}^{(a)}\rightarrow-\mathcal{E}^{(a)}$ for all replica indices $(a)$ simultaneously.}
{The 
{most}
RG relevant 
{perturbation}
{supported on the $\tau=0$ time-slice} 
{and}
{in the presence of}
{this 
{average (``weak") symmetry}}
is of
{the}
form $\mathcal{E}^{(a)}\mathcal{E}^{(b)}$ for replica indices $a\neq b$~{\footnote{Note that a term with equal replica indices of form $\mathcal{E}^{(a)}\mathcal{E}^{(a)}$, {can be evaluated}
using point splitting and the operator product expansion \cite{FriedanQiuShenker1984,BelavinPolyakovZamolodchikov1984,di1997conformal} (OPE)
\begin{equation*}
\mathcal{E}^{(a)}\times\mathcal{E}^{(a)}=I^{(a)}+\mathcal{E}'^{(a)}
 \end{equation*}
{where} {the operator}
{$\mathcal{E}'=$}
$:\mathcal{E} \mathcal{E}:\,
= \,:\phi^4:$, 
{in every replica $``a"$,}
is irrelevant 
{as an operator with support on the 1-dimensional $\tau=0$ time slice at the Ising tricritical point.}}}.}
{Finally{,} we must consider the sum of terms $\mathcal{E}^{(a)}\mathcal{E}^{(b)}$ over all possible pairs of {unequal} replica indices for the action to be symmetric under permutation of replica indices.}
{This gives us Eq. \ref{eqn:NthMomentReplicaFieldTheory} with $\Phi(x)$ in Eq. \ref{eqn:Phi(x)} back~\footnote{{Due to high scaling dimensions, terms with more than four $\mathcal{E}^{(a)}$ fields {and pairwise \textit{unequal} replica indices} are irrelevant under RG at the tricritical Ising 
point.
The term with exactly four $\mathcal{E}^{(a)}$ fields {(with pairwise unequal replica indices)} is relevant at the tricritical Ising point and is less relevant than $\Phi(x)$ in Eq. \ref{eqn:Phi(x)}.
Moreover, we {argue in App. \ref{Appendix:DetailsOfReplicaSummation} and \ref{app:AvoidedLevelCrossingsHigherCumulants} that} this term is {expected to} be irrelevant at the new IR 
fixed point. 
}}.
}

\subsection{{Controlled} Perturbative Renormalization Group Analysis \label{SubSecOnPerturbativeRG}}
{At the
{Ising tricritical point,}
the scaling dimension of the field $\mathcal{E}=\;:\phi^2:$ is
{$X_\mathcal{E} =\frac{1}{5}$}
\cite{FriedanQiuShenker1984}.}
{Thus{,} the RG eigenvalue of the coupling constant $\Delta$ in Eq. \ref{eqn:NthMomentReplicaFieldTheory} is $y_\Delta=1-2X_{\epsilon}=3/5>0$,}
{implying that the perturbation is relevant.}
{To study the effect of this perturbation, we will use}
{a perturbative RG 
analysis,
controlled by a small parameter $\epsilon$.}
{To obtain such a small parameter $\epsilon$, 
{we will consider}
the following 
generalization 
of the  
action 
in Eq. \ref{eqn:NthMomentReplicaFieldTheory},
{where}}
{
\begin{equation}\label{LabelEqGeneralmReplicaFieldTheory}
\begin{split}
-\mathbb{S}=
\sum_{a=1}^R
(-1)
S_*^{(a)}&
+
\Delta \int_{-\infty}^{+\infty} dx  \;\Phi(x)
\\
\Phi(x)=\sum_{\substack{a,b=1 \\ a\neq b}}^{R}
\mathcal{E}^{(a)}(x,&0)\mathcal{E}^{(b)}(x,0).
\end{split}
\end{equation}
{But now we consider, instead of Eq.~\ref{eqn:TricriticalIsingAction}, 
the more general fixed point described by the
action}
\begin{equation}\label{eqn:GeneralLGZAction}
S_*=\int d \tau \int d x \bigg\{\frac{1}{2}(\partial_{x} \phi)^2+\frac{1}{2}(\partial_{\tau} \phi)^2+
{g^*_{m-1}}
\phi^{2(m-1)}\bigg\},
\end{equation}
{where}
\begin{equation}
 \mathcal{E}=:\phi^{m-2}:\label{eqn:GeneralMEnergy},  
\end{equation}}
\noindent
{and {where}  $m\geq 4$ is an {even} integer.}

{Note that 
{setting}
$m=4$ in the above equations, we recover the problem at hand, i.e. the 
action 
and the field $\mathcal{E}$ given by Eq. \ref{eqn:TricriticalIsingAction} and \ref{eqn:EnergyFieldInTermsofPhi}, respectively}.
{For any  
integer $m\geq 3$ {(even or odd)}, the
fixed point action in Eq. \ref{eqn:GeneralLGZAction} 
{describes exactly~\cite{Zamolodchikov1986}}
the 
multi-critical points famously known as the $m^{\text{th}}$ 
unitary  minimal model conformal field theories (CFTs)
of  central 
charge~\cite{BelavinPolyakovZamolodchikov1984,FriedanQiuShenker1984}} 
\begin{eqnarray}
\label{LabelEqCentralChargeMinimal}
c(m)=1-{6\over m(m+1)}.
\end{eqnarray}
{(The same comment as in
footnote~\cite{Note16} applies to this Landau-Ginzburg action.)}
{We note that for arbitrary integer values $m \geq 3$, the operator $\mathcal{E}$ in Eq. \ref{eqn:GeneralMEnergy} is no longer the `energy' field of the Ising multi-critical point described by the  Landau-Ginzburg action in 
 Eq.~\ref{eqn:GeneralLGZAction} (which would be $:\phi^2:$).
However, we will {restrict ourselves to only \textit{even} values of $m$~\footnote{We will relax this restriction in Sect. \ref{Sec:IsingModel}, where we will consider an odd $m$ minimal model, namely the Ising CFT.}, and} keep using the symbol $\mathcal{E}$ for 
the field in Eq.~\ref{eqn:GeneralMEnergy}
for the following reason.
For the central charges $c(m)$, Eq.~\ref{LabelEqCentralChargeMinimal}, with {\it even} integer values of $m\geq 4${,} there is another critical model with the same central 
{charge, in addition to that described by the action in Eq.~\ref{eqn:GeneralLGZAction}.}
This is the {\it tricritical} $q$-state Potts model, where the value of $q$ is {given~\cite{DotsenkoFateev1984,DELFINO1999537,DengBloteNienhuis2004PRE,Nienhuis1982ExactTricrit,NienhuisBerkerRiedelSchick}
by}
\begin{equation}
\label{LabelEqTricriticalqPottsFctOfm}
\sqrt{q} = 2 \cos{\pi\over m},  \ \ m\geq 4  \  ({\it even}).
\end{equation}
When $q=2${,} this is of course the tricritical Ising model, which is described by the Landau-Ginzburg action
in Eq.~\ref{eqn:TricriticalIsingAction}
above, but for other values of the number $q$ of 
Potts states in Eq.~\ref{LabelEqTricriticalqPottsFctOfm}, e.g. for $q=3$,
it is a
slighly different theory than the one in Eq. \ref{eqn:GeneralLGZAction},
with the same central charge \cite{CardyOperatorContent1986,CappelliItzyksonZuber1987}.
This  will not be of relevance for the observables of interest to us{,} which 
turn out to be present in both theories (see also below). For example and of particular interest to us, when
$m\geq 4$ is even, the operator $\mathcal{E}$ from Eq.~\ref{eqn:GeneralMEnergy} is precisely the same operator as the energy (`thermal') operator in the 
tricritical $q-$state Potts model of the same central charge \cite{DengBloteNienhuis2004PRE,NienhuisBerkerRiedelSchick,Nienhuis1982ExactTricrit}. (In CFT language, that operator is the so-called Kac-Table primary operator $\varphi_{1,2}$ which has the scaling dimension listed in Eq.~\ref{EqScalingDimensionofEnergyNEW} below.)
When $m\to \infty$, the value of $q$ approaches $q=4${,} describing the $q=4$ state
tricritical Potts model{,} which turns out to be the same  as the {critical} (ordinary) $q=4$-state Potts model at central charge $c=1$~\cite{DengBloteNienhuis2004PRE,Nienhuis1982ExactTricrit,NienhuisBerkerRiedelSchick}.}
{Moreover, for even $m${,} all operators that appear when performing repeated operator product expansions of $\mathcal{E}$ with itself are operators present simultaneously 
in both, the 
tricritical q-state Potts model and the Landau-Ginzburg multicritical point described by Eq. \ref{eqn:GeneralLGZAction}~\footnote{{As already mentioned, the energy operator of the tricritical $q$-state Potts model is the so-called Kac-Table operator $\varphi_{1,2}$, and under repeated OPEs with itself, it generates the set of Kac-Table operators $\varphi_{1,n}$, all of which are common to both critical systems. (This set of operators forms an operator algebra closed under the operator product expansion.)}}, 
and all correlation functions of an arbitrary number of $\mathcal{E}$ operators are exactly the same in both systems. 
Since, as we will discuss shortly, we will be interested in computing the RG equation (beta function) for the coupling constant $\Delta$ in the generalized model 
Eq.~\ref{eqn:GeneralLGZAction}
for even values of $m$, 
which is 
uniquely 
determined~\footnote{{in a given RG scheme}} (to arbitrary loop order) by the set of  the correlation functions of an arbitrary number of $\mathcal{E}$ operators (which, as just mentioned, are the same for both systems), we can use either the Landau-Ginzburg 
formulation of Eq.~\ref{eqn:GeneralLGZAction} or  equivalently the tricritical q-state Potts model formulation, both yielding the same result for this  RG 
equation~\footnote{See also the discussion in App.~\ref{app:RGfromOPE}.}.}

{Specifically, we will proceed as follows. We are interested in the properties of the
replica
field theory in Eq.
{\ref{LabelEqGeneralmReplicaFieldTheory}~--~\ref{eqn:GeneralMEnergy}}
when $m=4$, describing the effects of the quantum mechanical measurements on the tricritical Ising ground state, as described in the previous sections. We will  study the generalization of this field theory to large even values of the parameter $m$ which, as already mentioned, provides an expansion parameter $\epsilon$ that is small when $m$ (even) is large. This is a pure field theory problem. 
We will find that for large even values of  $m${,} the field theory in
 Eq. {\ref{LabelEqGeneralmReplicaFieldTheory}}
 will exhibit a fixed point at a non-vanishing value 
 $\Delta_*$ of the coupling constant $\Delta$, controlled by  the parameter $\epsilon=3/(m+1)$, small when the even integer $m$ is large.
 At this fixed point{,} we compute universal  properties (including critical exponents) of a variety of observables
 perturbatively controlled by  the small parameter $\epsilon$. This is the same logic as in the familiar Wilson-Fisher $\epsilon=$
 $4-d$ expansion in dimensions $d$ smaller than $4$.  In contrast, here we always remain in $2=(1+1)$
 dimensions, but we  vary the central charge $c(m)$ by varying the even integer $m$~\footnote{{This is an expansion in $\sqrt{{3\over 2}(1-c(m))}$
 about $c(m)=1$, which can  equivalently  be viewed (as discussed above) as an expansion in
 ${3\over 2\pi} \sqrt{4-q}$ of the tricritical $q$ state Potts model about the $q=4$ state Potts model.}}. (This type of $\epsilon$-expansion within conformal perturbation theory in two dimensions was first performed in Refs. \cite{Zamolodchikov:1987--1/mExpansion,LudwigCardy1987687,LUDWIG198797}
 and subsequently used in many works.).
 This approach allows us to establish that at the 
 {finite-$\Delta_*$} fixed point the system has an
 extremely rich universal scaling behavior (to be discussed in subsequent sections){,} which we can access in a controlled manner perturbatively in $\epsilon$ (in the sense of an $\epsilon$-expansion).  
 Physically, this
 rich scaling behavior originates, as
 $m \to 4$,  from the intrinsic 
 {\it randomness} resulting from the 
 {\it indeterministic}  outcomes of  the quantum mechanical measurements performed on {the} ground state of {the} Ising tricritical point.}
{We now proceed to discuss the RG equation for the coupling constant $\Delta$ in Eq. {\ref{LabelEqGeneralmReplicaFieldTheory}}.
 For an arbitrary even integer 
 {$m\geq 4$}, 
 the scaling dimension of $\mathcal{E}$ in Eq. \ref{eqn:GeneralMEnergy} (with action $S_*$ in Eq. \ref{eqn:GeneralLGZAction}) is {\cite{FriedanQiuShenker1984,di1997conformal}},
\begin{equation}\label{EqScalingDimensionofEnergyNEW}
    X_{\mathcal{E}}={1\over 2} - {3 \over 2 (m+1)}{.}
    \quad 
\end{equation}
Thus{,} the RG eigenvalue of the coupling constant $\Delta$
{is}
\begin{eqnarray}
\label{LabelEqyDelta}
y_\Delta= 
1-X_\Delta
=1 - 2X_{\mathcal{E}} = {3\over (m+1)}\stackrel{\text{def}}{=} \epsilon
\end{eqnarray}
which  is greater than zero, implying that the perturbation is relevant and we will flow away from the
unperturbed fixed point at $\Delta=0$ for any given
{$m$}.}
{To obtain the 1-loop RG equation for the coupling constant $\Delta${,} we will need the OPE of the operator
  $\Phi(x)$ (from Eq.~\ref{LabelEqGeneralmReplicaFieldTheory}
  with $\mathcal{E}$ from Eq.~\ref{eqn:GeneralMEnergy}) with itself (see for example 
Refs.~\onlinecite{JLCardy_1986RGOPE,LUDWIG198797,cardy_1996,LUDWIGWIESE}).
 For any  conformal minimal model with even $m\geq 4$ in Eq. \ref{eqn:GeneralLGZAction}, the fusion rule~\footnote{{As already recalled in footnote {\cite{Note22}}, the operator $\mathcal{E}$ is the Kac-Table operator $\mathcal{E}=\varphi_{1,2}$}} 
of $\mathcal{E}=:\phi^{m-2}:$ with itself is given by,
\begin{equation}\label{eqn:varphi12OPE}
\mathcal{E}\times\mathcal{E}=I+\mathcal{E}'
 \end{equation}
where $\mathcal{E}'=:\phi^{2m-4}:$ is another scaling field in the $m^{\text{th}}$ 
Landau-Ginzburg theory
and it is irrelevant on
the 1-dimensional $\tau=0$ time slice,
for any $m\geq4$~\footnote{The scaling dimension of $\mathcal{E}'$ in 
the $m^{th}$ 
minimal model is $2\big(\frac{m-1}{m+1}\big)>1$ for {any even} $m\geq4$ {(see \cite{FriedanQiuShenker1984,di1997conformal})}
}.
In tricritical $q$-state Potts language where, as mentioned above, $\mathcal{E}$ is the (leading)
energy  operator, $\mathcal{E}'$ is simply the subleading energy  operator.
From Eq.~\ref{LabelEqGeneralmReplicaFieldTheory},\ref{eqn:varphi12OPE}, one obtains the OPE
\begin{eqnarray}\label{EqPhiPhiOPE}
 \nonumber
 &&
 \Phi(x_1) \Phi(x_2) \sim 
 {b \over |x_1 - x_2|} \Phi(x_2) +\dots\,,
 \\ \nonumber
 &&
 {\rm where}
 \\ 
 &&
 b = 4 (R-2),
 \end{eqnarray}
{and}
the 
ellipsis indicates  fields which are irrelevant when supported on the $\tau=0$ time-slice
for $m\geq 4$ (which includes the $m=4$ Ising tricritical case,  Eq. \ref{eqn:NthMomentReplicaFieldTheory}), and can be ignored.
To 1-loop order, the RG equation is then given by 
\cite{JLCardy_1986RGOPE},
\cite{LUDWIG198797},
\cite{cardy_1996}
\begin{eqnarray}\label{eqn:1loopRGdisorderstrength}
{d \Delta
\over d \ell}
= y_\Delta \Delta +  b \Delta^2 + {\cal O}(\Delta^3).
\end{eqnarray}
Thus, when the number of replicas is  $R<2$, there is a new fixed point at a non-vanishing {\it positive} value
\begin{eqnarray}\label{LabelEqOneLoopDelta*}
&&
    \Delta_*=
{-1 \over b} y_\Delta + {\cal O} (y_\Delta^2)
=
\frac{\epsilon}{4(2-R)} + {\cal O} (\epsilon^2) >0\;\;\;
\\ \nonumber
&&
\epsilon=\frac{3}{m+1}.
\end{eqnarray}
We are interested in the limit $R\to 1$ relevant for measurements
(satisfying the Born rule)
\cite{JianYouVasseurLudwig2019,BaoChoiAltman2019}{,}
and
since $\Delta$ describes a second cumulant, we are interested in  a non-negative value $\Delta \geq 0$.}
{Finally, we also note that  the RG analysis for the replica action 
    from Eq. \ref{LabelEqGeneralmReplicaFieldTheory}
    (with action {$S_*$} in Eq. 
    \ref{eqn:GeneralLGZAction} {and $m\geq 4$})
     has been 
     performed to two loop order~\footnote{{in a dimensional regularization  [by $\epsilon=3/(m+1)$] RG scheme, with minimal subtraction of poles in $\epsilon$}}
in Ref. \cite{JengLudwig}, and from 
this analysis we 
obtain the following fixed point coupling up to second order in $\epsilon$,
\begin{equation}\label{Eq:JengLudwig2LoopDelta*}
\begin{split}
\Delta_*=
   \frac{\epsilon}{4(2-R)}+\frac{\epsilon^2}{4(2-R)^2}+ {\cal O}(\epsilon^3),
\end{split}
\end{equation}
a result that will be used further below.
 Note that 
 the 1-loop results in Eq. \ref{LabelEqOneLoopDelta*} and Eq. \ref{Eq:JengLudwig2LoopDelta*} match 
 as they should because the  1-loop contribution $b$ to  the
 RG equation  is independent of the used RG scheme.}
 \vskip .8cm
\section{Correlation Functions\label{sec:CorrelationFunctions}}
{In this section{,} we will discuss measurement-averaged moments of various correlation functions in the ground state of the tricritical O'Brien-Fendley chain.
Following the logic of the preceding section, we will first express these measurement-averaged moments at the Ising tricritical point as correlation functions  of  corresponding fields in  the replica  Landau-Ginzburg field theory describing the  Ising tricritical point as in 
Eqs.~\ref{eqn:TricriticalIsingAction},
\ref{eqn:NthMomentReplicaFieldTheory}, 
\ref{eqn:Phi(x)}. 
Subsequently, we will consider the generalization of these correlation functions to the sequence of field theories with parameter $m\geq 4$ (even) in Eqs.~\ref{LabelEqGeneralmReplicaFieldTheory}, \ref{eqn:GeneralLGZAction}, \ref{eqn:GeneralMEnergy}{,}
which possess a small expansion parameter $\epsilon$ when $m$ is large. We will identify the generalization of the fields of the Ising tricritical point to general values of (even) $m$, and calculate their correlation functions in the replica field theory at the RG fixed {point} controlled by $\epsilon$ (discussed in the preceding section). This provides, analogous to the ordinary Wilson-Fisher $\epsilon = (4-d)$-expansion, a controlled expansion of critical exponents and other universal properties in an expansion in $\epsilon$. Just as in the case of Wilson-Fisher expansion the  case of $d=2$ and $d=3$ dimensions is of  particular interest, of  particular interest to us is the case of $m=4$.}

\vskip .5cm

\subsection{{Measurement-averaged moments of the spin-spin correlation function}
\texorpdfstring{$\overline{\langle \hat{\sigma}^{z}_{i}\hat{\sigma}^{z}_{j}\rangle^N}$}{Lg}\label{SubSecOnSpinCorrelation}}
If we consider two lattice spin operators $\hat{\sigma}^{z}$
{at the Ising tricritical point}
at
sites $i$ and $j$ {(also see footnote \cite{Note13})}, 
we can use Eqs.~\ref{eqn:TricriticalIsingAction}, 
\ref{LabelEqReplicaPathInt},
\ref{eqn:NthMomentReplicaFieldTheory}, 
\ref{eqn:Phi(x)}
to 
{express the measurement-averaged}
$N^{\text{th}}$ moment of their
{ground state}
correlation 
{function in replica field theory language:}
{
At long wavelengths and to {leading} order in scaling dimension, the lattice operator $\hat{\sigma}^{z}$} {is represented~\cite{ZouVidal} at the Ising tricritical point of the O'Brien-Fendley chain
by the spin field $\sigma(x)=\phi$ 
{appearing in the action} Eq.~\ref{eqn:TricriticalIsingAction}.}
{{The $N$-th moment} average $\overline{\langle \hat{\sigma}^{z}_{i}\hat{\sigma}^{z}_{j}\rangle^{N}}$
is then given {in continuum language} by the following correlation function in the replica field theory} {for the Ising tricritical point,
discussed in 
Eqs.~\ref{eqn:TricriticalIsingAction}, 
\ref{LabelEqReplicaPathInt},
\ref{eqn:NthMomentReplicaFieldTheory}, 
\ref{eqn:Phi(x)}}
{
{
\begin{equation}\label{LabelEqSpinMomementsToReplica}
\overline{\langle\hat{\sigma}^{z}_{i}\hat{\sigma}^{z}_{j}\rangle^{N}}
\sim \langle
{\mathfrak{S}}^{{\bm \{}\alpha_i{\bm \}}}(x, 0)\,{\mathfrak{S}}^{{\bm \{}\alpha_i{\bm \}}}(y, 0)\rangle,
\end{equation}
{where}
we {have} defined {${\mathfrak{S}}^{{\bm \{}\alpha_i{\bm \}}}(x, \tau)$ as}
\begin{equation}\label{LabelEqDefMathFrakS}
{\mathfrak{S}}^{{\bm \{}\alpha_i{\bm \}}}(x, \tau)
:=
\left [ \prod_{i=1}^N\sigma^{(\alpha_{i})}(x,\tau)\right ],
\end{equation}}
{and}
$1 \leq \alpha_{i}\leq R$ are  {pairwise distinct 
replica indices in the  $R$-replica theory.}}
{{As $R\rightarrow 1$,}
the physics at long distances is determined by the 
{new, measurement-dominated} fixed point discussed in {the previous} section, {and} {the}
correlation function in Eq. \ref{LabelEqSpinMomementsToReplica} will  asymptotically exhibit power law behavior,}
{
\begin{eqnarray}
    &&\langle
{\mathfrak{S}}^{{\bm \{}\alpha_i{\bm \}}}(x, 0)\,{\mathfrak{S}}^{{\bm \{}\alpha_i{\bm \}}}(y, 0)\rangle\propto
\frac{1}{|x-y|^{2\boldsymbol{X^{(\sigma),R=1}_N}
}}, 
\\ \nonumber
   &&  {\rm as} \ \ |x-y|\rightarrow\infty \;{.}
\end{eqnarray}}
{{Here}
the power law exponent 
{$\boldsymbol{X^{(\sigma),R=1}_N}$}
characterizes the scaling behavior of the measurement averaged $N^{\text{th}}$ moment of the $\langle\hat{\sigma}^{z}_{i}\hat{\sigma}^{z}_{j}\rangle$ correlation function {at the Ising tricritical point}.}\par
{We can evaluate the power law exponent 
$\boldsymbol{X^{(\sigma),R}_N}$
{in an} expansion in $\epsilon=\frac{3}{m+1}$ 
{at the new fixed point $\Delta_*$ 
discussed in the previous Section~\ref{SubSecOnPerturbativeRG},}
by computing the above correlation function 
{with  the}
generalized replica action in 
Eq.~\ref{LabelEqGeneralmReplicaFieldTheory}~--~\ref{eqn:GeneralMEnergy}}.
{To this end, we use the  spin field of the generalization of the Ising tricritical point to the tricritical $q$-state Potts model for
$\sqrt{q} = 2 \cos{\pi\over m}$ with $m\geq 4$ {\it even}, which has scaling dimension~\cite{DengBloteNienhuis2004PRE} 
\begin{equation}
X_{\sigma}=\frac{m^2-4}{8m(m+1)},
\ \ (m\geq 4, \ {\rm even}),
\label{Eq:ScalingDimensionSpinNEW}
\end{equation}
and which we denote by  the same symbol $\sigma(x)$
{as that used above in the tricritical Ising case.} The tricritical Potts spin field is known~\footnote{{
As already mentioned, for even $m\geq 4$
the tricritical $q$-state Potts energy operator
$\mathcal{E}$
corresponds to the Kac-table primary field $\varphi_{1,2}$, 
while the leading tricritical Potts spin field $\sigma$
corresponds to the Kac-Table primary field $\varphi_{m/2, m/2}$~\cite{DengBloteNienhuis2004PRE}. The OPE of these two fields reads $\varphi_{1,2} \times \varphi_{m/2, m/2}=$
$\varphi_{m/2, m/2+1} + \varphi_{m/2, m/2-1}=$
$\varphi_{m/2, m/2} + \varphi_{m/2, m/2+2}$, where in the last equality we have used the symmetry of the Kac Table, $h_{r,s}=$
$h_{m-r, m+1-s}$, and $\varphi_{m/2, m/2+2}$ denotes the subleading tricritical Potts spin field $\sigma'$.}} 
to have a natural OPE with the tricritical Potts energy operator,
\begin{equation}\label{eqn:spinenergyOPEnew}
\sigma \times \mathcal{E} = \sigma + \sigma',
\end{equation}
where the subleading tricritical Potts spin field $\sigma'$ has scaling dimension
$X_{\sigma'}=${$\frac{9m^2-4}{8m(m+1)}$}.
We note 
in passing that the
scaling dimensions of the
tricritical $q$-state Potts spin and subleading spin fields also
match those of the following fields 
in {the} Landau-Ginzburg formulation~\footnote{{The only difference for even $m \geq 6$ is that in the Potts formulation there are two different spin and subleading spin fields, degenerate in scaling dimension. {\cite{CappelliItzyksonZuber1987,di1997conformal}}
This doubling of spin fields in the Potts formulation will be of no relevance for us since we will only be interested in the two-point function of the tricritical $q$-state Potts spin fields.}},
\begin{equation}\label{eqn:GeneralMSpinNEW}
\sigma(x)=:\phi^{\frac{m}{2}-1}:(x)
 \qquad
 \sigma'(x)=\;:\phi^{3(\frac{m}{2}-1)}:(x) .
\end{equation}
At $m=4$, this gives back {the} spin field and {the} subleading spin field  of the tricritical Ising CFT,
$\sigma(x)=\phi$
{and} $\sigma'(x)=\;:\phi^{3}:$.}

Let us 
{denote, for general even values of $m$,}
the coefficient of {the field} $\sigma$ in the OPE of $\mathcal{E}$ and $\sigma$ 
{in the tricritical $q$-state Potts model}
by $c_{\sigma\mathcal{E}\sigma}$.
{Then, 
the OPE of $\mathfrak{S}^{\{\alpha_{i}\}}$
{(recalling its definition in Eq.~\ref{LabelEqDefMathFrakS})}
with $\Phi(x)$ is given by
{\begin{eqnarray}
    \Phi(x) \cdot \mathfrak{S}^{\{\alpha_{i}\}}(y,0)
 \sim
\frac{N(N-1)c^2_{\sigma\mathcal{E}\sigma}}{|x-y|^{2X_{\mathcal{E}}}} \ \ 
\mathfrak{S}^{\{\alpha_{i}\}}(x,0)+ \dots\nonumber\\ \label{eqn:spinPhiOPEnew}
\end{eqnarray}} 
where the ellipsis indicates} 
{fields 
which, at $\epsilon=0$ ($m=\infty$),
have scaling dimensions different from those of
$\mathfrak{S}^{
\{\alpha_{i} \}
}(x,0)$.}
{{Upon}
{using
{the}
OPE in 
Eq.~\ref{eqn:spinPhiOPEnew},} the 1-loop RG equation for {$\mathfrak{S}^{\{\alpha_{i}\}}(x,0)$}
is given by
{
\begin{align}
    {d g_{\{\alpha_{i}\}}
\over d \ell}
=(1-N&X_{\sigma}) g_{\{\alpha_{i}\}} +   2N(N-1)c^2_{\sigma\mathcal{E}\sigma} \Delta g_{\{\alpha_{i}\}}+\dots, 
\label{LabelEqRG forgalpha}
\end{align}}}
{{where the scaling dimension  $X_{\sigma}$ of the tricritical Potts spin field $\sigma(x)$  was 
recalled in Eq.~\ref{Eq:ScalingDimensionSpinNEW},
and}
we have defined $g_{\{\alpha_{i}\}}$ as the coupling constant for the term 
{$\int dx\,\mathfrak{S}^{\{\alpha_{i}\}}(x,0)$}
when added to the 
{action.}
The 
{ellipsis in the above equation indicates}
not only the higher order terms but also quadratic terms involving couplings other than $g_{\{\alpha_{i}\}}$ and $\Delta$.
{This yields the  decay exponent for the 
{correlation function of the replicated product of tricritical Potts spins (i.e. $\mathfrak{S}^{\{\alpha_{i}\}}(x,0)$)} at the new fixed point $\Delta_*$ of the field theory
Eq.~\ref{LabelEqGeneralmReplicaFieldTheory}~--~\ref{eqn:GeneralMEnergy}
to first order in $\epsilon$,}}
\begin{multline}\label{eqn:spinexp}
     {\langle\mathfrak{S}^{\{\alpha_{i}\}}(x,0)\,
    \mathfrak{S}^{\{\alpha_{i}\}}(y,0)\rangle}\propto
\frac{1}{|x-y|^{2\boldsymbol{X^{(\sigma),R}_N}}},\\
 \boldsymbol{X^{(\sigma),R}_N}=    
     NX_{\sigma} -   
    \frac{N(N-1)c^2_{\sigma
    \mathcal{E}\sigma}}{2(2-R)}  \ \epsilon+
    \mathcal{O}\left(\epsilon^2\right) .      
\end{multline}
{The OPE coefficient $c_{\sigma\mathcal{E}\sigma}$} can also be expanded in powers of 
$\epsilon$ about $\epsilon=0$ {(}corresponding to $m=\infty${)}.
The $m=\infty$ {conformal} minimal model corresponds to the
{4-state}
tricritical Potts model, 
and we can replace the 
{OPE coefficient}
$c_{\sigma\mathcal{E}\sigma}$ in the {above equation} by 
{its value
in the 
 $4$-state
tricritical Potts model \cite{cmp/1104161089}; any 
$\epsilon$-dependence of the OPE coefficient $c_{\sigma\mathcal{E}\sigma}$ will only yield contributions of order
$\mathcal{O}(\epsilon^2)$}
to Eq.~\ref{eqn:spinexp}.
{The OPE coefficient
in the tricritical 4-state Potts model is 
{equal~\cite{cmp/1104161089}
to $c_{\sigma \mathcal{E}\sigma}=$
$\frac{1}{\sqrt{2}}$, yielding}}
\begin{equation}\label{eqn:spin}
\boldsymbol{X^{(\sigma),R}_N}=
    \frac{N}{8}
\bigg[
1-{1\over 3} \left(1+\frac{6(N-1)}{(2-R)}\right ) \ \epsilon
+\mathcal{O}\left ( \epsilon^2\right )
\bigg].
\end{equation}
{
In the limit $R\to 1$ the above 
expression 
provides, as $m \to 4$,  an expansion  in $\epsilon=3/(m+1)$
(in the sense of the $\epsilon$-expansion) of}
the scaling
{dimension}
of the measurement averaged $N^{th}$ moment of  
{the}
$\langle \hat{\sigma}^{z}_{i}\hat{\sigma}^{z}_{j}\rangle$ correlation 
{function at the Ising tricritical point,}
\begin{equation}\label{eqn:spinRone}
\boldsymbol{X^{(\sigma),R=1}_N}=
    \frac{N}{8}
\bigg[
1-{1\over 3} \left(1+6(N-1)\right ) \ \epsilon
+\mathcal{O}\left ( \epsilon^2\right )
\bigg].
\end{equation}
{
Note that this expression shows that for the first moment, $N=1$, the first order correction in $\epsilon$
to the exponent $X_\sigma$  
{(}observed without measurements{)} vanishes, consistent with the
expected absence of corrections to $X_\sigma$ arising from {Born-rule} measurements in the tricritical Ising case, $m=4$, to any order in $\epsilon$ due to the result in Eq.~\ref{eqn:bornruleavgN=1}.
}

\vskip 0.5cm

\subsection{{Measurement averaged moments of {the}  energy-energy correlation function} \texorpdfstring{$\overline{\langle\hat{\boldsymbol{E}}_{i+\frac{1}{2}}\hat{\boldsymbol{E}}_{j+\frac{1}{2}}\rangle^N}$}{Lg}\label{SubSecOnEnergyCorrelation}}
\vskip .2cm
Let us now consider{, again at the Ising tricritical point,} two energy operators $\hat{\boldsymbol{E}}$ (given in Eq. \ref{eqn:boldsymbol{E}}) on links $i+\frac{1}{2}$ and $j+\frac{1}{2}$ of the lattice
{(also see footnote \cite{Note13})}.
We can {again} use 
Eqs.~\ref{eqn:TricriticalIsingAction}, 
\ref{LabelEqReplicaPathInt},
\ref{eqn:NthMomentReplicaFieldTheory}, 
\ref{eqn:Phi(x)}
to 
{express}
the measurement averaged $N^{\text{th}}$ moment of the correlation function $\langle\hat{\boldsymbol{E}}_{i+\frac{1}{2}
}\hat{\boldsymbol{E}}_{j
+\frac{1}{2}
}\rangle$
{in field theory language.}
In continuum {language}, as discussed in Section \ref{section:Measurements}, the lattice operator $\hat{\boldsymbol{E}}$ is given by the energy scaling field $\mathcal{E}$
{at the Ising tricritical point}.
Thus,
from Eq. \ref{eqn:NthMomentReplicaFieldTheory}, the measurement averaged $N^{\text{th}}$ moment 
{of $\langle\hat{\boldsymbol{E}}_{i}\hat{\boldsymbol{E}}_{j
}\rangle$ (as before, we have dropped the lattice offset $+1/2$ on the $\hat{\boldsymbol{E}}$ operators for notational simplicity)
can be written as}
{
\begin{equation}\label{LabelEqEnergyMomementsToReplica}
\overline{\langle\hat{\boldsymbol{E}}_{i}\hat{\boldsymbol{E}}_{j}\rangle^{N}}
\sim \langle
{\mathfrak{E}}^{{\bm \{}\alpha_i{\bm \}}}(x, 0)\,{\mathfrak{E}}^{{\bm \{}\alpha_i{\bm \}}}(y, 0)\rangle
\end{equation}
where we {have} defined,
\begin{equation}\label{LabelEqDefMathFrakE}
{\mathfrak{E}}^{{\bm \{}\alpha_i{\bm \}}}(x, \tau)
:=
\left [ \prod_{i=1}^N\mathcal{E}^{(\alpha_{i})}(x,\tau)\right ]
\end{equation}}
{and $\alpha_{i}$ are pairwise distinct replica indices in the R-replica field theory.} Again, as $R\rightarrow 1$,
the physics at long distances is determined by the 
new, measurement-dominated fixed point discussed in {the previous} section, {and} {the}
correlation function in Eq.~\ref{LabelEqEnergyMomementsToReplica}
will  asymptotically exhibit power law behavior,
{
\begin{eqnarray}
&&
\langle
{\mathfrak{E}}^{{\bm \{}\alpha_i{\bm \}}}(x, 0)\,{\mathfrak{E}}^{{\bm \{}\alpha_i{\bm \}}}(y, 0)\rangle
\propto
\frac{1}{|x-y|^{2\boldsymbol{X^{({\cal E}),R=1}_N}
}}, 
\\ \nonumber
   &&  {\rm as} \ \ |x-y|\rightarrow\infty \;{.}
\end{eqnarray}}
{{Here}
the power law exponent 
{$\boldsymbol{X^{({\cal E}),R=1}_N}$}
characterizes the scaling behavior of the measurement averaged $N^{\text{th}}$ moment of the
$\langle\hat{\boldsymbol{E}}_{i
}\hat{\boldsymbol{E}}_{j
}\rangle$
correlation function {at the Ising tricritical point}.}

{{Analogous
to the discussion of the   moments of the
spin operator
in the preceding subsection, we can evaluate
the power law exponent 
$\boldsymbol{X^{({\cal E}),R}_N}$
 in an expansion in $\epsilon=\frac{3}{m+1}$ 
 at the new fixed point $\Delta_*$ 
discussed in the previous Section~\ref{SubSecOnPerturbativeRG},
by computing the above correlation function 
 with  the
generalized replica action in 
Eq.~\ref{LabelEqGeneralmReplicaFieldTheory}~--~\ref{eqn:GeneralMEnergy}}.}
{Unlike the product 
{$\mathfrak{S}^{\{\alpha_{i}\}}$}}
of replicated spin fields {in Eq.~\ref{LabelEqDefMathFrakS}}, 
{the product $\mathfrak{E}^{\{\alpha_{i}\}}$
of replicated energy fields} 
{doesn't turn out  to be
\cite{LUDWIG1990infinitehierarchy,JengLudwig}}
a scaling operator at the new RG fixed point ({even to 1-loop order}).
The {corresponding} scaling operators at the new fixed point 
{are instead}
given by a linear
{superposition of}
{$\mathfrak{E}^{\{\beta_{i}\}}$ with different possible sets of replica indices $\{\beta_{1},\,\cdots,\,\beta_{N}\}$.}
{These} scaling operators at the {new fixed point} transform in irreducible representations 
{of the permutation group $S_R$ of the $R$ replicas}
{(as introduced in Ref. \cite{LUDWIG1990infinitehierarchy}).}
Thus, the correlation function 
{$\langle
{\mathfrak{E}}^{{\bm \{}\alpha_i{\bm \}}}(x, 0)\,{\mathfrak{E}}^{{\bm \{}\alpha_i{\bm \}}}(y, 0)\rangle$}
will be expressed as a sum of power laws
{which, at large distances,
turn out to be dominated~\cite{LUDWIG1990infinitehierarchy,JengLudwig}
by}
the leading {(smallest)}
scaling dimension in the sum. 
Details are provided in App. \ref{appendix:IrreducibleRepresentation}. In a theory with $R$ replicas, we
{obtain}
\begin{equation}
    {\langle
{\mathfrak{E}}^{{\bm \{}\alpha_i{\bm \}}}(x, 0)\,{\mathfrak{E}}^{{\bm \{}\alpha_i{\bm \}}}(y, 0)\rangle}\propto\frac{1}{|x-y|^{2\boldsymbol{X^{(\mathcal{E}),R}_N}}}, \;\;\; |x-y|\rightarrow \infty
\end{equation}
{where} for the case of Born rule measurements ($R\rightarrow 1$),
\begin{subequations}\label{eq:EnergyMoments} 
\begin{eqnarray}
&& \boldsymbol{X^{(\mathcal{E}),R=1}_{N=1}}=\frac{1}{2}-\frac{\epsilon}{2}+\mathcal{O}(\epsilon^3)\label{eqn:bornmeasurementvarphi12I}\\
&&\boldsymbol{X^{(\mathcal{E}),R=1}_{N>1}}=\frac{N}{2}\bigg[1+\epsilon-(3N-5)\epsilon^2+\mathcal{O}(\epsilon^3)\bigg]. \qquad\label{eqn:bornmeasurementvarphi12II}
\end{eqnarray}
\end{subequations}
{Analogous to the case of {the moments of}
the spin operator, as $m\rightarrow 4$, the above {expression for} $\boldsymbol{X^{(\mathcal{E}),R=1}_N}$ provides an expansion
{in $\epsilon=3/(m+1)$ (in the sense of the $\epsilon$-expansion) of}
the scaling dimension of the measurement averaged $N^{\text{th}}$ moment of the $\langle\hat{\boldsymbol{E}}_{i
}\hat{\boldsymbol{E}}_{j
}\rangle$ 
correlation function at the Ising tricritical point.} 

{We note that for the first moment, $N=1$, 
the above scaling dimension
matches with the  scaling dimension of {the} {energy operator} 
$X_{\mathcal{E}}$ (from 
      Eq.~\ref{EqScalingDimensionofEnergyNEW})
at the unperturbed fixed point up to second order in $\epsilon=3/(m+1)$.}
  {This is again consistent with  Eq. \ref{eqn:bornruleavgN=1} which implies that, in the tricritical Ising case, $m=4$, there should be no corrections 
      {to $X_{\mathcal{E}}$
      at any order in $\epsilon$  arising from measurements following the Born-rule.}}

{We close this section by noting that it turns out 
\cite{DavisCardy2000}
that  the $N$th moments of the 
correlation function of the {\it subtracted} energy operator describing the deviation from its expectation value in a fixed quantum trajectory, 
\begin{eqnarray}
\label{LabelEqDEFDeltaE}
&&\delta \hat{\boldsymbol{E}}_{i}
:= \hat{\boldsymbol{E}}_{i} -
\langle 
\hat{\boldsymbol{E}}_{i}
\rangle, 
\\ \nonumber
&&
 \langle \delta \hat{\boldsymbol{E}}_{i} \ 
\delta \hat{\boldsymbol{E}}_{j}\rangle=
 \langle \hat{\boldsymbol{E}}_{i} \ 
\hat{\boldsymbol{E}}_{j}\rangle 
-
\langle\hat{\boldsymbol{E}}_{i} \rangle \ 
\langle \hat{\boldsymbol{E}}_{j}\rangle,
 \end{eqnarray}
decay with a {\it single} power law, and not with a sum of different power laws as the $N$th moments listed in Eq.~\ref{LabelEqEnergyMomementsToReplica}. That is, at the Ising tricritical point, the moments of the resulting {\it connected} correlation function of energy operators decay with a single power law,
\begin{eqnarray}
\label{LabelEqScalingMomentsDeltaE}
&&\overline{
{\langle \delta \hat{\boldsymbol{E}}_{i} \ 
\delta \hat{\boldsymbol{E}}_{j}\rangle}^N}
=\overline{
\left [ 
\langle \hat{\boldsymbol{E}}_{i}\hat{\boldsymbol{E}}_{j}\rangle
-
\langle \hat{\boldsymbol{E}}_{i}\rangle 
\langle\hat{\boldsymbol{E}}_{j}\rangle \right ]^N}
\propto
\\ \nonumber
&&
\propto 
{1 \over 
    |x-y|^{
        2\boldsymbol{\tilde{X}^{({\cal E}),R=1}_N}}},
\ \ N \geq 1,
\end{eqnarray}
where the expression for $\boldsymbol{\tilde{X}^{({\cal E}),R=1}_N}$ is that on the right hand side of
Eq.~\ref{eqn:bornmeasurementvarphi12II}, however now valid  for all positive integers $N$, including $N=1$.
In the language of the replica field theory, this is written in the form
\begin{align}
\label{LabelEqMomentsDeltaE}
& {\langle
{\mathfrak{E}}_{NNR}(x, 0)\,{
\mathfrak{E}}_{NNR}(y, 0)\rangle}\propto\frac{1}{|x-y|^{2\boldsymbol{\tilde{X}^{(\mathcal{E}),R=1}_N}}},\\ & \text{ as }\;|x-y|\rightarrow \infty \nonumber
\end{align}
where the field ${\mathfrak{E}}_{NMR}(x, 0)$, transforming in a specific irreducible representation of the permutation group $S_R$ of the replicas, is defined in Eq.~\ref{EqScalingOperatorsIrreducible} of 
App.~\ref{appendix:IrreducibleRepresentation}.
}

\subsection{Multifractality of Scaling Dimensions\label{SubSecOnMultiFractality}}
{We know 
from the POVM condition, Eq. \ref{eqn:bornruleavgN=1},
that the measurement-averaged}
{first moment of} 
{correlation functions 
{such as}
$\langle \hat{\sigma}^{z}_{i}\hat{\sigma}^{z}_{j}\rangle$  and $\langle \hat{\boldsymbol{E}}_{i}\hat{\boldsymbol{E}}_{j}\rangle$ 
{exhibits}
the same power law behavior as 
in the unmeasured ground state of the tricritical Ising Hamiltonian.}
{(This has also been verified within the $\epsilon$ expansion to the order we have evaluated  it - see 
Eqs.~\ref{eqn:spinRone},
~\ref{eqn:bornmeasurementvarphi12I}  above.)}
From Eqs. \ref{eqn:spinRone} and ~\ref{eqn:bornmeasurementvarphi12II},
{however,} 
{one sees}
that for
{each of the two operators $\hat{\sigma}^{z}_i$ and $\delta\hat{\boldsymbol{E}}_{i}$}
we have obtained an infinite number of independent scaling dimensions {which are associated 
{with} {the measurement-averaged} higher
{integer} moments of {their} correlation functions}.
{This is a signature of multifractality.}
{In a fixed quantum trajectory both correlation functions 
$\langle \hat{\sigma}^{z}_{i}\hat{\sigma}^{z}_{j}\rangle$  and $\langle \delta \hat{\boldsymbol{E}}_{i} \delta \hat{\boldsymbol{E}}_{j}\rangle$ are  bounded from above and {are} non-negative
for sufficiently weak measurement strength}~\footnote{{
Since the operators $\hat{\sigma}^{z}_{i}$ and $\hat{\boldsymbol{E}}_{i}$ have eigenvalues $\pm 1$,
in any quantum state
the correlation functions 
$\langle \hat{\sigma}^{z}_{i}\hat{\sigma}^{z}_{j}\rangle$  and $\langle \delta \hat{\boldsymbol{E}}_{i} \delta \hat{\boldsymbol{E}}_{j}\rangle$ are
{both} bounded 
from above and below.
[E.g., the correlation function $\langle \hat{\sigma}^{z}_{i}\hat{\sigma}^{z}_{j}\rangle$ 
{lies in the interval}
$[-1,1]$ 
{(consider $(\hat{\sigma}^{z}_{i}-\hat{\sigma}^{z}_{j})^2$),} and analogously $\langle \delta \hat{\boldsymbol{E}}_{i} \delta \hat{\boldsymbol{E}}_{j}\rangle=\langle \hat{\boldsymbol{E}}_{i}\hat{\boldsymbol{E}}_{j}\rangle-\langle \hat{\boldsymbol{E}}_{i}\rangle \langle\hat{\boldsymbol{E}}_{j}\rangle$ 
{lies in the interval} $[-2,2]$).]
If the given correlation function (either $\langle\hat{\sigma}^{z}_{i}\hat{\sigma}^{z}_{j}\rangle$ or $\langle \delta \hat{\boldsymbol{E}}_{i} \delta \hat{\boldsymbol{E}}_{j}\rangle
$) in the 
{critical} ground state $\ket{0}$ is oscillating between positive and negative values
{with a period consisting  of a certain number of lattice sites}
as we vary $j$, we can choose to look at a subset of sites $j$ for which the correlation function is strictly positive.
Now, if we consider performing weak measurements (small $\lambda$) on the ground state $\ket{0}$, a particular quantum trajectory will be given by $\ket{\psi_{\vec{m}}}=
{{\hat {\bm K}}_{\vec m} {\ket 0}/
\sqrt{p_{0}(\vec{m})
}}$ (see Eq. \ref{LabelEqKrausOperator} and Eq. \ref{EqProductofLocalKraus}). The correlation function 
$\langle\hat{\sigma}^{z}_{i}\hat{\sigma}^{z}_{j}\rangle$ (or $\langle \delta \hat{\boldsymbol{E}}_{i} \delta \hat{\boldsymbol{E}}_{j}\rangle=\langle \hat{\boldsymbol{E}}_{i}\hat{\boldsymbol{E}}_{j}\rangle-\langle \hat{\boldsymbol{E}}_{i}\rangle \langle\hat{\boldsymbol{E}}_{j}\rangle
$) in this quantum trajectory and for a given value of $|j-i|$ will be an analytic function of the measurement strength $\lambda$. 
Since $\lambda=0$ corresponds to performing no measurements at all and given that the correlation function  $\langle\hat{\sigma}^{z}_{i}\hat{\sigma}^{z}_{j}\rangle$ (or $\langle \delta \hat{\boldsymbol{E}}_{i}\delta \hat{\boldsymbol{E}}_{j}\rangle$) was positive-valued in the ground state $\ket{0}$, for sufficiently small values of $\lambda$ the correlation function will also be positive in a given quantum trajectory obtained upon measurements. 
{For sufficiently weak measurement strength $\lambda$, we can thus} 
restrict ourselves to correlation functions $\langle\hat{\sigma}^{z}_{i}\hat{\sigma}^{z}_{j}\rangle$ and $\langle \delta \hat{\boldsymbol{E}}_{i}\delta \hat{\boldsymbol{E}}_{j}\rangle$ which are bounded from above and {are}  non-negative.}}.
{This is in line with 
the  field theory calculation
which represents a controlled perturbative RG calculation in the 
(weak) strength of measurements.}
Then, the $N$th moments  
of these correlation functions
{for integer values of $N$} are known to determine the  entire probability distribution 
{of these correlation functions}
(and are analytic functions of $N$). 
(See e.g. Refs. \onlinecite{LUDWIG1990infinitehierarchy,fellerintroduction,Witten2019,Boas1954EntireFunction,Titchmarsh1939}.)
Thus, the exponents $\boldsymbol{X^{(\sigma),R=1}_N}$
and $\boldsymbol{\tilde{X}^{({\cal E}),R=1}_N}$, while initially defined for integer values of $N$, are in fact defined for 
real values of $N$ (by analytic continuation).
Hence, to
each of the two operators
$\hat{\sigma}^{z}_i$ and $\delta\hat{\boldsymbol{E}}_{i}$ is associated  a continuous spectrum of scaling dimensions, obtained by continuously varying $N$. 

{Moreover, physically, while correlation functions represent random 
observables which are not self-averaging (as also  reflected by the non-linear dependence of
Eqs. \ref{eqn:spinRone} and ~\ref{eqn:bornmeasurementvarphi12II}
on the moment order $N$), their logarithm {\it is} self-averaging, and a cumulant expansion in the logarithm of the correlation function corresponds to a Taylor expansion of $\boldsymbol{X^{(\sigma),R=1}_N}$
and $\boldsymbol{\tilde{X}^{({\cal E}),R=1}_N}$ in $N$ about $N=0$~(see, e.g. Refs. \onlinecite{DERRIDA-PhysRepts1984,LUDWIG1990infinitehierarchy,ZabaloGullansWilsonVasseurLudwigGopalakrishnanHusePixley,LiVasseurFisherLudwig,JianShapourianBauerLudwig}).
{This provides the {\it typical} scaling exponents},} 
\begin{eqnarray}
\label{Eq:SpinEnergyTypicalDefinationNEW}
&&
\boldsymbol{X^{(\sigma),R=1}_{\text{typ}}}=
\lim_{N\rightarrow 0} \frac{\boldsymbol{X^{(\sigma),R=1}_N}}{N},
\\ \nonumber
&&
\boldsymbol{\tilde{X}^{({\cal E}),R=1}_{\text{typ}}}=
\lim_{N\rightarrow 0} \frac{\boldsymbol{\tilde{X}^{({\cal E}),R=1}_N}}{N}
\end{eqnarray}
where
\begin{eqnarray}
\nonumber
&&  
\overline{\log \langle \hat{\sigma}^{z}_{i}\hat{\sigma}^{z}_{j}\rangle}=
-2\boldsymbol{X^{(\sigma),R=1}_{\text{typ}}}\log|x-y|,
\\ \nonumber
&&
\overline{\log \langle 
\delta \hat{\boldsymbol{E}}_{i}
\delta \hat{\boldsymbol{E}}_{}
\rangle}=
    -2\boldsymbol{\tilde{X}^{({\cal E}),R=1}_{\text{typ}}}\log|x-y|.
\end{eqnarray}
{Specifically,}
Eq.~\ref{eqn:spinRone},
\begin{equation}\label{Eq:SpinTypical}
\boldsymbol{X^{(\sigma),R=1}_{\text{typ}}}=
   \frac{1}{8}\left[1+\frac{5}{3}\epsilon+\mathcal{O}(\epsilon^2)\right],
\end{equation}
and 
Eq.~\ref{eqn:bornmeasurementvarphi12II},
\begin{align}
& \boldsymbol{\tilde{X}^{(\mathcal{E}),R=1}_{\text{typ}}}
=\frac{1}{2}\left[1+\epsilon+5\epsilon^2+\mathcal{O}(\epsilon^3)\right],
\label{Eq:EnergyTypical}
\end{align}
{provide, as $m\to 4$, an expansion in $\epsilon=3/(m+1)$ of the scaling exponents of,
respectively, the typical spin-spin and  the typical {(connected)} energy-energy correlation function with Born-rule measurements on the ground state of the tricritical Ising Hamiltonian.}
\subsection{Logarithmic Correlation Functions\label{SubSec:LogarithmicCorrelationFunctions}}
   Until now, we have used the replica field theory formalism with
   an arbitrary 
   number $R$ of replicas and its $R\rightarrow1$ limit to evaluate various quantities averaged with Born-rule probabilities (see Eq. \ref{eqn:Nthmomentsetup}).
   {In taking the $R \to 1$ limit, the scaling dimensions of two operators which are distinct when $R \not = 1$ 
   {may become equal} at $R=1$.}
   Such a collision of scaling dimensions while taking replica limits can give rise to ``logarithmic correlation functions'' at the measurement-dominated fixed point. {The corresponding
   logarithmic factors multiplying the power law decay of certain correlation functions at criticality are a hallmark of so-called logarithmic CFTs, which are  a class of non-unitary CFTs.}
   (See e.g. Ref. 
   \onlinecite{Gurarie1993,GurarieLudwig2005,VasseurJacobsenSaleur,CardyLogarithm1999,DavisCardy2000,CardyLogarithm2013}).
{We demonstrate in the present subsection that the indeterministic (random) nature of measurement outcomes performed on a critical ground state generates critical states that carry these hallmarks of logarithmic CFTs.}
    In this section, we address these logarithmic CFT features of the measurement-dominated fixed point, and
    in particular, highlight correlation functions which contain {multiplicative} logarithms of distance on top of a power law decay. \par
    As discussed in Sect. \ref{SubSecOnEnergyCorrelation}, the scaling operators at the new fixed point $\Delta_{*}(\epsilon)$ (Eq. \ref{Eq:JengLudwig2LoopDelta*}) transform in irreducible representations of the symmetric group $S_R$ and the correlation functions of such operators exhibit a pure power law decay at the new fixed point.
    Let us consider two operators $\mathcal{O}$ and $\tilde{\mathcal{O}}$ both {transforming} in irreducible representations of {the} symmetric group $S_R$ s.t. the two correlation functions $\langle \mathcal{O} \mathcal{O}\rangle$ and $\langle \tilde{\mathcal{O}} \tilde{\mathcal{O}}\rangle$ are pure power law decaying at the new fixed point $\Delta_{*}(\epsilon)$, 
    i.e.
\begin{eqnarray}
    \langle \mathcal{O}(x,0) \mathcal{O}(y,0)\rangle=\frac{A(R)}{|x-y|^{2X(R)}},\\
    \langle \tilde{\mathcal{O}}(x,0) \tilde{\mathcal{O}}(y,0)\rangle=\frac{\tilde{A}(R)}{|x-y|^{2\tilde{X}(R)}}.
\end{eqnarray}
{Let us} suppose that the correlators have colliding scaling dimensions in the replica limit $R\rightarrow 1$, 
\begin{equation}
    \lim_{R\rightarrow 1} (\tilde{X}(R)-X(R))=0 \;\;\;(\text{but }\tilde{X}(R)\neq X(R) \text{ if } R\neq1 ),
\label{EqCriterionForLogarithmicCorrelator1}
\end{equation}
and that the 
{amplitudes} 
$A(R)$ and $\tilde{A}(R)$ of the correlators can be normalized s.t.
\begin{eqnarray}
    \lim_{R\rightarrow 1} (\tilde{A}(R)+A(R))=\mathcal{K},\label{EqCriterionForLogarithmicCorrelator2}
\end{eqnarray}
{where $\mathcal{K}$ is a constant}.\\
Then as $R\rightarrow 1$, we can write,
\begin{eqnarray}
    \tilde{A}(R)&&=-A(R)+
    \mathcal{K}\label{LabelEqNormalizationLogCorr}\\
    \tilde{X}(R)&&=X(R)+a(R-1)\label{LabelEqCollidingScalingDim}
\end{eqnarray}
where 
$a\neq 0$ 
{is also a}
constant independent of $R$.\\
Then in the limit $R\rightarrow 1$,
\begin{eqnarray}
   && \langle \tilde{\mathcal{O}}(x,0) \tilde{\mathcal{O}}(y,0)\rangle+ \langle \mathcal{O}(x,0) \mathcal{O}(y,0)\rangle=\nonumber\\&&=\frac{\tilde{A}(R)}{|x-y|^{2\tilde{X}(R)}}+\frac{A(R)}{|x-y|^{2X(R)}}\nonumber\\
   &&=\frac{-A(R)+
   \mathcal{K}
   }{|x-y|^{2X(R)+2a(R-1)}}+\frac{A(R)}{|x-y|^{2X(R)}}\nonumber\\
     &&=\frac{1}{|x-y|^{2X(R)}}\big[
     \mathcal{K}
     +2aA(R)(R-1)\log |x-y|+\nonumber\\&&\;\;\;\;\;\;\;\;\;\;\;\;\;\;\;\;\;\;\;\;\;\;\;\;\;\;\;\;\;\;\;\;\;\;\;\;\;\;\;\;\;\;\;\;\;\;\;\;\;+\mathcal{O}((R-1))\big]
\end{eqnarray}
{
Now the last term in the above equation clearly vanishes as $R\rightarrow1$ and therefore we are left with,}
\begin{eqnarray}\label{EqLogarithmicCorrelator}
    \lim_{R\rightarrow 1}\; [\langle \tilde{\mathcal{O}}(x,0) \tilde{\mathcal{O}}(y,0)\rangle+ \langle \mathcal{O}(x,0) \mathcal{O}(y,0)\rangle]=\nonumber\\=\lim_{R\rightarrow 1}\frac{\mathcal{K}+2aA(R)(R-1)\log |x-y|}{|x-y|^{2X({R=1})}}.
\end{eqnarray}
Therefore, in addition to Eq. \ref{EqCriterionForLogarithmicCorrelator1} and \ref{EqCriterionForLogarithmicCorrelator2}, if we have 
\begin{equation}
    \lim_{R\rightarrow 1}A(R)(R-1)=\text{finite}
\end{equation}
or equivalently,
\begin{equation}
    A(R)=\mathcal{O}\bigg(\frac{1}{R-1}\bigg)\label{EqCriterionForLogarithmicCorrelator3}
\end{equation}
we see that the following correlation function will be logarithmic at the new fixed point,
\begin{equation}
     \lim_{R\rightarrow 1}\; [\langle \tilde{\mathcal{O}}(x,0) \tilde{\mathcal{O}}(y,0)\rangle+ \langle \mathcal{O}(x,0) \mathcal{O}(y,0)\rangle].
\end{equation}
\par
In App. \ref{appendix:IrreducibleRepresentation}, we identify two such operators $\mathcal{O}$ and $\tilde{\mathcal{O}}$ which transform irreducibly under the symmetric group $S_R$,
\begin{equation}\label{EqLogarithmicOperatorConstructionO}
    \mathcal{O}=\frac{1}{R-1}\mathfrak{E}_{20R}=\frac{1}{R-1}\sum_{\substack{a,b=1\\a\neq b}}^{R}\mathcal{E}^{a}\mathcal{E}^{b}
\end{equation}
and
\begin{equation}\label{EqLogarithmicOperatorConstructionTildeO}
\begin{split}
\tilde{\mathcal{O}}&=\frac{1}{(R-1)(R-2)}\mathfrak{E}_{22R}\\&=\frac{1}{(R-1)(R-2)}\sum_{\substack{a,b=1\\a\neq b;\,a,b\neq 1,2}}^{R}(\mathcal{E}^{a}-\mathcal{E}^{1})(\mathcal{E}^{b}-\mathcal{E}^{2}).
\end{split}
\end{equation}
Here the field ${\mathfrak{E}}_{NMR}$, transforms in a specific irreducible representation of the {symmetric} group $S_R$ and is defined in Eq.~\ref{EqScalingOperatorsIrreducible} of 
App.~\ref{appendix:IrreducibleRepresentation}.
The normalization factors in front of $\mathfrak{E}_{20R}$ and $\mathfrak{E}_{22R}$ are chosen such that they satisfy  the criterion in Eqs. \ref{EqCriterionForLogarithmicCorrelator2} and \ref{EqCriterionForLogarithmicCorrelator3}.
(See App.~\ref{appendix:IrreducibleRepresentation}.)\\
In App. \ref{appendix:IrreducibleRepresentation}, we show that the scaling dimensions of above two operators (which have unequal scaling dimensions at a generic $R\neq 1$) become equal to each other at $R=1$.
Then, following our analysis above, the correlator {(see App.~\ref{appendix:IrreducibleRepresentation} for its detailed form)}
\begin{equation}
\begin{split}\label{LabelEqLogarithmicCorrelationGeneralm}
    \lim_{R\rightarrow 1}\; [\langle \tilde{\mathcal{O}}(y,0) \tilde{\mathcal{O}}(x,0)&\rangle+ \langle \mathcal{O}(y,0) \mathcal{O}(x,0)\rangle]=\\
  &4\langle \mathcal{E}^{1}(y,0)\mathcal{E}^{1}(x,0)\mathcal{E}^{2}(y,0)\mathcal{E}^{3}(x,0)\rangle\\&-3\langle\mathcal{E}^{1}(y,0)\mathcal{E}^{2}(y,0)\mathcal{E}^{3}(x,0)\mathcal{E}^{4}(x,0)\rangle  
\end{split}
\end{equation}
is logarithmic at all fixed point{s} $\Delta_{*}(\epsilon)$ parameterized by {even} $m$. In particular, as $m\rightarrow 4$, we obtain
{the result} that the following correlator averaged over 
measurements
{with Born-rule probability} should be logarithmic at the Ising tricritical 
point,
\begin{equation}\label{Eq:LogarithmicCorrelationFinal}
    \overline{4\langle\hat{\boldsymbol{E}}_{i}\hat{\boldsymbol{E}}_{j}\rangle \langle \hat{\boldsymbol{E}}_{i}\rangle\langle \hat{\boldsymbol{E}}_{j}\rangle-3\langle \hat{\boldsymbol{E}}_{i}\rangle^2\langle \hat{\boldsymbol{E}}_{j}\rangle^2} \propto \frac{\log|j-i|+\mathcal{O}(1)}{|j-i|^{2\boldsymbol{X^{(\mathcal{E}),R=1}_{N=2}}}},
\end{equation}
where $\boldsymbol{X^{(\mathcal{E}),R=1}_{N=2}}$ is given by Eq. \ref{eqn:bornmeasurementvarphi12II} 
{and $\mathcal{O}(1)$ denotes a constant.}\\
{Finally, we note that in logarithmic CFTs the dilation operator of the scale transformations is not diagonalizable, and the logarithmic correlation functions are associated to Jordan cells of the dilation operator \cite{Gurarie1993,GurarieLudwig2005,CardyLogarithm2013}.
In particular, the Jordan cell or the `logarithmic pair $(C,D)$' associated to the logarithmic correlation function in Eq. \ref{LabelEqLogarithmicCorrelationGeneralm} is formed by the following scaling operators in the $R\rightarrow1$ replica limit,
\begin{equation}
\begin{split}
D=&\mathcal{O}+\tilde{\mathcal{O}}
\\
    \rightarrow\;\;&\mathcal{E}^{1}\mathcal{E}^{2}+\sum_{\substack{\alpha\neq 1}}^{R}\mathcal{E}^{\alpha}\mathcal{E}^{1}+\sum_{\substack{\alpha\neq 2}}^{R}\mathcal{E}^{\alpha}\mathcal{E}^{2}-\sum_{\substack{\alpha,\beta=1\\\alpha\neq \beta}}^{R}\mathcal{E}^{\alpha}\mathcal{E}^{\beta},
\end{split}
\end{equation}
and 
\begin{equation}
    C=(X(R)-\tilde{X}(R))\mathcal{O}\rightarrow-a\times\sum_{\substack{\alpha,\beta=1\\\alpha\neq \beta}}^{R}\mathcal{E}^{\alpha}\mathcal{E}^{\beta}
\end{equation}
}{where {the universal constant $a$ 
is defined in Eq. \ref{LabelEqCollidingScalingDim}}.}
{The correlation functions of the operators $C$ and $D$  are given by,
\begin{equation}
\begin{split}
    \langle D(x,0)D(y,0)\rangle&=\frac{\tilde{A}(R)}{|x-y|^{2\tilde{X}(R)}}+\frac{A(R)}{|y-x|^{2X(R)}},\\
     \langle C(x,0)D(y,0)\rangle&=\frac{A(R)(X(R)-\tilde{X}(R))}{|x-y|^{2X(R)}
     },\\
      \langle C(x,0)C(y,0)\rangle&=\frac{A(R)(X(R)-\tilde{X}(R))^2}{|x-y|^{2X(R)}},
\end{split}
\end{equation}
and thus in the limit $R\rightarrow1$ (see Eq. \ref{EqLogarithmicCorrelator}),
\begin{equation}
\begin{split}
    \langle D(x,0)D(y,0)\rangle
    &\longrightarrow \frac{2\zeta\times(\log|x-y|+\mathcal{O}(1))}{|x-y|^{2\boldsymbol{X^{(\mathcal{E}),R=1}_{N=2}}}},\\
     \langle C(x,0)D(y,0)\rangle&\longrightarrow \frac{-\zeta}{|x-y|^{2\boldsymbol{X^{(\mathcal{E}),R=1}_{N=2}}}},\\
      \langle C(x,0)C(y,0)\rangle&\longrightarrow 0
\end{split}
\end{equation}
where $\zeta:=\lim_{R\rightarrow1}a(R-1)A(R)$.
}
\vskip .8cm
\section{Entanglement Entropies\label{sec:EntanglementEntropy}}
\begin{figure}
\centering
\includegraphics[width=8cm]{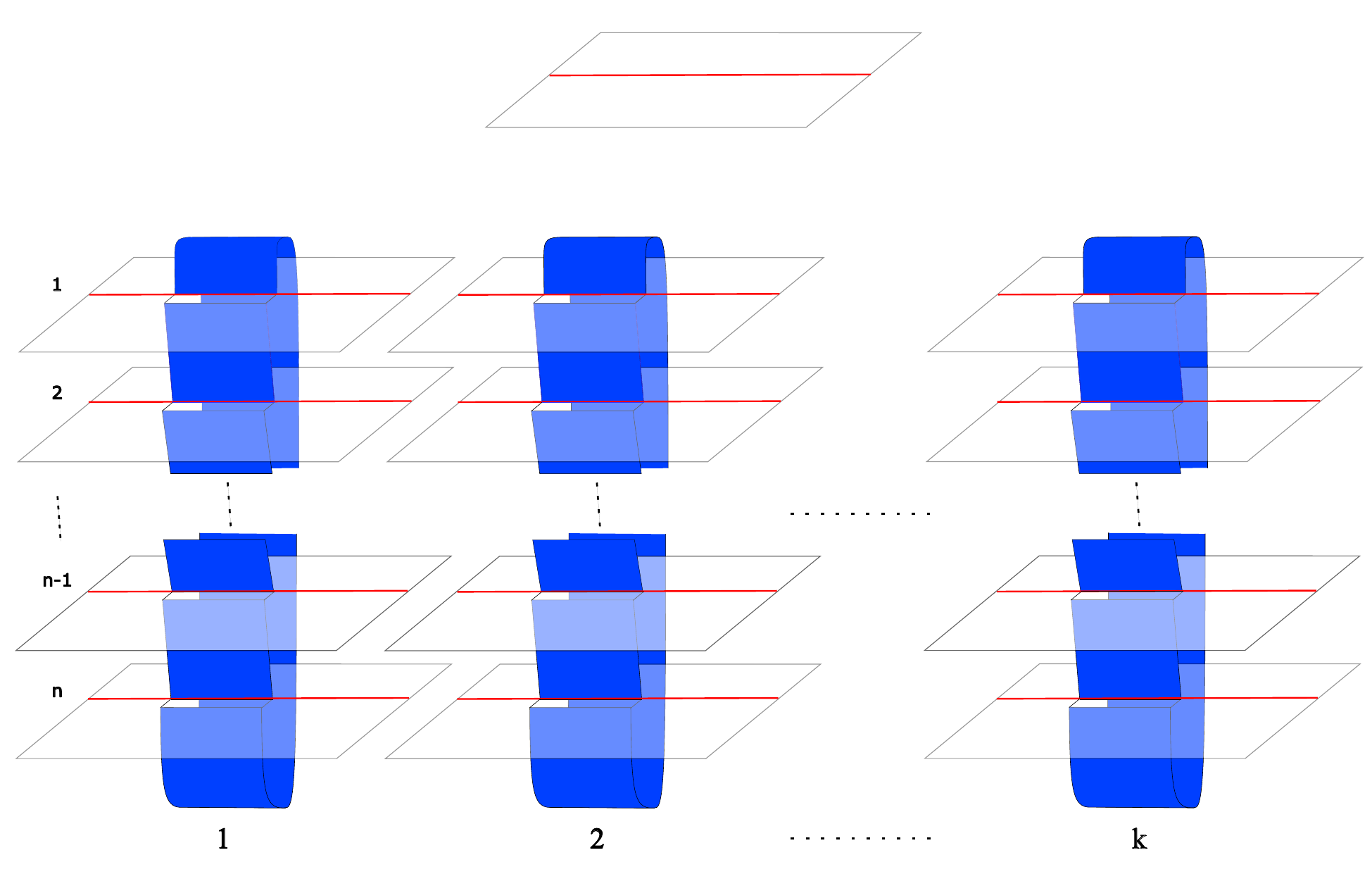}
\caption{Comparing with the Riemann surfaces that appear in the calculation of entanglement entropies for translationally-invariant (non-random) CFTs (see Ref. {\cite{calabrese2004,*calabrese2009entanglement} or \cite{CardyAlvaredoDoyon}), we have $k$ $n-$sheeted Riemann surfaces for the calculation of the measurement averaged $n^{\text{th}}$ R\'enyi entropy.}
In total, we have $R=nk+1$ copies of the theory, where the standalone copy comes from Born-rule probability factor, and all $nk+1$ copies are interacting with each other (but not with themselves) on the defects shown in red lines. 
The explicit form of the interaction is given in App. \ref{appendix:EntanglementEntropy} (Eq. \ref{eqn:eerepl}).\label{fig:EntanglementGluing}}
\end{figure}
{At long wavelengths,} the $n^\text{th}$ R\'enyi entanglement entropy 
{of the ground state of a translationally invariant,
i.e. non-random
$(1+1)d$ CFT}
can be expressed {as the logarithm} of 
{the} correlation function of 
{two $n$-\text{twist fields}}
by considering $n$ copies of the corresponding 2D CFT \cite{calabrese2004,*calabrese2009entanglement}. {(See also {Ref.}~\onlinecite{HolzheyLarsenWilczek1994}.)}
Following our calculations of measurement averaged moments of correlation functions {and, in particular, the measurement averaged logarithm of correlation functions 
in Sect. \ref{SubSecOnMultiFractality}}, 
we 
{will now also}
calculate the {average of {the} logarithm} of the twist field correlation functions 
(which are 
the R\'enyi entropies) using {{the controlled}
perturbative RG expansion}.
{This will involve calculating the correlation function of multiple copies of twist fields
in the generalized replica action given in Eq. \ref{LabelEqGeneralmReplicaFieldTheory}~--~\ref{eqn:GeneralMEnergy}. 
Since the twist fields are geometrical in nature, their generalization to higher $m$ theories appearing in Eq. \ref{eqn:GeneralLGZAction} is natural.}
{This will allow us to evaluate the 
{universal coefficient of the logarithm of subsystem size of the measurement averaged R\'enyi entanglement entropies at the tricritical Ising point in an expansion in $\epsilon=3/(m+1)$.}}
\par
To calculate the $n^{\text{th}}$ R\'enyi entropy, we will have to consider $n$ copies of a state, corresponding to a given set of measurement outcomes, and `glue' the copies to form a $n-$sheeted Riemann surface (see App. \ref{appendix:EntanglementEntropy}).
Moreover, to be able to 
{perform the average over measurement outcomes of the {\it logarithm} of the $n$-twist field correlation function,}
we will have to introduce another replica index $k$. {Overall we need to introduce $R=nk+1$ replica copies, where the additional copy comes due to the Born rule probability factor $p_{0}(m)$
{(analogous to Eq. \ref{eqn:SetupForMeasurementAvg}),}
and the limit $k \rightarrow 0$ or $R \rightarrow 1$ will correspond to the Born-rule measurement averaged R\'enyi entropy(s) \cite{JianYouVasseurLudwig2019}.}
The details of these calculations are given in  
 App. \ref{appendix:EntanglementEntropy} and we 
 {obtain {the following expression}
 for the measurement averaged
$n^{\text{th}}$~R\'enyi entropy} {$\overline{S_{n,A}}$}
\begin{equation}\label{eqn:RenyiEntropyAverageFinal}
\overline{S_{n,A}}=\frac{1}{1-n}\frac{\text{d}}{\text{d}k}\bigg|_{k=0}\langle {\prod_{j=1}^{k}\bigg(\Tau^{(j)}_{n}(u,0)(\Tau^{(j)}_{n})^{-1}(v,0)\bigg)\rangle_{\Delta_*}} .
\end{equation}
{Here, $\Tau^{(j)}_{n}(u,0)$ denotes the twist field,  where}
the superscript $j$ 
{specifies}
which copy of 
{the} Riemann surface (out of $k$ {copies}) the twist field corresponds to{, and} the subscript $n$ signifies that we are dealing with twist fields for a 
{$n$-sheeted}
Riemann surface
{(compare Fig.~\ref{fig:EntanglementGluing})}{.}
{The subscript}
{$\Delta_*$} indicates that the correlation function is evaluated at the 
{measurement-dominated}
fixed point.
\par
{We now evaluate the scaling dimension of the replicated twist field,
occurring in the correlation function {in} Eq.~\ref{eqn:RenyiEntropyAverageFinal},
{in the replica field theory Eq.~\ref{LabelEqGeneralmReplicaFieldTheory}, \ref{eqn:GeneralLGZAction}, \ref{eqn:GeneralMEnergy}
to 1-loop order  in the $\epsilon$ expansion using the OPE.}}
The coefficient 
{of the twist field 
in the OPE of the twist field
with the perturbation $\Phi(x)$}
is calculated in App. \ref{appendix:OPEtwistsPhi} and is equal to $kI_{n}$, where $I_{n}$ is 
{defined as}
\begin{equation}\label{eqn:In}
 I_{n}=\frac{n}{2\pi}\int_{0}^{\infty}\text{d}s\;\frac{1-s^{n-1}}{(1-s)(1+s^n)}+
 {\mathcal{O} \left (\epsilon \right )}.
\end{equation}
Then the 1-loop RG equation for the scaling dimension of {the} twist fields $\big[ \prod_{j}^{k}\Tau^{(j)}_{n}\big ]$
{can be obtained as usual, e.g., from the RG equation for  a coupling constant $g_{(n,k)}$ for the term  $\int dx \big[\prod_{j}^{k}\Tau^{(j)}_{n}\big]$ when added to the action. This yields}
\begin{equation}
    \frac{\text{d}g_{(n,k)}}{\text{d}l}=(1-kd_{n})g_{n,k}+2k\Delta I_{n}g_{n,k}+\dots
\end{equation}
where $d_{n}:=\frac{c}{12}(n-\frac{1}{n})$ is the scaling dimension of twist field  $\mathcal{T}_{n}$ 
{
in the unperturbed CFT
\cite{KniznikTwistFields}, and the 
 ellipsis represents terms  that will end up contributing only to corrections of second order in $\epsilon$.}
Then~\footnote{{We note that the above result for the scaling dimension {of  the $k^{th}$ moment} 
of the $n$-twist field at the new $\Delta_{*}$-fixed point is linear in $k$ to 1-loop order.
{However, in analogy with observations made in 
Ref.~\onlinecite{LiVasseurFisherLudwig} in a related context,
we expect non-linearities in $k$ to appear in higher orders in $\epsilon$.
We plan to address the calculation of these non-linearities  in future work.}}}
\begin{equation}\label{LabelEqReplicatedTwistFieldCorrelationFunction}
\begin{split}
     &\langle \prod_{j=1}^{k}\big(\Tau^{(j)}_{n}(u,0)(\Tau^{(j)}_{n})^{-1}(v,0)\big)\rangle_{\Delta_{*}} \propto\bigg(\frac{u-v}{a}\bigg)^{-2d^{*}_{n,k}}\\
     &\text{where, \ \  }d^{*}_{n,k}=kd_{n}-2k\Delta_{*} I_{n}+\mathcal{O}((\Delta_*)^2). 
\end{split} \qquad 
\end{equation}
{Here,}
we have inserted 
{a}
short distance cutoff $a$ 
to make the final result dimensionless.
The 
{measurement-averaged}
$n^{\text{th}}$ R\'enyi entanglement entropy for an interval 
{of length $l=|u-v|$ is then given, using Eq. \ref{eqn:RenyiEntropyAverageFinal}, by}
\begin{eqnarray}
\overline{S_{n,A}}&&=\frac{c^{(\text{eff})}_{n}}{3} \ln\frac{l}{a}+{\mathcal{O}(1)} \;\;\text{where,} \label{eqn:RenyiEntropyAverageFI}\\
c^{(\text{eff})}_{n}&&= 
c(m) \bigg(\frac{1+1/n}{2}\bigg)-\frac{3 I_{n}}{(n-1)}\epsilon+\mathcal{O}(\epsilon^2),\label{eqn:RenyiEntropyAverageFII}
\end{eqnarray}
{where $c(m)$ is the central charge of the unperturbed theory,
Eq.~\ref{LabelEqCentralChargeMinimal}, 
while $I_n$ is the integral in   Eq.~\ref{eqn:In}.}
To obtain the 
{measurement-averaged}
von Neumann entanglement entropy{,} we 
take {the} limit $n\rightarrow 1$ in the above equation. From Eq. \ref{eqn:In},
{one obtains}
\begin{eqnarray}\label{eqn:derivativeIn}
    \frac{\text{d}I_n}{\text{d}n}\bigg|_{n=1}=\frac{1}{2\pi}\int_{0}^{\infty}\text{d}s\;\frac{\ln(s)}{s^2-1}=\frac{\pi}{8}.
\end{eqnarray}
Thus{,} using Eqs. \ref{eqn:RenyiEntropyAverageFI}, \ref{eqn:RenyiEntropyAverageFII} and  \ref{eqn:derivativeIn}, the 
{measurement-averaged}
von Neumann entropy is
\begin{equation}
\begin{split} \overline{S_{1,A}}=\lim_{n\rightarrow1} \overline{S_{n,A}}
=\frac{
{c^{(\text{eff})}_{n=1}}
}{3}\ln\frac{l}{a}+{\mathcal{O}(1)} ,
\end{split}
\end{equation}
 where ${c^{(\text{eff})}_{n=1}}$  at the
 {measurement-dominated fixed point $\Delta_*$}
 is given by
\begin{equation}
\;\;\;\;\;\;{c^{(\text{eff})}_{n=1}}=
c(m)
-\frac{3\pi}{8}\epsilon+\mathcal{O}(\epsilon^2). 
\end{equation}
{This provides, as $m \to 4$, an expansion in $\epsilon=3/(m+1)$ of the ``effective central charge'' 
at the measurement-dominated fixed point  of the tricritical Ising ground state.}
Note that, unlike 
{in translationally invariant (i.e., unmeasured and thus non-random) CFTs}
\cite{calabrese2004,*calabrese2009entanglement}, 
here the measurement averaged 
{$n^{\rm th}$}
R\'enyi entanglement entropies  $\overline{S_{n,A}}$ are \textit{not} simply
{a $\frac{1}{2}(1+\frac{1}{n})$ 
multiple}
of the von Neumann entropy. 
{Rather, the universal coefficients of the logarithm
{of subsystem size}
are all independent 
{of each other}
for different measurement averaged R\'enyi entanglement entropies.}
{This feature is similar to having a hierarchy of 
{independent} (``multifractal'')  scaling dimensions 
{for measurement-averaged moments of correlation functions of operators,}
as discussed in Section \ref{SubSecOnMultiFractality}.}
{Note that the $n^{\text{th}}$ R\'enyi entropy being 
$\frac{1}{2}(1+\frac{1}{n})$ times the von Neumann entanglement entropy plays a central role in the calculation of entanglement spectrum shown in Ref. \onlinecite{CalabreseLefevre2008} for the usual {(unmeasured)} 1d critical ground states.}
{Thus, we expect 
the entanglement spectrum of the present system under measurements 
to
exhibit qualitatively different universal features when compared to the (unmeasured) 1d critical ground states.}
 \par
If we have our system at finite temperature, the unmeasured state is given by Gibbs 
{state $\rho=\frac{e^{-\beta H}}{Z}$, in contrast to the (pure) critical ground state.}
Following
{steps that parallel}
the pure state 
{calculation above (using twist fields), we find}
that the measurement averaged $n^{\text{th}}$ R\'enyi entanglement entropy can be written as 
\begin{equation}\label{eqn:RenyiEntropyAverageFiniteTemp}
\overline{S_{n,A}}=\frac{1}{1-n}\bigg(\frac{\text{d}}{\text{d}k}\bigg|_{k=0}\frac{\mathcal{Z(\beta)}_{A}}{\mathcal{Z}(\beta)_{\varnothing}}\bigg)
\end{equation}
where now,
\begin{equation}\label{eqn:RenyiEntropyAverageFiniteTempI}
    \begin{split}
    \mathcal{Z_{A}}(\beta)=&\sum_{\vec{\mathbf{m}}}\text{Tr}((K_{\vec{\mathbf{m}}}e^{-\beta H} K_{\vec{\mathbf{m}}}^{\dagger})^{\otimes nk+1}\mathscr{S}^{k}_{n,A})\\
 \mathcal{Z}_{\varnothing}(\beta)=&\sum_{\vec{\mathbf{m}}}\text{Tr}((K_{\vec{\mathbf{m}}}e^{-\beta H}K_{\vec{\mathbf{m}}}^{\dagger})^{\otimes nk+1}).
    \end{split}
\end{equation}
{(See Eq. \ref{EqPermutationOperatorEE} for the definition of $\mathscr{S}^{k}_{n,A}$).} Then {following the discussion in App. \ref{appendix:EntanglementEntropy}},
one 
{verifies}
that the average entanglement entropy is still given by Eq. \ref{eqn:RenyiEntropyAverageFinal}{,} but
{now the twist field correlation function}
in this equation is calculated on a cylinder
{of}
{finite circumference $\beta$} instead of a plane. 
{Since we are interested in evaluating this correlation function 
{of twist-fields}
at the 
{new fixed point $\Delta_*$, which is conformally invariant,}
we can use the 
conformal transformation 
\begin{equation}
    w\rightarrow z=\frac{\beta}{2\pi}\log w
\end{equation}
to map the twist field correlation on the plane (given in Eq. \ref{LabelEqReplicatedTwistFieldCorrelationFunction}) to the twist field correlation function on the cylinder.}
{Thus}{,} the measurement averaged $n^\text{th}$ Renyi entanglement entropy for a region of length $l\equiv|u-v|$ at finite {(inverse)} temperature {$\beta$} is given by
\begin{equation}\label{eqn:RenyiEntropyAverageFFiniteTemp}
\begin{split}
\overline{S_{n,A}}=&\frac{c^{(\text{eff})}_{n}}{3} \ln\bigg(\frac{\beta}{\pi a}\sinh{\frac{\pi l}{\beta}}\bigg)+{\mathcal{O}(1)}, 
\end{split}
\end{equation} 
where {the universal coefficient }$c^{(\text{eff})}_{n}$ {is} given 
{by Eqs.~\ref{eqn:In}, \ref{eqn:RenyiEntropyAverageFII}.}
At {inverse}  temperature {$\beta<<l$},
{this reduces to}
\begin{equation}\label{eqn:HighTempRenyi}
    \frac{\overline{S_{n,A}}}{l}\;{\sim}\;  c^{(\text{eff})}_{n}\frac{\pi}{3}\frac{1}{\beta}.
\end{equation}
\par  
If we take the subsytem size $l$ to approach
{the length $L$ of the total system,}
the universal coefficients $c_{n}^{\text{eff}}$ 
{will} also appear in the  measurement averaged (\textit{not} entanglement) R\'enyi entropies calculated for the {full} mixed state of the system at finite temperatures~\footnote{Since the subsystem size approaches the system size $L$, the twist fields will be sitting at the ends of the one dimensional quantum system and such correlator will no longer be a pure power law, and will depend on the full operator content of the theory. However, the leading order piece proportional to the system size will not depend on the boundary condition and will be the same as in Eq. \ref{eqn:HighTempRenyi}.}. 
At finite temperatures {satisfying the condition} {$\beta<<L$}, the measurement averaged R\'enyi entropy of the  mixed state of the system is
{then} 
given by
\begin{equation}\label{eqn:FullSystemRenyi}
    \frac{\overline{R_{n}}}{L}\sim c^{(\text{eff})}_{n}\frac{\pi}{3}\frac{1}{\beta}+\mathcal{O}\bigg(\frac{1}{\beta^2}\bigg),
\end{equation}
{which is extensive in 
{the total system size $L$.}}
We also note that, since the R\'enyi entropy of the full system is a self-averaging quantity,
{it is thus represented by its average in Eq.~\ref{eqn:FullSystemRenyi}.}
This expression should be contrasted with
{the R\'enyi-index $n$ dependence of 
}
the 
extensive R\'enyi entropies
for 
{the unmeasured, i.e. non-random}
1d quantum critical 
{system}
at thermal equilibrium,
{which is} given by 
\cite{AffleckUniversalFreeEnergy,*CardyNightingaleBlote},
\cite{calabrese2004,*calabrese2009entanglement}
\begin{equation}\label{Eq:FullSystemRenyiClean}
    \frac{R_n}{L}=
{c(m) \cdot  \left [{1\over 2}  (1 + {1\over n})\right] \ \frac{\pi}{3}
\frac{1}{\beta}}
    +\mathcal{O}\bigg(\frac{1}{\beta^2}\bigg),
\end{equation}
{where $c(m)$, Eq.~\ref{LabelEqCentralChargeMinimal},
is the central charge of the corresponding 
{unmeasured} $2D$ CFT.}
{We note that due to the $n$-dependence of the universal coefficient $c_{n}^{\text{eff}}$
{(from Eq.~\ref{eqn:In}, \ref{eqn:RenyiEntropyAverageFII})}
in Eq.~\ref{eqn:FullSystemRenyi},
the leading order finite temperature behavior of the measurement averaged $n^{\text{th}}$ R\'enyi entropy of the \textit{full} mixed state of the system does \text{not} satisfy the simple relation in Eq. \ref{Eq:FullSystemRenyiClean}, which is valid for translationally invariant ({unmeasured,} non-random) CFTs.
}
\vskip .8cm
{\section{Effective ``ground-state degeneracy" \texorpdfstring{$g_{\text{eff}}$}{Lg}\label{Sec:EffectiveGroundStateDegeneracyGeff}}}
{At measurement-induced phase transitions of deep  random quantum circuits with measurements in the bulk of the space-time of the circuit there exists a universal quantity known as the ``effective central charge'' which is defined in terms of  the replica limit $R\to 1$ of the derivative with respect to $R$ of  the universal finite-size correction of the free energy of the circuit on a cylinder or strip of finite circumference or width. In a sense, it replaces the notion of the central charge  which is zero at $R=1$, in the non-unitary CFT describing the transition. The ``effective central charge'' has been shown in Ref.~\onlinecite{ZabaloGullansWilsonVasseurLudwigGopalakrishnanHusePixley,KumarKemalChakrabortyLudwigGopalakrishnanPixleyVasseur}
to represent the universal finite-size scaling behavior of the Shannon-entropy of the measurement record of the circuit, the latter providing an
expression for 
the logarithm of the partition function in the language of the measurements performed on the circuit.
The ``effective central charge'' is   {\it not} equal to the universal coefficient of the logarithm of subsystem size of the  entanglement entropy at measurement-induced transitions in these deep random quantum circuits.}

{In  CFTs with boundary (or defect, before folding
\cite{FendleyLudwigSaleur1995}, \cite{LeclairLudwig}) there is a quantity referred to as  the ``ground-state degeneracy $g$" or ``zero-temperature entropy" $S:= \ln g$ which is a universal constant associated with any specific conformally invariant boundary RG fixed point. It plays a role for boundary (defect) CFTs analogous to the role played by the central charge in a bulk CFT. In particular, in unitary CFTs  it decreases upon boundary RG flows, a property often referred to as the ``g-theorem'' 
\cite{AffleckLudwig1991},\cite{FriedanKonechny}.
The {\it defect} (boundary, after folding) piece 
$\ln Z_{d}$ 
(subscript $d$ standing for ``defect'')
of the logarithm of the partition function of a CFT on a
cylinder
of large length {$\beta$} and  
circumference
{$L\ll\beta$}
has~\footnote{{With periodic boundary conditions in 
{the spatial direction of size $L$,}
we have to consider the CFT with a defect on a torus with radii $\beta$ and $L$, where $\beta$ denotes the inverse temperature (see Fig. \ref{fig:cylinder}). 
{After folding the torus at $\tau=\beta/2$ and at $\tau=0$ (the location of the defect), we obtain a finite 
{`double-sheeted'} cylinder of circumference $L$ and length $\beta/2$. 
The $\tau=0$ boundary of this {`doubled-sheeted'}  cylinder is associated with the defect, while the boundary at $\tau=\beta/2$ is `trivial' and it moves off to infinity in the limit $\beta\rightarrow\infty$ {of interest}, 
{as} we are interested in the ground state of the system.
Thus, the defect free-energy can be thought of as being associated 
with the boundary free-energy of this semi-infinite {
(`double-sheeted'})  cylinder}}}
a non-universal contribution $f_{d}$ per unit 
length 
{$L$}
of the defect, plus a universal constant, 
{length-$L$}
independent contribution $S=\ln g$,}
\begin{eqnarray}
\nonumber
\ln Z_{d}= f_{d} \cdot 
{L}
+ S,
\ \ S=\ln g.
\end{eqnarray}
In the 
defect (boundary, after folding) CFT of interest in the present paper the universal quantity $
S(R)= \ln g(R)$ depends on the number $R$ of replicas and must vanish in the $R\to 1$ limit due to the POVM condition Eq.~\ref{LableEq-POVM-Condition}.

{
In general we obtain for the type of measurement problems on a critical ground state discussed in the present paper, analogous to the logic used in Ref.~\onlinecite{ZabaloGullansWilsonVasseurLudwigGopalakrishnanHusePixley} 
to obtain the Shannon entropy of a deep circuit with bulk measurements,
an expression for the partition function $Z_{R}$ of our system from Eq.~\ref{eqn:Nthmomentsetup} and Eq.~\ref{LabelDefBornRuleProb} upon  setting all operators to the identity,
\begin{eqnarray}
\nonumber
&& 
Z_{R=1+r}=
\sum_{\vec m}
p({\vec m}) \ [p({\vec m})]^r
\\ \nonumber
&& {\rm or:}
\\ \nonumber
&&
\ln Z_{R=1+r}=
\\ \nonumber
&&
=\ln \left \{
\sum_{\vec m} [p({\vec m})
+ r p({\vec m}) \ln (p({\vec m})) + {\cal O}(r^2)
]
\right \}
\\ \nonumber
&& {\rm and:}
\\ \nonumber
&&{d \over dr}_{|r=0} \ln Z_{R=1+r}=
\sum_{\vec m} p({\vec m}) \ln (p({\vec m}))=
\\ \label{LabelEqShannonEntropyUniversal}
&&
= - S_{Shannon}(\{p({\vec m})\}
= f_{d,\text{eff}}
\cdot 
{L}
 \ \ + \  \  S_{\text{eff}},
 \qquad
\end{eqnarray}
where
\begin{eqnarray}
\label{LabelEqDefSShannonMeasurementRecord}
S_{Shannon}(\{p({\vec m})\}
=
-
\sum_{\vec m} p({\vec m}) \ln (p({\vec m}))
\end{eqnarray}
is the Shannon entropy of the measurement record, while
\begin{eqnarray}
\nonumber
&&
f_{d,\text{eff}} := \left ( {d \over dr}_{|r=0} f_d(R=1+r) \right ) =
{\rm non-universal}
\\ \nonumber
&& {\rm and} \\ \nonumber
&&S_{\text{eff}}:= {d \over dr}_{|r=0} S(R)=
 {d \over dR}_{|R=1} \ln g(R)=
 \\ \nonumber
 &&=
  \left ( {d \over dR}_{|R=1} g(R) \right )/g(R=1) := \ln g_{\text{eff}}
  = {\rm universal}
\end{eqnarray}
The ``effective boundary entropy'' $S_{\text{eff}}$, the logarithm  of the ``effective boundary degeneracy'' $g_{\text{eff}}$, therefore characterizes the universal constant, i.e. system-size-$L$ independent part of the Shannon entropy of the measurement record on the critical ground state.
The Shannon entropy thus expresses the Born-rule averaged {\it defect} free energy in terms of a quantity directly related to the measurements.
The ``effective boundary entropy'' $S_{\text{eff}}$
plays a role in our problem of measurements on the quantum critical ground states
analogous to the role played by the ``effective central charge" in the deep  circuits with bulk measurements where, as already mentioned, the latter 
{describes}
the universal finite-size scaling information contained in the measurement record in the space-time bulk of the circuit. 
}

{The value of $g_0$ at our  ultraviolet  
defect (boundary) fixed point $\Delta=0$ is 
$S_0=\ln g_0=0$
since there the space-time has no defect  at all and thus there is no 
{defect (boundary)}
contribution to the free energy of the space-time.}
{We have calculated the universal boundary entropy $S(R)=\ln g(R)$ at our new fixed point $\Delta_*$, 
Eq.~{\ref{Eq:JengLudwig2LoopDelta*}},
using our $\epsilon$-expansion.
{Following} {Ref. \onlinecite{AffleckLudwig1991} and \onlinecite{JengLudwig}}{, we obtain}
\begin{eqnarray}
\nonumber
&&
g(R) = g_0 + \delta g(R), \ {\rm where} \ g_0=1, \\ \nonumber
&&
\ln g(R) =  \ln [1 + \delta g(R)]
= \delta g(R) + {\cal O} \left (\delta g(R)\right )^2,
\\ \nonumber
&&\delta g(R)=
-
 \frac{\pi^2}{24} {R(R-1)\over (R-2)^2} \epsilon^3 + {\cal O}(\epsilon^4).
\end{eqnarray}
We note that $\ln g(R)$, the  universal constant  contribution from the boundary to the logarithm of the  partition function vanishes as $R\to 1$ to the order in $\epsilon$ we are considering,  consistent with requirement from the POVM condition enforcing a partition function equal to unity in this limit.
The  ``{\it effective ground state degeneracy}''
$g_{\text{eff}}$
of the 
{defect (boundary)}
fixed point $\Delta_*$ is thus found to lowest non-vanishing order in the $\epsilon$ expansion to be
\begin{eqnarray}
\nonumber
&&
g_{\text{eff}} = 1 + \delta g_{\text{eff}},
\\ \nonumber
&&
\delta g_{\text{eff}} :=
\left ({d 
\over dR}\right)\bigg|_{R=1} \delta g(R)
= -\frac{\pi^2}{24}\epsilon^3 + {\cal O}(\epsilon^4).
\end{eqnarray}
Thus,
the  universal constant contribution to the Shannon entropy of the   measurement record on the tricritical Ising ground state is, owing to  Eq.~\ref{LabelEqShannonEntropyUniversal}, given by the following expansion
\begin{eqnarray}
\nonumber
&&
\Bigl [ S_{Shannon}(\{p({\vec m})\}) \Bigr ]_{\rm universal \ part}
=
\\ 
\label{LabelEqShannonEntropyUniversalExpansion}
&&
=-S_{\text{eff}}= - \ln g_{\text{eff}} = \frac{\pi^2}{24} \epsilon^3 + {\cal O}(\epsilon^4),
\end{eqnarray} 
as $m\rightarrow4$.}

We close this section by noting that the `boundary entropy'~\cite{AffleckLudwig1991} has recently also been discussed in the different context  of decoherence in Refs. 
\onlinecite{ZouSangHsieh,AshidaFurukawaOshikawa}.
\vskip .8cm
{\section{Measurements of \texorpdfstring{$\hat{\sigma}^z_i$}{Lg} in the Quantum Ising Model\label{Sec:IsingModel}}}

{(a): So far we addressed in this paper the problem of 
measurements with the energy operator
$\hat{\boldsymbol{E}}_{j+\frac{1}{2}}$, 
Eq.~\ref{eqn:boldsymbol{E}},  performed on the ground state of the {\it tri}critical quantum Ising model (using the formulation by O'Brien and Fendley). We developed an $\epsilon=3/(m+1)$ expansion for the resulting rich and novel critical properties, where $m \geq 4$ is an {\it even} integer characterizing a subset of minimal model  CFTs which can be represented by tricritial $q$-state Potts models ($\sqrt{q}=2 \cos{\pi\over m}$, Eq.~\ref{LabelEqTricriticalqPottsFctOfm}). The tricritical Ising model itself, the aim of our study, corresponds to the smallest {\it even} value $m=4$.}

{(b):  In the present section we address the problem of 
measurements with the Pauli spin operator ${\hat \sigma}^z_i$, performed on the ground state of the {\it critical} ({\it not} tricritical) quantum Ising model. We will again develop an $\epsilon=3/(m+1)$ expansion for corresponding resulting rich and novel critical properties of this system. Here, however, $m\geq 3$ is an {\it odd} integer that characterizes another (complementary) subset of minimal model CFTs, which can be represented as a subset of Ising multicritical points (namely those with $m=$ odd). The Ising model itself, the aim of our study, corresponds to the smallest {\it odd} value $m=3$.}

{There is a common principle unifying problems (a) and (b), which allows us to apply the logic we have already developed for (a) also to (b) - with some modifications:}

{As described above, the continuum field  ${\cal E}$ describing the measurement operator $\hat{\boldsymbol{E}}_{j+\frac{1}{2}}$
Eq.~\ref{eqn:ContinuumLimitofTwoLatticeOperators}, \ref{eqn:boldsymbol{E}}
in the 
{O'Brien-Fendley model}
is the energy operator 
{of the tricritical Ising point.}
This, as already mentioned above,
corresponds to the so-called Kac-Table primary field $\varphi_{1,2}$ of the minimal model CFT
[see,  e.g., Refs. \onlinecite{BelavinPolyakovZamolodchikov1984}, \onlinecite{di1997conformal}]
with $m=4$ describing the tricritical Ising model. Moreover, for arbitrary {\it even} values of $m \geq 4$, the same
Kac-Table primary field $\varphi_{1,2}$ of the minimal model CFT corresponds to the energy operator $\mathcal{E}$ in the corresponding
tricritical $q$-state Potts model that we use to define our $\epsilon$ expansion, i.e. $\mathcal{E}=\varphi_{1,2}$ for $m\geq 4$ {\it even}. At the same time, this operator is represented in the Landau-Ginzburg description with action Eq.
\ref{eqn:GeneralLGZAction}
by 
\begin{eqnarray}
\nonumber
&& {\rm tricritical \ Ising:}
\\ \label{LabelEqmathcalEvarphi12Phimminustwo}
&& {\cal E} = \varphi_{1,2} = {:}\phi^{m-2}{:},
 \ \ {\rm when} \ m\geq 4, \  m={\rm  even},
 \qquad
\end{eqnarray}
(see Eq.~\ref{eqn:GeneralMEnergy}).}

{
On the other hand, for the $m=3$ conformal minimal model, describing the critical Ising model, the continuum field representing the spin operator ${\hat \sigma}^z_i$ is also represented by the Kac-Table $\varphi_{1,2}$ operator at that value $m=3$. 
{Moreover, for arbitrary {\it odd} values of $m \geq 3$, the same
 Kac-Table primary field $\varphi_{1,2}$ of the minimal model CFT is 
 represented in the Landau-Ginzburg description with action Eq.~\ref{eqn:GeneralMEnergy} again by $\varphi_{1,2} = {:}\phi^{m-2}{:}$, with the crucial difference that since now $m=$ {\it odd}, this field  now changes sign under the Ising $Z_2$ symmetry $\phi \to -\phi$.}
 {We will use this field 
 as the generalization of the spin operator of the Ising model ($m=3$ minimal model CFT) to the minimal model CFT at {\it odd} $m \geq 3$, {and will denote it by the symbol
 \begin{eqnarray}
 \nonumber
 && {\rm critical \ Ising:}
 \\
 \label{LabeEqmathcalSvarphi12phimminustwo}
 &&
 \mathcal{S} := \varphi_{1,2} ={\bf :}\phi^{m-2}{:},
 \ \ \ {\rm when} \ m\geq 3, \ m= {\rm odd}.
 \qquad
\end{eqnarray}
The operator $\mathcal{S}$ plays, for the $\epsilon=3/(m+1)$ expansion with {\it odd} $m\geq 3$ in the Ising case, exactly the same role that the operator $\mathcal{E}$ played for the $\epsilon=3/(m+1)$ expansion with {\it even} $m \geq 4$ in the tricritical Ising case that {we} have already described in this paper. 
{Various
discussions and calculations} discussed so-far for the tricritical Ising case can literally be taken over to the Ising case by simply replacing $\mathcal{E}$ by $\mathcal{S}$, and by replacing  ``$m\geq 4$, $m=$ {\it even}'' by ``$m\geq 3$, $m=$ {\it odd}''.
}}}\\
{However, before proceding with this,}
we 
need to understand a subtlety of the Ising case, $m=3$. This will be done below after briefly reviewing again our $\epsilon$ expansion approach. 

{
Let us recapitulate that in  Sect.~\ref{section:Measurements} we}
started with a quantum critical ground state in the universality class of {the}
{Ising} tricritical 
point ($m=4^{\text{th}}$ minimal model) and considered the problem of performing measurements
{of the  lattice energy operator $\hat{\boldsymbol{E}}_{j+\frac{1}{2}}$ in Eq.~\ref{eqn:boldsymbol{E}} (which 
{is odd}
under Kramers-Wannier duality)}
on it.
In the process of analyzing this 
{problem within an $\epsilon$ expansion,} 
we introduced 
{the}
generalized replica action in Eq. \ref{LabelEqGeneralmReplicaFieldTheory}{,} with action $S_{*}$  corresponding to the $m^{\text{th}}$ minimal model CFT (Eq. \ref{eqn:GeneralLGZAction}) and a field $\mathcal{E}$ in this CFT.
We motivated the choice of {the} field $\mathcal{E}$ in these higher minimal model {CFTs} by restricting
to only the {\it even} $m$ minimal model {CFTs}
and using the tricritical $q-$state Potts formulation of {the {\it even}  $m$} minimal model
{CFTs,}
{where the field} $\mathcal{E}$ in Eq. \ref{eqn:GeneralMEnergy} corresponds to the 
energy {operator in the tricritical q-state Potts model}, {which for $q=2$ is the same as the tricritical Ising point}.
{This provided the basis for the $\epsilon$
expansion, where the control parameter $\epsilon=$
$3/(m+1)$ is small when the {\it even} integer $m\geq 4$ is large.}

Until now, {for us} the
{generalized replica field theory
{in 
Eqs. \ref{LabelEqGeneralmReplicaFieldTheory}~--~\ref{eqn:GeneralMEnergy}}
with} {\it  even}
$m\geq 4$ minimal model 
CFTs
only served as a tool to
{perturbatively study the replica field theory for measurements at the Ising tricritical point (i.e. the replica field theory at  $m=4$) in 
{the}
small parameter $\epsilon=3/(m+1)$ {via the $\epsilon$ expansion we developed  in Sect. \ref{SubSecOnPerturbativeRG}
above.}}
However{,} as {purely} a defect field theory problem with replica action in Eqs. \ref{LabelEqGeneralmReplicaFieldTheory}~--~\ref{eqn:GeneralMEnergy}, it is clear that the perturbative RG analysis 
{presented}
in Sect. \ref{SubSecOnPerturbativeRG}
applies to {\it any} $m^{\text{th}}$ minimal model {with $m\geq 4$},
{\it including also}
the minimal model
{CFTs}
with {\it odd} $m$.
This is because {the} {field $\mathcal{E}$, Eq.~\ref{LabelEqmathcalEvarphi12Phimminustwo},
being} {the}
{energy operator}
of {the} tricritical $q-$state Potts models, {when $m \geq 4$ is {\it even}}, was only used to motivate the choice of field $\mathcal{E}$ in the
minimal models {with larger parameter $m>4$}{,} and the field theory problem is perfectly well defined with just Eqs. \ref{LabelEqGeneralmReplicaFieldTheory}~--~\ref{eqn:GeneralMEnergy} 
{also for {\it odd} values of $m$}, as far as the RG analysis is concerned~\footnote{{
Unlike the even $m$ minimal model CFTs, there does \textit{not} exist a {tri}critical $q-$state Potts model with the same central charge as the odd $m$ minimal model CFTs.
However, this is immaterial to the RG analysis of the replica action in Eq. \ref{LabelEqGeneralmReplicaFieldTheory} for the multicritical point given by an odd $m$ minimal model CFT in Eq. \ref{eqn:GeneralLGZAction} and with the symbol $\mathcal{E}$ replaced by $\mathcal{S}=:\phi^{m-2}:$.
We note that for the odd $m$ minimal model CFTs the field $\mathcal{S}=:\phi^{m-2}:$ is one of the spin fields of the multicritical point and \textit{not} an energy field, and this distinction is also inconsequential to the RG analysis of the replica action presented in Sect. \ref{SubSecOnPerturbativeRG}.  As already mentioned above, for both odd $m$ and even $m$ minimal model CFTs, the field $:\phi^{m-2}:$ is the so-called Kac-Table operator $\varphi_{1,2}$. See also footnote \cite{Note22}
}}.
{Excluding}
Sect. \ref{SubSecOnSpinCorrelation} on the {(replicated) tricritical Potts} spin correlation functions, the {results from} perturbative RG analysis for the entanglement entropies in Sect. \ref{sec:EntanglementEntropy} and the {(replicated)} {$\mathcal{E}=:\phi^{m-2}:$}  field correlation functions in Sect.~\ref{SubSecOnEnergyCorrelation} also {readily} extend to all 
{$m\geq 4$} ({both even and odd})
minimal model CFTs
in Eq.~\ref{eqn:GeneralLGZAction}. {However, the physical meaning of this replica field theory in 
Eq.~\ref{LabelEqGeneralmReplicaFieldTheory}, and the corresponding $\epsilon=3/(m+1)$ expansion, is completely different in the case of $m \geq 3$ with $m=$ {\it odd}, as compared to the case of $m\geq 4$ with $m=$ {\it even}, discussed in Sect.~\ref{SubSecOnPerturbativeRG}: The former case 
{will serve to describe}
the $\epsilon=3/(m+1)$ expansion for the problem of $\hat{\sigma}^z_i$ measurements on the ground state of the critical quantum Ising Hamiltonian. In this case, it is useful to rename the operator
$\mathcal{E}$, Eq.~\ref{LabelEqmathcalEvarphi12Phimminustwo} as $\mathcal{S}$, Eq.~\ref{LabeEqmathcalSvarphi12phimminustwo}, which is now {\it odd} under the Ising $Z_2$  symmetry. (In the Ising case,
$m=3$, $\mathcal{S}$ is the continuum field representing the
Pauli $\hat{\sigma}^z_i$ operator that is measured in this case.) 
{For general $m\geq 3$ with $m=$
 {\it odd}
the replica field theory action is the same as in Eqs. \ref{LabelEqGeneralmReplicaFieldTheory}~--~\ref{eqn:GeneralLGZAction}, but with the field $\mathcal{E}$ replaced by $\mathcal{S}$, i.e. 
\begin{align}
-\mathbb{S}&=
\sum_{a=1}^R
(-1)
S_*^{(a)}
+
\Delta \int_{-\infty}^{+\infty} dx  \;\Phi(x)\label{LabelEqGeneralmReplicaFieldTheoryOddm}
\\
\Phi(x)&=\sum_{\substack{a,b=1 \\ a\neq b}}^{R}
\mathcal{S}^{(a)}(x,0)\mathcal{S}^{(b)}(x,0),\label{LabelEqPhiformathalSmodd}\\
S_*&=\int d \tau \int d x \bigg\{\frac{1}{2}(\partial_{x} \phi)^2+\frac{1}{2}(\partial_{\tau} \phi)^2+
{g^*_{m-1}}
\phi^{2(m-1)}\bigg\},\nonumber\\\label{eqn:GeneralLGZActionOddm}
\end{align}
with the notation $\mathcal{S}$ as defined in Eq.~\ref{LabeEqmathcalSvarphi12phimminustwo}.}
}

{For the $m=3$ minimal model, 
{the lowest value of $m=$ {\it odd}, describing the Ising critical point,  however}, there is an 
additional subtlety that arises in the RG analysis of the replica action in Eqs. \ref{LabelEqGeneralmReplicaFieldTheoryOddm}~--~\ref{eqn:GeneralLGZActionOddm}.}
{For minimal models $m\geq 4${,} the 
{terms denoted by the ellipsis}  
``$\dots$" in the OPE {in Eq. \ref{EqPhiPhiOPE}}
{of the perturbation $\Phi$ {(Eq.~\ref{LabelEqGeneralmReplicaFieldTheory}
and
Eq.~\ref{LabelEqPhiformathalSmodd} for  $m=$ {\it even} and {\it odd}, respectively),} with itself} contained only irrelevant fields {localized 
on the $\tau=0$ time-slice. Hence} 
we obtained
Eq. \ref{eqn:1loopRGdisorderstrength} at 1-loop order, which described the RG flow of $\Delta$.
{Precisely at the Ising critical point  $m=3$} however, for an arbitrary number of replicas $R${,} the OPE {in Eq. \ref{EqPhiPhiOPE}}
{of the perturbation $\Phi$,
Eq.~\ref{LabelEqPhiformathalSmodd},
with itself} 
contains a term
which is 
{\it  exactly}
marginal. 
}{In particular, {in the case of $m=3$,} the following term 
\begin{equation}\label{EqLinearReplicaEnergy}
    \sum_{\alpha}\mathfrak{e}^{(\alpha)}
    \;\;\text{where }\;\;\mathfrak{e}=:\phi^2:=:\phi^{2m-4}:|_{m=3}
\end{equation}
 appears on the RHS of Eq. \ref{EqPhiPhiOPE}. {Here,}
 $:\phi^{2m-4}:$ is irrelevant {on the $\tau=0$  time-slice} for minimal models with $m\geq 4$
 {but it} 
 is {\it exactly} marginal at $m=3${,} i.e. 
  {at}
 the Ising critical point.
This term, however, 
{turns out to come {(see Eq. \ref{EqPhiPhiApp} in App. \ref{app:HigherLoopIsing})}} 
with a coefficient $(R-1)$ in the OPE in Eq. \ref{EqPhiPhiOPE}, and hence it 
vanishes
in the limit 
$R\rightarrow1$.
Moreover,
we show
{in App. \ref{app:HigherLoopIsing}}
that in the limit $R\rightarrow1$ such a term 
{cannot}
{be generated under the
RG in any order of the
coupling constant $\Delta$ of the perturbation $\Phi$.
Since the  exactly  marginal term in Eq. \ref{EqLinearReplicaEnergy} cannot be produced under {the} RG in the $R\rightarrow 1$ limit {at $m=3$, the Ising case~\footnote{{Recall from the discussion above that, when $m>3$, this term is replaced by an irrelevant operator on the time-slice which can be ignored.}}}, 
the RG analysis performed in Sect. \ref{SubSecOnPerturbativeRG}, \ref{SubSecOnEnergyCorrelation} and \ref{sec:EntanglementEntropy} will also
provide an expansion in large {\it odd}  $m$ (small $\epsilon=$ $3/(m+1)$)
for the generalized replica theory 
in Eqs. \ref{LabelEqGeneralmReplicaFieldTheoryOddm}~--~\ref{eqn:GeneralLGZActionOddm} {all the way}  down to
$m=3$, i.e. down to the Ising critical point.}}
{
Thus in the replica limit $R \to 1$, the 1-loop RG equation derived in Eq. \ref{eqn:1loopRGdisorderstrength} also applies to an expansion in $\epsilon$ and $m=$ odd,  down to  
the Ising critical point.
}

{
{To discuss this case in detail,} let us consider the $1d$ quantum Ising model at its 
{\it critical}
{({\it not} {\it tri-}critical)}
point described by the Hamiltonian in Eq. \ref{EqIsingCriticalHamiltonian}, which lies in the universality class of the $m=3$ minimal model CFT.}
{Let us consider performing (weak-) measurements with operator $\hat{\sigma}_{i}^{z}$ at \textit{all} sites $i$ 
{on the ground state of the {\it critical} Ising model (Eq. \ref{EqIsingCriticalHamiltonian}).}
Since $(\hat{\sigma}_{i}^{z})^2=1$ and $\hat{\sigma}_{i}^{z}$ at different sites commute with each other, one 
{immediately verifies}
that the details from sections \ref{subsec:MeasProto1} and \ref{SubSecReplicaTrick}, 
{where the 
O'Brien-Fendley chain at its Ising {\it tri}critical point was discussed,}
generalize straightforwardly 
{to the case of}
measurements with $\hat{\sigma}_{i}^{z}$
{on the ground state of the {\it critical} quantum Ising model, where the measurement operator 
{$\hat{\boldsymbol{E}}_{i}$}
of the former is now replaced by $\hat{\sigma}_{i}^{z}$}.}
In particular, the measurement averaged moments
{of correlation functions} for this measurement protocol are given by
\begin{equation}
\begin{split}
\overline{[\langle{\hat {\cal O}}_1\rangle_{\vec m} ...
  \langle{\hat {\cal O}}_N\rangle_{\vec m}]}&\propto  \lim_{R\rightarrow 1}\text{Tr}\bigg(\hat{\mathcal{O}}_{1}^{(1)}\hat{\mathcal{O}}_{2}^{(2)}\dots\hat{\mathcal{O}}_{N}^{(N)}\times\\
 \times (\ket{0}\bra{0})^{\otimes R} &\exp{\bigg\{4\tilde{\Delta}\sum_{i=\text{integer}}\sum_{\substack{a,b=1\\a\neq b}}^{R}(\hat{\sigma}^{z}_{i})^{(a)}(\hat{\sigma}^{z}_{i})^{(b)}\bigg\}}\bigg).
 \label{eqn:NthmomentsetupIsing} 
\end{split}
\end{equation}
{Here, the state $\ket{0}$ 
{now denotes}
the ground state of the 
{\it critical} Ising Hamiltonian {listed} in Eq. \ref{EqIsingCriticalHamiltonian}.}
In the above equation, in contrast to
{the corresponding equation} Eq. \ref{eqn:NthmomentsetupFINALProtocol1}
{of the tricritical point for the O'Brien-Fendley chain}{,} we have a sum over \textit{all} sites $i$ and 
operators $(\hat{\sigma}^{z}_{i})^{(a)}$.
This is because in 
{the present}
section we are performing measurements at all sites $i$ with operator $\hat{\sigma}^{z}_{i}$,
{instead of with operator $\hat{\boldsymbol{E}}_{i}$ 
{for even}
$i$.}
{Just as in the tricritical Ising case, Eqs.~\ref{EqProductofLocalKraus}, \ref{LabelEqKrausKti},
we have gone over to a formulation using continuous (``softened'') measurement outcomes $t_i$ with a symmetric distribution $P(t_i)$, which we again for now first assume to be a zero-mean Gaussian Eq.~\ref{eqn:tcumulants} (only second cumulant non-vanishing).} \\

{In continuum language, one 
{now sees} that 
equation
{Eq.~\ref{eqn:NthmomentsetupIsing} above}
reduces to
Eq.~\ref{eqn:NthmomentsetupFINALProtocol1}}
{but where  the action {$S_{*}$} in 
Eq.~\ref{eqn:TricriticalIsingAction}
is now}
the 
Landau-Ginzburg(-Zamolodchikov) action of the Ising {\it critical} point, i.e.
\begin{eqnarray}
    S_*=\int d \tau \int d x \bigg\{\frac{1}{2}(\partial_{x} \phi)^2+\frac{1}{2}(\partial_{\tau} \phi)^2+
{g^{*}_{\text{2}}\phi^4}\bigg\}{,}\label{LabelEqLGZActionIsing}
\end{eqnarray}
{as opposed to that of the Ising {\it tri}critical point in Eq.~\ref{eqn:TricriticalIsingAction}.}
{{Also, in contrast to Eq. \ref{eqn:Phi(x)},} the perturbation} $\Phi(x)$ is {now}
given by
\begin{eqnarray}\label{EqIsingPhi}
&&\Phi(x) \;{=}
\sum_{\substack{a,b=1 \\ a\neq b}}^{R}
\mathfrak{s}^{(a)}(x,0) \mathfrak{s}^{(b)}(x,0),\label{EqIsingDefect}
\end{eqnarray}
{where the field $\mathfrak{s}(x,\tau)$ is the the continuum field corresponding to the  lattice operator $\hat{\sigma}^{z}_{i}$ at the Ising critical point with
the scaling dimension $1/8$ \cite{BelavinPolyakovZamolodchikov1984, BaxterBook}}.
{As discussed at the start of this section, the field $\mathfrak{s}$ is given by the Kac's table field $\varphi_{1,2}$ at the Ising critical point, and thus can be obtained by taking $m\rightarrow3$ limit of the field $\mathcal{S}$ in Eq. \ref{LabeEqmathcalSvarphi12phimminustwo}. 
The special symbol $\mathfrak{s}(x,\tau)$ for the field $\mathcal{S}$ in the Ising case $m=3$ (and only for $m=3$) is used to stress the additional subtlety arising in this case.}
{In particular, substituting $m=3$ in the generalized replica field theory in Eqs. \ref{LabelEqGeneralmReplicaFieldTheoryOddm},~\ref{LabelEqPhiformathalSmodd}, 
and \ref{eqn:GeneralLGZActionOddm}, we precisely recover the replica field theory for measurements performed {with the operator $\hat{\sigma}^{z}_{i}$} at {the} Ising critical point.}
{We note that the role played by the  
{average (``weak")}
Kramers-Wannier symmetry mentioned in the last paragraph of Sect.~\ref{sec:FieldTheoryandFixedPoint} is now played by the 
{average (``weak")}
Ising $Z_2$ symmetry. Thus, the perturbation
in Eq.~\ref{EqIsingPhi} represents the most RG relevant perturbation invariant under 
the (``average'') Ising $Z_2$ and replica permutation symmetries. Less relevant interaction terms which can be thought of as being associated with higher cumulants of the distribution $P(t_i)$ are discussed in App.~\ref{LabelAppB1HigherCumulantsCriticalIsing} and {are }found to be irrelevant at the new fixed point $\Delta_*$. This means that our result will also  be valid for 
{weak}
measurements where the distribution $P(t_i)$ is a (normalized) sum of delta functions.}
\par
{{Since the replica field theory for measurements performed with the operator $\hat{\sigma}^{z}_{i}$ at} the Ising critical point corresponds to the limit $m\rightarrow 3$ of the generalized 
replica theory in 
{
Eq. \ref{LabelEqGeneralmReplicaFieldTheoryOddm} with {\it odd} $m\geq 3$,
we see,  
{following the discussion on the analogy between replica field theories in Eqs. \ref{LabelEqGeneralmReplicaFieldTheory}~--~\ref{eqn:GeneralLGZAction} for even $m$ and replica field theories in Eqs. \ref{LabelEqGeneralmReplicaFieldTheoryOddm}~--~\ref{eqn:GeneralLGZActionOddm} for odd $m$,} that the expansion in $\epsilon=3/(m+1)$}
}
\begin{equation}\label{EqTwoLoopIsingDelta}
    \Delta_*=
   \frac{\epsilon}{4}+\frac{\epsilon^2}{4}+ {\cal O}(\epsilon^3)
\end{equation}
{(obtained by taking $R\rightarrow1$ in Eq. \ref{Eq:JengLudwig2LoopDelta*}) provides, 
as $m \to 3$, the location (in coupling constant space) of the
{measurement-dominated} fixed point which}
governs the IR physics 
{of measurements with the spin-operator ${\hat \sigma}_i^z$ performed  at all sites  on the ground state of the critical quantum Ising model}~\footnote{In App. \ref{Appendix:DetailsOfReplicaSummation}{,} we will show that {the} higher ($>2$) cumulants of $P(t_{i})$ {(}analogous to the case of tricritical Ising point{)} {are inconsequential to} the IR physics of measurement averaged quantities at Ising critical point. 
In particular, {the} higher cumulants generate terms where an even number {($\geq 2$)} of pairwise unequal replica copies of spin field $s(x,\tau)$ interact with each other on the $\tau=0$ {time slice}.
Out of these the $4-$replica and $6-$replica terms are relevant, the $8-$replica term is marginal, and all the other higher replica terms are irrelevant {\it at} the $m=3$ {minimal model CFT, i.e.} the Ising critical point.
{Moreover}, the {aforementioned} relevant and marginal terms at the Ising critical point {$(m=3)$}
are irrelevant at fixed points described by the large$-m$ minimal model {CFTs}.
Then following the standard reasoning used in the case of the $\phi^6$ interaction at the Wilson-Fisher fixed point in $d=4-\epsilon$ dimensions, the relevant and marginal terms at the Ising critical point are expected to be irrelevant at the 
{new}
fixed point $\Delta_*$ even at $m=3$. This is discussed in more detail in App. \ref{LabelAppB1HigherCumulantsCriticalIsing}.
{Also see App. \ref{app:AvoidedLevelCrossingsHigherCumulants} for a general argument for the irrelevance of higher cumulants ($2k\geq 4$) based on avoided level crossings.}}. 
{{By reasoning completely parallel to the tricritical case discussed above,}
 the measurement-averaged $n^{\text{th}}$ R\'enyi entanglement entropy $S_{n,A}$ for an interval of length $l$ of the {ground state of the} critical Ising chain
 {subjected to ${\hat \sigma}^z_i$ measurements at all sites} is given by the $\epsilon-$expansion from Eq. \ref{eqn:RenyiEntropyAverageFI}, 
\begin{equation}\label{eqn:RenyiEntropyAverageIsing}
\begin{split}&\overline{S_{n,A}}=\frac{c^{(\text{eff})}_{n}}{3} \ln\frac{l}{a}+{\mathcal{O}(1)} \;\;\text{where,}\\
&c^{(\text{eff})}_{n}= 
{c(m)}
\bigg(\frac{1+1/n}{2}\bigg)-\frac{3 I_{n}}{(n-1)}\epsilon+\mathcal{O}(\epsilon^2)\\
&I_{n}=\frac{n}{2\pi}\int_{0}^{\infty}\text{d}s\;\frac{1-s^{n-1}}{(1-s)(1+s^n)}+
 {\mathcal{O} \left (\epsilon \right )}{,}
\end{split}
\end{equation} 
but \textit{now} 
{in the limit} $m\rightarrow 3$ 
{with} $\epsilon=3/(m+1)$.
Finally, since the field $\mathfrak{s}(x,\tau)$ is 
{the} $m\rightarrow3$ limit of the 
{field $\mathcal{S}$ 
{defined 
in Eq. \ref{LabeEqmathcalSvarphi12phimminustwo}} 
for general odd-$m$ minimal model CFTs (which in turn forms the analogue of the field $\mathcal{E}$ Eq. \ref{LabelEqmathcalEvarphi12Phimminustwo} defined for even-$m$ minimal model CFTs),}
the measurement averaged $N^{\text{th}}$ moment of the correlation function of {the} $\hat{\sigma}_{i}^{z}$ operator at the {Ising critical} point will be given by
$m\rightarrow 3$
limit of Eq. \ref{eq:EnergyMoments},
\begin{subequations}\label{eq:SIGMAZMOMENTSISING} 
\begin{align}
&\boldsymbol{X^{(\hat{\sigma}_ \text{Is}),R=1}_{N=1}}=\frac{1}{2}-\frac{\epsilon}{2}+\mathcal{O}(\epsilon^3)\\
& \boldsymbol{X^{(\hat{\sigma}_ \text{Is}),R=1}_{N>1}}=\frac{N}{2}\bigg[1+\epsilon-(3N-5)\epsilon^2+\mathcal{O}(\epsilon^3)\bigg]. \qquad
\end{align}
\end{subequations}
}
{This shows that at the Ising critical point, the scaling dimensions of the measurement averaged moments of the $\hat{\sigma}_{i}^{z}$ correlation function 
exhibit
{multifractal scaling}
{(see Sect. \ref{SubSecOnMultiFractality})}. 
In particular, 
{at the Ising critical point with measurements, the typical
{connected}
correlation function of the $\hat{\sigma}_{i}^{z}$ operator} will be given by the following power law exponent
\begin{equation}
    \boldsymbol{\tilde{X}^{(\hat{\sigma}_\text{Is}), R=1}_{\text{typ}}}=\frac{1}{2}\left[1+\epsilon+5\epsilon^2+\mathcal{O}(\epsilon^3)\right]
\end{equation}
{as $m\rightarrow3${, where the definition of
$\boldsymbol{\tilde{X}^{(\hat{\sigma}_\text{Is}), R=1}_{\text{typ}}}$
is completely analogous to the definition of
$\boldsymbol{\tilde{X}^{(\mathcal{E}), R=1}_{\text{typ}}}$ in the
tricritical Ising case.~\footnote{{I.e., $\boldsymbol{\tilde{X}^{(\hat{\sigma},\text{Is}), R=1}_{\text{typ}}}$
is obtained from the moments of the {\it subtracted} Pauli spin operator describing the deviation from its expectation value in a fixed quantum trajectory, 
$\delta \hat{\sigma}^z_{i}
:=$ $\hat{\sigma}^z_{i} -
\langle 
\hat{\sigma}^z_{i}
\rangle$, and 
$ \langle \delta \hat{\sigma}^z_{i} 
\delta \hat{\sigma}^z_{j}\rangle=$
$ \langle \hat{\sigma}^z_{i} \hat{\sigma}^z_{j}\rangle 
-
\langle\hat{\sigma}^z_{i} \rangle
\langle \hat{\sigma}^z_{j}\rangle$.}
}}}}

\vskip .9cm
\section{Conclusions and Discussion\label{Sec:Conclusion}}
We have demonstrated that performing weak measurements on relatively simple quantum critical ground states can give rise to critical states with highly complex and novel scaling behavior described by novel universality classes. 
We started our study with the critical ground state in the universality class of the
 Ising tricritical
point in the lattice formulation by O'Brien and Fendley, and subjected it to weak measurements with a lattice
operator which corresponds to the continuum energy operator at the Ising tricritical point.
The described weak measurements turn out to be a relevant perturbation at the 
 Ising tricritical point and the critical properties of the states obtained upon measurements are no longer dictated by the 
Ising tricrticial
point itself.
We showed that the critical behavior of the tricritical Ising ground state subjected to the described weak measurements is governed by a new, measurement-dominated fixed point, which occurs at a finite strength of measurements. 
We presented a controlled perturbative RG analysis, i.e. an $\epsilon$ expansion, to study the universal critical properties of this measurement-dominated fixed point and
we calculated a variety of universal quantities (described below) in this $\epsilon$ expansion.
\par
We found the first manifestation of the novel scaling properties of the measurement-dominated fixed point in the scaling properties of the 
 measurement-averaged
moments of correlation functions. 
In particular, we showed that the measurement averaged $N$th moments of both the spin and the energy correlation function at the tricritical Ising point decay with independent power-law exponents for each $N$.
Thus, there exists an infinite number of independent scaling exponents associated with each correlation function.
 Moreover, noninteger moments $N$ of the correlation functions  
also exhibit
scaling behavior, resulting in a continuous spectrum of scaling exponents for each operator, spin and energy.
Each continuous spectrum of scaling exponents is related to 
a
universal scaling form of the probability distribution of the given correlation function in states obtained upon measurements, and we determined the typical scaling behavior of the spin and the connected energy correlation function.
We also
demonstrated the presence of
logarithmic CFT features 
at
the measurement-dominated fixed point, 
in particular
the presence of logarithmic correlation functions.
We showed that, unlike in usual (unitary) CFTs
where all correlation functions are power law decaying, 
measurement-averaged correlations functions may possess
a multiplicative logarithm of distance on top of a power law decay. Such logarithmic correlation functions are associated with the non-diagonalizability of the dilation operator, and we also identified the `logarithmic pair' of scaling operators that span the $2\times 2$ Jordan cell of the dilation operator corresponding to the obtained logarithmic correlation function.\par
Another novel feature of the finite measurement strength fixed point was found in the universal coefficients $\frac{1}{3}c_{n}^{(\text{eff})}$ of the logarithm of subsystem size in the measurement averaged $n$th R\'enyi entanglement entropies. 
We found that similar to the infinite hierarchy of scaling exponents in the case of moments of the correlation functions, the universal coefficients $c_{n}^{(\text{eff})}$ associated with the $n$th measurement averaged R\'enyi entanglement entropies are also independent of each other for different $n$.
This is 
in contrast
to the unmeasured 1d quantum critical ground states 
 (and all unitary CFTs)
where the universal coefficients of the logarithm of subsystem size for all $n$th R\'enyi entropies are 
all related
solely to a single number,
the central charge of the corresponding 2D CFT.
We showed that $c_{n}^{(\text{eff})}$ also appears in the coefficient of the leading order finite temperature correction to the measurement averaged \textit{extensive} $n$th R\'enyi entropy of the full (thermal) mixed 
Gibbs state
of the system.\par
The problem of performing weak measurements on a 1d quantum critical ground state can be formulated as a field theory problem with a one dimensional defect at the zero-time slice of the corresponding (replicated) (1+1)d CFT.
We showed, generally, that for a given 1d quantum critical ground state, the universal ``Affleck-Ludwig" effective boundary entropy associated with this defect (boundary, after folding) appears as a constant, system size independent piece in the Shannon entropy of the measurement record on the ground state. 
In the case of the tricritical Ising ground state subjected to weak measurements with the energy operator, we calculated this universal contribution to the Shannon entropy to leading order in
the $\epsilon$ expansion.
We note that the role of the effective boundary entropy in the case of a 1d critical ground state subjected to measurements is analogous to that of the `effective central charge' 
at the measurement-induced transition of a deep quantum circuit, 
where the latter characterizes the finite size 
dependence of
the Shannon entropy of the measurement record on the bulk of the deep quantum circuit at the measurement-induced transition.
\par
Finally, we also studied the ground state of the quantum critical Ising model subjected to weak measurements with the spin operator $\hat{\sigma}_{i}^{z}$.
By appropriately generalizing the controlled perturbative RG analysis, i.e. the $\epsilon$ expansion, developed in the case of the tricritical Ising point, we demonstrated that the critical behavior of the critical Ising ground state subjected to weak measurements with the $\hat{\sigma}_{i}^{z}$ operator is also governed by another measurement-dominated fixed point, which occurs
at a finite strength of measurements.
At the Ising critical point, we determined the power-law exponents of the measurement averaged moments of the $\hat{\sigma}_{i}^{z}$ correlation function to two-loop order in the $\epsilon$ expansion and 
found these exponents to be independent of each other.
We also obtained the power law exponent of the typical connected correlation function of the $\hat{\sigma}_{i}^{z}$ at the Ising critical point in the $\epsilon$ expansion.
Lastly, we also calculated, to leading order in the $\epsilon$ expansion, the universal coefficients $\frac{1}{3}c_{n}^{(\text{eff})}$ of the logarithm of subsystem size in the measurement averaged $n$th R\'enyi entanglement entropies at the Ising critical point.
Again, analogous to the case of the tricritical Ising point, the universal coefficients $c_{n}^{(\text{eff})}$ at the Ising critical point are also found to be independent of each other for different values of $n$.

\vskip 1.5cm

\begin{acknowledgments}
One of us (AWWL) thanks Sam Garratt for an inspiring discussion on Ref. \onlinecite{GarrattWeinsteinAltman2022} in Fall 2022, as well as  especially Romain Vasseur and Chao-Ming Jian for collaboration on several previous works in the related area of measurement induced phase transitions.

\end{acknowledgments}
\vskip .8cm

\appendix

 \section{{
Higher 
{Cumulants}
and {a Comment on} `Non-Local' Fields
}\label{Appendix:DetailsOfReplicaSummation}}

\subsection{{Higher Cumulants: Ising tricritical point}
\label{LabelAppB1HigherCumulantsTricriticalIsing}}
{We begin by discussing the Ising tricritical point.}
{In Eq. \ref{Eq:MeasurementsSummedonLattice}, we averaged over measurement outcomes by assuming that only the second cumulant of the distribution $P(t_{i})$ is non-zero.
In this Appendix, we will provide justification for why {the} higher even cumulants of $P(t_{i})$~\footnote{{The distribution} $P(t_{i})$ is taken to be an even function of $t_{i}$ to satisfy Eq. \ref{LabelEqPOVMKt_i}.} {cannot} change the
{critical} behavior of the system at long distances.}\par
{
{A 
{non-vanishing} $(2n)^{\text{th}}$ cumulant ($\tilde{\Delta}_{2n}$)} of $P(t_{i})$ will give rise to {the following term}
\begin{equation}
\tilde{\Delta}_{2n}\sum_{i=\text{even}}4\left(\sum_{a=1}^{R}\hat{\boldsymbol{E}}_{i}^{(a)}\right)^{2n}
\end{equation}
in the 
{exponential}
on the RHS of Eq. \ref{Eq:MeasurementsSummedonLattice}.}
{The above expression can be 
{simplified, using $\left ( \hat{\boldsymbol{E}}_{i}^{(a)}\right)^2=1$, which shows that it corresponds to}
a superposition of terms of {the following} form (summed over all even $i$)
\begin{equation}\label{Eq:HigherCumulantsSimplified}
{\sum_{\substack{a_{j_1},a_{j_2},\dots,a_{j_{2k}}}}^{\substack{\text{all indices are}\\\text{pairwise distinct}}}}\hat{\boldsymbol{E}}_{i}^{(a_{j_1})}\hat{\boldsymbol{E}}_{i}^{(a_{j_2})}\cdots\hat{\boldsymbol{E}}_{i}^{(a_{j_{2k}})},
\end{equation}
where $k$ is an integer less than $n$.}
{In continuum language, we can replace each $\hat{\boldsymbol{E}}^{(a_{j_{l}})}_{i}$ in the above expression with {the} corresponding continuum field $\mathcal{E}^{(a_{j_{l}})}(x,0)$ in {the} replica copy $``a_{j_{l}}"$.}
{
{Going over to}
continuum language, 
{the} above expression 
{thus reads}
\begin{equation}\label{EqHigherCumulantsContinuum}
{\sum_{\substack{a_{j_1},a_{j_2},\dots,a_{j_{2k}}}}^{\substack{\text{all indices are}\\\text{pairwise distinct}}}}(\mathcal{E}^{(a_{j_{1}})})(\mathcal{E}^{(a_{j_{2}})})\cdots(\mathcal{E}^{(a_{j_{2k}})}).
\end{equation}
}
{In principle, in each of the 
{parentheses}
in the above expression we can have a contribution from the 
subleading energy field $\mathcal{E}''$ 
which (just like the leading energy field $\mathcal{E}$) is also odd under K-W duality 
and can appear in the continuum representation of {the} lattice operator 
{[compare Eqs.~\ref{eqn:ContinuumLimitofTwoLatticeOperators},\ref{eqn:boldsymbol{E}}].}
However, $\mathcal{E}''$ (scaling dimension $=3$) is highly irrelevant as a field with support on the 1-dimensional time slice $\tau=0$, and we can drop it.}
{Coming back to Eq. \ref{EqHigherCumulantsContinuum}, since the scaling dimension of {the} field $\mathcal{E}$
{at the Ising tricritical point} is $1/5$, for all $k>2$ the term appearing in Eq. \ref{EqHigherCumulantsContinuum} is irrelevant as a field with support on the 1-dimensional time slice $\tau=0$, and again we can ignore it.}
{The only relevant term coming from {the} higher cumulants
{$k\geq 2$}
(apart from the {$k=1$ term} already appearing in Eq. \ref{eqn:Phi(x)}) is
{the $k=2$ term,}
\begin{equation}\label{EqFourUnEqualReplica}
    {\sum_{a,b,c,d}^{\substack{\text{all indices are}\\\text{pairwise distinct}}}}\mathcal{E}^{(a)}\mathcal{E}^{(b)}\mathcal{E}^{(c)}\mathcal{E}^{(d)},
\end{equation}
{which arises}
{from all cumulants higher than or equal to the four.}
This term has scaling dimension $4/5 < 1$
{(while being less relevant than 
$\Phi(x)$ in Eq. \ref{eqn:Phi(x)}.)}}

{The discussion above of scaling dimensions of the operators in
Eq.~\ref{EqHigherCumulantsContinuum}
arising from higher cumulants
was referring to the  Ising tricritical point where  the measurement strength is $\Delta=0$. However, since we are in fact
interested in the measurement-dominated fixed point at which $\Delta_*\not =0$, which we control within  the $\epsilon=3/(m+1)$-expansion, we are really concerned with the relevance/irrelevance of these operators at the 
$\Delta_* \not =0$ fixed point. Now, all ``higher-cumulant'' ($k\geq 2$) operators
in
Eq.~\ref{EqHigherCumulantsContinuum}
are highly irrelevant at the 
$\Delta_* \not =0$ fixed point when $\epsilon=3/(m+1)$ is small, i.e. when $m$ is large:
At $\epsilon=0$ (where $1/m=0$)
they have scaling dimensions $=2k \times (1/2)>1$, i.e. are irrelevant (by integers)
on the one-dimensional time-slice when $k\geq 2$, and for small $\epsilon$ those scaling dimensions only change  by small amounts (of order $\epsilon$, $\epsilon^2$, ... etc.) when going to the finite-$\Delta_*$ fixed point of order $\epsilon$. 
More specifically, one can show explicitly~\cite{JengLudwig}
that 
the operators  in 
Eq.~\ref{EqHigherCumulantsContinuum}
become even more irrelevant at the $\Delta_* \not =0$  fixed point within the 
{1-loop}
epsilon expansion, as compared to their dimensions 
$=2k \times (1/2)$ at the $\Delta=0$ fixed point.
(I.e. the order $\epsilon$ 
shifts  
of
their scaling dimensions away from their already highly irrelevant $\epsilon=0$ values are all positive.)}
{This is a familiar feature of the epsilon expansion, well  known
already from that of $\phi^4$ Landau-Ginzburg theory in $d=4-\epsilon$ dimensions
where, although the $\phi^6$ perturbation is relevant at the {\it Gaussian} fixed point when $d<3$ (while being irrelevant when $d>3$), it is
{\it irrelevant} at the {\it Wilson-Fisher} fixed point of physical interest for  {\it all} dimensions $d \geq 2$.
Analogously, in the epsilon expansion from
Sect.~\ref{SubSecOnPerturbativeRG} of interest in this paper, while
the  operator in Eq.~\ref{EqHigherCumulantsContinuum} associated with the fourth cumulant $k=2$ is, at the {\it unperturbed} fixed point $\Delta=0$ (analogous to the Gaussian fixed
point in $\phi^4$ Landau-Ginzburg theory),
relevant  when $m=4$ and irrelevant for all even values  $m >4$,
it is analogously expected to be irrelevant at the  $\Delta_*\not = 0$ fixed point of interest, 
Eq.~\ref{Eq:JengLudwig2LoopDelta*}, for all even values 
$m \geq 4$, i.e. including $m \to 4$. (A general argument for the irrelevance of all higher cumulants ($k\geq 2$) based on avoided level crossings is presented in 
App.~\ref{app:AvoidedLevelCrossingsHigherCumulants}.)}

{Hence, we do not expect the higher 
{cumulants}
of  the distribution $P(t_{i})$ to change the long distance behavior of the system. This implies in particular that in  the case weak measurements with discrete measurement outcomes 
(Eqs.~\ref{LabelEqKrausOperator}, \ref{LabelEqKrausOperatorProto1}, \ref{LabelEqKrausKti}), where $P(t_{i})$ is a (normalized) sum of delta functions [compare the discussion below Eq.~\ref{LabelEqKrausOperatorProto1}] and thus contains even cumulants higher than the second, the same critical behavior results as in the case of a zero-mean Gaussian distribution $P(t_i)$.
}\\

\subsection{{Higher  Cumulants: Ising Critical Point}
\label{LabelAppB1HigherCumulantsCriticalIsing}}

{Following the above discussion for the tricritical Ising case, we will now  
{provide a justification for why the higher even cumulants of $P(t_i)$ are not expected to change the critical long-wavelength properties}
in the case of measurements with {the} $\hat{\sigma}^{z}_{i}$ 
{operator on the ground state of the critical quantum Ising model.}
Analogous to Eq. \ref{Eq:HigherCumulantsSimplified}, {the} higher cumulants for the $\hat{\sigma}^{z}_{i}$ measurements will give rise to terms of form
\begin{equation}\label{HigherCumulantsSimplifiedIsing}
   {\sum_{\substack{a_{j_1},a_{j_2},\dots,a_{j_{2k}}}}^{\substack{\text{all indices are}\\\text{pairwise distinct}}}}(\hat{\sigma}^{z}_{i})^{(a_{j_1})}(\hat{\sigma}^{z}_{i})^{(a_{j_2})}\cdots(\hat{\sigma}^{z}_{i})^{(a_{j_{2k}})},  
\end{equation}
which will appear in the 
{exponentional} 
in Eq. \ref{eqn:NthmomentsetupIsing}}.
{In continuum language, we can replace each $(\hat{\sigma}^{z}_{i})^{(a_{j_l})}$  {operator}
in the above equation by the continuum field $\mathfrak{s}^{(a_{j_{l}})}(x,\tau)$.  Since
the scaling dimension of $\mathfrak{s}^{(a_{j_{l}})}(x,\tau)$ is $1/8$, all the terms with $k>4$ appearing in the above equation are irrelevant at the
{(unmeasured)}
Ising critical point.
{Thus, at the Ising critical point, the}
$4$-replica and $6$-replica terms (corresponding to $k=2$ and $k=3$ in 
Eq.~\ref{HigherCumulantsSimplifiedIsing}) are {relevant}, 
while the $8$-replica term (corresponding to $k=4$) is {marginal
(while these terms are all less relevant than the perturbation $\Phi(x)$ in 
Eq.~\ref{EqIsingPhi}
corresponding to $k=1$).}}

{In analogy with the tricritical Ising case in the preceding subsection, the discussion above of the scaling dimensions of the operators in Eq.~\ref{HigherCumulantsSimplifiedIsing} arising from higher cumulants
was referring to the Ising critical point where the measurement strength is $\Delta=0$. However, again, as we are interested in the measurement-dominated fixed point at which $\Delta_* \not = 0$, which we control within the $ \epsilon = 3/(m+1)$-expansion [where now $m=$ odd], we are really interested in the relevan{ce}/irrelevance
of these operators at this 
{new}
fixed point. Again, all ``higher-cumulant'' operators $(k\geq 2)$ in Eq.~\ref{HigherCumulantsSimplifiedIsing} are highly irrelevant at the
$\Delta_* \not =0$ fixed point when $\epsilon = 3/(m+1)$ is small~\footnote{{For odd $m>3$ minimal models, we denote the generalization of the field $\mathfrak{s}$ by the symbol $\mathcal{S}$ defined in Eq. \ref{LabeEqmathcalSvarphi12phimminustwo}}}, i.e. when $m=$ odd is large: At $\epsilon=0$ ($1/m=0$) they have again  scaling dimensions $=2k \times (1/2)>1$, i.e. are again irrelevant (by integers) on the one-dimensional
time-slice when $k \geq 2$. And again, for small $\epsilon$ those scaling dimensions
only change by small amounts when going  to the finite-$\Delta_*$ fixed point of order $\epsilon$. Again,  specifically, within the 
{1-loop}
epsilon expansion~~\cite{JengLudwig} these operators become more {\it irrelevant} as compared to their already irrelevant scaling dimensions
$=2 k \times (1/2)$ at $\Delta=0$. Again, in analogy with the discussion of the $\phi^6$ term in the $d=4{-}\epsilon$ expansion of $\phi^4$ Landau-Ginzburg theory, the operators in Eq.~\ref{HigherCumulantsSimplifiedIsing} with $k=2, 3, 4$ are expected to be irrelevant at the $\Delta_* \not =0$ fixed point of interest for all odd values of $m \geq 3$, i.e. including $m=3$. 
(For a general argument for the irrelevance of all higher cumulants ($k\geq 2$) based on avoided level crossings  we refer again to  
App.~\ref{app:AvoidedLevelCrossingsHigherCumulants}.)}

{Hence, again,
we do not expect the higher 
cumulants
of  the distribution $P(t_{i})$ to change the long distance behavior of the system. 
This  implies again in particular  that in  the case of  weak measurements
with discrete measurement outcomes 
(Eqs.~\ref{LabelEqKrausOperator}, \ref{LabelEqKrausOperatorProto1}, \ref{LabelEqKrausKti}),
where $P(t_{i})$ is a (normalized) sum of delta functions (compare discussion below Eq.~\ref{LabelEqKrausOperatorProto1}) and thus contains even cumulants higher than the second, the same critical behavior results as in the case of a zero-mean Gaussian distribution $P(t_i)$.} \\

\par

\subsection{{Locality of Observables}
\label{LabelAppB3LocalityOfFields}}

{We will close this section by discussing the significance 
{of `locality' of}
{fields}
$\mathcal{O}_{i}$
{associated with lattice operators $\hat{\mathcal{O}}_i$ and}
{which appear}
in Eq. \ref{eqn:NthMomentReplicaFieldTheory}.}
{Note that in deriving 
Eq. \ref{eqn:NthMomentReplicaFieldTheory},}
{we assumed that the operators {$\hat{\mathcal{O}}_{i}$} in their continuum representation correspond to `local' fields {$\mathcal{O}_{i}$,} i.e. $\mathcal{O}_{i}$ can be expressed in terms of local combinations
{involving}
the {Landau-Ginzburg} field $\phi(x,\tau)$ {and its normal ordered higher powers}.
Important differences in the gluing of field configurations $\{\phi^{(a)}(x,0^-)\}_{j=1}^{R}$ and $\{\phi^{(a)}(x,0^+)\}_{j=1}^{R}$ {(appearing in Fig. \ref{fig:cylinder} and Eq. \ref{eqn:vaccumpathint})} could occur {if the fields $\mathcal{O}_{i}$ are} \textit{non-local}.}
{An example {of this} is seen in the case of measurements performed on the Tomonaga-Luttinger liquids (TLLs), studied in Ref. \cite{GarrattWeinsteinAltman2022}, when calculating the correlations functions of phase $e^{i\theta(x)}$. The phase $\theta(x)$ is termed as a `non-local' field in the bosonic theory of {the} field $\phi(x)$~\footnote{$\partial_{x} \phi(x)$ is proportional to the density of the TLLs} as they satisfy the {following} equal time commutator
\begin{equation}
    [\phi(x),\theta(x')]=i\pi H(x-x')=\begin{cases} 
      i\pi & x\geq x' \\
    0  & x<x'
   \end{cases}.
\end{equation}
In the calculation of phase correlation functions, as noted in {Ref.} \cite{GarrattWeinsteinAltman2022}, the two field configurations  $\{\phi^{(a)}(x,0^-)\}_{j=1}^{R}$ and $\{\phi^{(a)}(x,0^+)\}_{j=1}^{R}$differ with each other on an interval of values of position $x$, and are not `identified/glued' on this interval.}
{In this work, we have only considered correlation functions (and their moments) {of lattice} operators $\hat{\mathcal{O}}_{i}$ {which correspond to `local' fields in continuum, i.e. they can be expressed as local combinations of Landau-Ginzburg field $\phi$} {and normal ordered higher powers of $\phi$}.}

\vspace{1cm}

\section{Irreducible Representations of the Symmetry Group\label{appendix:IrreducibleRepresentation}}
{Unlike the unperturbed critical theory ($\Delta=0$),} the operators $\prod_{i=1}^{N}\mathcal{E}^{(\alpha_{i})}(x,0)$
 ({$1\leq\alpha_{i}\leq R$ 
 in a theory
 with $R$ replica})
 are {no longer} scaling operators {at the new fixed point ($\Delta=\Delta_{*}$).} 
 Rather, as discussed in Ref. \cite{LUDWIG1990infinitehierarchy} and \cite{JengLudwig}, the scaling operators at the new fixed point {are formed out of} linear {superposition of operators} $\prod_{i=1}^{N}\mathcal{E}^{(\alpha_{i})}(x,0)$ for different choices of replica indices $\{\alpha_i\}$, and {they }transform in irreducible representations of the symmetric group $S_R$. Following Ref. \cite{JengLudwig}, the 
 corresponding 
 scaling operators at the {new} fixed point are given by 
 \begin{multline}\label{EqScalingOperatorsIrreducible}  
\mathfrak{E}_{NMR}=
\\\sum_{\substack{\alpha_{i}\neq\alpha_j\\1\leq \alpha_i\leq \\R-M}}(\mathcal{E}^{(\alpha_1)}-\mathcal{E}^{(R)})\dots(\mathcal{E}^{(\alpha_M)}-\mathcal{E}^{(R-M+1)})\mathcal{E}^{(\alpha_{M+1})}\dots\mathcal{E}^{(\alpha_{N})}\\
({0}\leq M\leq N).
\end{multline}
The scaling dimensions of the above operators are 
{calculated to two-loop order in Ref. \cite{JengLudwig} in a dimensional regularization [by $\epsilon=3/(m+1)$] RG scheme, with minimal subtraction of poles in $\epsilon$. From their analysis, the scaling dimension of the operator in Eq. \ref{EqScalingOperatorsIrreducible} is given by}

\begin{align}
&\boldsymbol{X^{(\mathcal{E}),R}_{NM}}=NX_{\mathcal{E}}-\gamma(\Delta_{*})\;\dots\;\bigg(X_{\mathcal{E}}=\frac{1}{2}-\frac{3}{2(m+1)}\bigg)\label{eq:XNMScalingDimensions}\\&\Delta^{*}=\frac{\epsilon}{4(2-R)}+\frac{\epsilon^2}{4(2-R)^2}+ {\cal O}(\epsilon^3)\;\;\dots\;\bigg(\epsilon=\frac{3}{m+1}\bigg)\nonumber\\
&\gamma(\Delta)=2\tilde{b}_{NMR}\Delta-8(N(R-N)+(N-1)\tilde{b}_{NMR})\Delta^2\nonumber\\&\hspace{7cm} + O(\Delta^3)\nonumber\\
 &\tilde{b}_{NMR}=2((N -M)R-N^2 +M(M -1))\nonumber
\end{align}
{In table \ref{LabelTableScalingDimensions}, 
we list the scaling dimensions $\boldsymbol{X^{(\mathcal{E}),R}_{NM}}$ for $R=0,1$ {and}
$N=1,2,3$.}
{Out of the $N$ scaling dimensions{,} corresponding to different values of $M$ in Eq. \ref{eq:XNMScalingDimensions}, the smallest one dictates the 
{power law behavior}
of $\overline{\langle \mathcal{E}(x,0)\mathcal{E}(y,0)\rangle^N}$ when $|x-y|\rightarrow \infty$.} 
{
 Setting $R=1$ and minimizing the scaling dimension $\boldsymbol{X^{(\mathcal{E}),R}_{NM}}$ in Eq. \ref{eq:XNMScalingDimensions} over possible values of $M$ 
 ,}
we obtain Eqs. \ref{eqn:bornmeasurementvarphi12I} and \ref{eqn:bornmeasurementvarphi12II}. \\
\begin{table}
\begin{center}
\caption{Scaling Dimensions $\boldsymbol{X^{(\mathcal{E}),R}_{NM}}$
{for $N=1,2,3$ and $R=0,1$.}}
\label{LabelTableScalingDimensions}
\begin{tabular}{| m{1.25cm} | m{3.25cm}| m{3.5cm} | } 
\hline
$(N,M)$& $\boldsymbol{X^{(\mathcal{E}),R=0}}_{NM}$ & $\boldsymbol{X^{(\mathcal{E}),R=1}_{NM}}$ \\
\hline
$(1,0)$ &$\frac{1}{2}+\frac{\epsilon^2}{8}+\mathcal{O}(\epsilon^3)$ &$\frac{1}{2}-\frac{\epsilon}{2}+\mathcal{O}(\epsilon^3)$ \\ 
$(1,1)$
&$\frac{1}{2}+\frac{\epsilon^2}{8}+\mathcal{O}(\epsilon^3)$ &  $\frac{1}{2}+\frac{\epsilon}{2}+\epsilon^2+\mathcal{O}(\epsilon^3)$\\
\hline
$(2,0)$ & $1+\epsilon-\frac{\epsilon^2}{2}+\mathcal{O}(\epsilon^3)$&$1+\epsilon-\epsilon^2+\mathcal{O}(\epsilon^3)$ \\ $(2,1)$
 & $1+\epsilon-\frac{\epsilon^2}{2}+\mathcal{O}(\epsilon^3)$ & $1+2\epsilon-\epsilon^2+\mathcal{O}(\epsilon^3)$ \\ $(2,2)$
 & $1-\frac{\epsilon^2}{2}+\mathcal{O}(\epsilon^3)$ & $1+\epsilon-\epsilon^2+\mathcal{O}(\epsilon^3)$\\
\hline
$(3,0)$ & $\frac{3}{2}+3\epsilon-\frac{27\epsilon^2}{8}+\mathcal{O}(\epsilon^3)$ &$\frac{3}{2}+\frac{9\epsilon}{2}-9\epsilon^2+\mathcal{O}(\epsilon^3)$ \\ $(3,1)$
 & $\frac{3}{2}+3\epsilon-\frac{27\epsilon^2}{8}+\mathcal{O}(\epsilon^3)$ & $\frac{3}{2}+\frac{11\epsilon}{2}-10\epsilon^2+\mathcal{O}(\epsilon^3)$ \\ $(3,2)$
 & $\frac{3}{2}+2\epsilon-\frac{23\epsilon^2}{8}+\mathcal{O}(\epsilon^3)$ & $\frac{3}{2}+\frac{9\epsilon}{2}-9\epsilon^2+\mathcal{O}(\epsilon^3)$\\ $(3,3)$
& $\frac{3}{2}-\frac{15\epsilon^2}{8}+\mathcal{O}(\epsilon^3)$ & $\frac{3}{2}+\frac{3\epsilon}{2}-6\epsilon^2+\mathcal{O}(\epsilon^3)$\\ 
 \hline
\end{tabular}
\end{center}
\end{table}
\par
{As an aside, we note that in Ref. \cite{JengLudwig} they were in interested in the limit $R\rightarrow0$ (which corresponds to quenched disorder), and in this limit the smallest scaling dimension for a fixed $N$ in Eq. \ref{eq:XNMScalingDimensions} is given by,}
\begin{equation}\label{EqEnergyMomentsR=0}
  \boldsymbol{X^{(\mathcal{E}),R=0}_{N}}=\frac{N}{2}(1-\frac{\epsilon^2}{4}(3N-4)+\mathcal{O}(\epsilon^2)).
\end{equation}
\vskip 0.6cm
\subsection*{Colliding Scaling Dimensions in Replica Limit \texorpdfstring{$R\rightarrow1$}{Lg}}
As discussed in Sect. \ref{SubSec:LogarithmicCorrelationFunctions}, scaling dimensions of operators with unequal scaling dimensions at generic replica number $R\neq1$ can become equal to each other at $R=1$.
To see this collision of scaling dimensions in replica limit $R\rightarrow1$, we consider two operators $\mathfrak{E}_{20R}$ and $\mathfrak{E}_{22R}$ from Eq. \ref{EqScalingOperatorsIrreducible}.
We note that the correlation function of the $\mathfrak{E}_{20R}$ operator is given by,
\begin{equation}
\begin{split}
    &\langle \mathfrak{E}_{20R}(r,0)\mathfrak{E}_{20R}(0,0)\rangle = 2R(R-1)\times\\&\bigg(\langle \mathcal{E}^{1}(r)\mathcal{E}^{1}(0)\mathcal{E}^{2}(r)\mathcal{E}^{2}(0)\rangle+\\&\;\;\;\;\;\;\;\;\;\;\;\;\;\;\;\;\;\;+\frac{(R-2)(R-3)}{2}\langle\mathcal{E}^{1}(r)\mathcal{E}^{2}(r)\mathcal{E}^{3}(0)\mathcal{E}^{4}(0)\rangle\\&\;\;\;\;\;\;\;\;\;\;\;\;\;\;\;\;\;\;\;\;\;\;\;\;\;\;\;\;\;\;\;+2(R-2)\langle \mathcal{E}^{1}(r)\mathcal{E}^{1}(0)\mathcal{E}^{2}(r)\mathcal{E}^{3}(0)\rangle\bigg),\label{EqE20R}
\end{split}
\end{equation}
while for $\mathfrak{E}_{22R}$ operator,
\begin{equation}
\begin{split}\label{EqE22R}
    &\langle \mathfrak{E}_{22R}(r,0)\mathfrak{E}_{22R}(0,0)\rangle = (R-3)(R-2)^2(R-1)\times \\&\bigg(\langle \mathcal{E}^{1}(r)\mathcal{E}^{1}(0)\mathcal{E}^{2}(r)\mathcal{E}^{2}(0)\rangle+ \langle\mathcal{E}^{1}(r)\mathcal{E}^{2}(r)\mathcal{E}^{3}(0)\mathcal{E}^{4}(0)\rangle\\&\;\;\;\;\;\;\;\;\;\;\;\;\;\;\;\;\;\;\;\;\;\;\;\;\;\;\;\;\;\;\;\;\;\;\;\;\;\;\;\;\;\;\;\;\;-2\langle \mathcal{E}^{1}(r)\mathcal{E}^{1}(0)\mathcal{E}^{2}(r)\mathcal{E}^{3}(0)\rangle\bigg).
\end{split}
\end{equation}
Ignoring the overall $R$ dependent constants, clearly, the expressions in parentheses in Eqs. \ref{EqE20R} and \ref{EqE22R} are identical to each other in $R\rightarrow1$ limit.
Thus, the two operators $\mathfrak{E}_{20R}$ and $\mathfrak{E}_{22R}$ have colliding scaling dimensions in the replica limit $R\rightarrow1$,
i.e. the scaling dimensions of the two operators are equal to each other in the limit $R\rightarrow1$ at the new fixed point $\Delta_{*}(\epsilon)$ ($\epsilon=3/(m+1)$) for all {even} values of $m$. 
(This can also be verified using the $\epsilon$-expansion
{for the scaling dimensions of the two operators using Eq.~\ref{eq:XNMScalingDimensions}.)}
Moreover, with the given normalization for operator $\mathcal{O}$ (Eq. \ref{EqLogarithmicOperatorConstructionO}) and operator $\tilde{\mathcal{O}}$ (Eq. \ref{EqLogarithmicOperatorConstructionTildeO}), it can be easily verified that the criterion in  Eq. \ref{EqCriterionForLogarithmicCorrelator2} is satisfied by the {amplitudes of} correlators  $\langle \mathcal{O}(r) \mathcal{O}(0)\rangle$ and $\langle \tilde{\mathcal{O}}(r) \tilde{\mathcal{O}}(0)\rangle$.
Finally, since the correlation functions in the parentheses of Eqs. \ref{EqE20R} and \ref{EqE22R} are physical correlators, we expect to get a finite answer for them in the $R\rightarrow1$ 
limit, and thus {operators $\mathcal{O}$ and $\tilde{\mathcal{O}}$ also satisfy the criterion in Eq. \ref{EqCriterionForLogarithmicCorrelator3}.
As discussed in Sect. \ref{SubSec:LogarithmicCorrelationFunctions}, such colliding of scaling dimensions give rise to logarithmic correlation functions at the new fixed point.
\vskip .8cm
\section{Details of Entanglement Entropy Calculation\label{appendix:EntanglementEntropy}}
{Given a set of measurement outcomes 
{${\vec m}=$}
$\{m_{j}\}$, the state obtained after measurements 
is}
\begin{equation}
    \ket{\Psi_{\{m_{j}\}}}=\frac{\hat{K}_{\vec{{\mathbf{m}}}}\ket{0}}{\sqrt{\bra{0}(\hat{K}_{\vec{{\mathbf{m}}}})^{\dagger}\hat{K}_{\vec{{\mathbf{m}}}}\ket{0}}}.
\end{equation}
The $n^{\text{th}}$ R\'enyi entanglement entropy of a spatial region $A=[u,v]$ in this state is given by
\begin{multline}\label{Eqn:EEmeas}
    S_{n,A}(\ket{\Psi_{\{m_{j}\}}})=\frac{1}{1-n}\ln \left \{\text{Tr}_{A}\left [\rho_{A}\left (\ket{\Psi_{\{m_{j}\}}}\right)\right]
^n\right \}{,}
\end{multline}
where the reduced density matrix is
\begin{equation}
\begin{split}
    \rho_{A}(\ket{\Psi_{\{m_{j}\}}})=&\text{Tr}_{\bar{A}}\left(\ket{\Psi_{\{m_{j}\}}}\bra{\Psi_{\{m_{j}\}}}\right)\\=&\frac{\text{Tr}_{\bar{A}}\left(K_{\vec{\mathbf{m}}}\ket{0}\bra{0}(K_{\vec{\mathbf{m}}})^{\dagger}\right)}{\text{Tr}(K_{\vec{\mathbf{m}}}\ket{0}\bra{0}(K_{\vec{\mathbf{m}}})^{\dagger})}
\end{split}
\end{equation}
and $\bar{A}$ is complement of spatial region $A$. Then
\begin{widetext}
\begin{eqnarray}
    S_{n,A}(\ket{\Psi_{\{m_{j}\}}})=&&\frac{1}{1-n}\big\{\ln(\text{Tr}_{A}(\text{Tr}_{\bar{A}}(K_{\vec{\mathbf{m}}}\ket{0}\bra{0}(K_{\vec{\mathbf{m}}})^{\dagger}))^{n})
    -\ln(\text{Tr}(K_{\vec{\mathbf{m}}}\ket{0}\bra{0}(K_{\vec{\mathbf{m}}})^{\dagger}))^n\big\}\nonumber\\
    =&&\frac{1}{1-n}\{\ln\text{Tr}(\mathscr{S}_{n,A}(K_{\vec{\mathbf{m}}}\ket{0}\bra{0}(K_{\vec{\mathbf{m}}})^{\dagger})^{\otimes n})
    -\ln\text{Tr}((K_{\vec{\mathbf{m}}}\ket{0}\bra{0}(K_{\vec{\mathbf{m}}})^{\dagger})^{\otimes n})\}\nonumber\\
    =&&\frac{1}{1-n}\lim_{k\rightarrow 0}\frac{1}{k}\times\big\{\text{Tr}(\mathscr{S}^{k}_{n,A}(K_{\vec{\mathbf{m}}}\ket{0}\bra{0}(K_{\vec{\mathbf{m}}})^{\dagger})^{\otimes nk})-\text{Tr}((K_{\vec{\mathbf{m}}}\ket{0}\bra{0}(K_{\vec{\mathbf{m}}})^{\dagger})^{\otimes nk})\big\}{.}\label{Eq:REEinAQuantumTrajectory}
\end{eqnarray}
Here the permutation operator $\mathscr{S}_{n,A}$ {
is defined~\cite{JianYouVasseurLudwig2019}} as
\begin{equation}\label{EqPermutationOperatorEE}
    \mathscr{S}_{n,A}=\underset{x}{\Pi} \scalebox{1.2}{$\mathcal{\chi}$}_{g_{x}} \text{and } g_{x}=\begin{cases}
   (1,2,\dots,n )& x \in A\\
   \text{identity}=e & x \in \bar{A}\\
   \end{cases}{,}
\end{equation}
where $g_{x}$ labels the permutation on site $x$, and $\scalebox{1.2}{$\chi$}_{g_{x}} =\sum_{[i]} \ket{i_{g_{x}(1)}i_{g_{x}(2)}\dots i_{g_{x}(n)}}\bra{i_1i_2\dots i_n}$ is its representation on the replicated on-site Hilbert space. 
{Since the operator $(K_{\vec{\mathbf{m}}})^{\otimes nk}=K_{\vec{\mathbf{m}}}\otimes K_{\vec{\mathbf{m}}}\cdots \otimes K_{\vec{\mathbf{m}}}$
commutes with the permutation operator $\mathscr{S}_{n,A}$, 
using cyclicity of trace we can write Eq. \ref{Eq:REEinAQuantumTrajectory} as
\begin{equation}
    S_{n,A}(\ket{\Psi_{\{m_{j}\}}})=\frac{1}{1-n}\lim_{k\rightarrow 0}\frac{1}{k}\times\big\{\text{Tr}(\mathscr{S}^{k}_{n,A}(\ket{0}\bra{0})^{\otimes nk}({\hat {\bm K}}_{\vec m}^{\dagger}{\hat {\bm K}}_{\vec m})^{\otimes nk})-\text{Tr}((\ket{0}\bra{0})^{\otimes nk}({\hat {\bm K}}_{\vec m}^{\dagger}{\hat {\bm K}}_{\vec m})^{\otimes nk})\big\}
\end{equation}
}
{Then }the average of the $n^{\text{th}}$ R\'enyi entropy over the measurement outcomes with Born rule is given by,
\begin{equation}
 \begin{split}
\overline{S_{n,A}}=&\sum_{\vec{\mathbf{m}}}p_{0}(\vec{\mathbf{m}})S_{n,A}(\ket{\Psi_{\{m_{j}\}}})\\=&\lim_{k\rightarrow 0}\frac{1}{(1-n)k}\sum_{\vec{\mathbf{m}}}p_{0}(\vec{\mathbf{m}})\big\{\text{Tr}(\mathscr{S}^{k}_{n,A}(\ket{0}\bra{0})^{\otimes nk}({\hat {\bm K}}_{\vec m}^{\dagger}{\hat {\bm K}}_{\vec m})^{\otimes nk})-\text{Tr}((\ket{0}\bra{0})^{\otimes nk}({\hat {\bm K}}_{\vec m}^{\dagger}{\hat {\bm K}}_{\vec m})^{\otimes nk})\big\}{.}
 \end{split}   
\end{equation}
Since $p_{0}(\vec{\mathbf{m}})=\text{Tr}(K_{\vec{\mathbf{m}}}\ket{0}\bra{0}(K_{\vec{\mathbf{m}}})^{\dagger})${$=\text{Tr}(\ket{0}\bra{0}K_{\vec{\mathbf{m}}}^{\dagger}K_{\vec{\mathbf{m}}})$}
{we obtain}
{
$\overline{S_{n,A}}=$~$\frac{1}{1-n}\lim_{k\rightarrow 0}
{1\over k} \left [ \mathcal{Z}_{A} -\mathcal{Z}_{\varnothing}\right ]$,}
where
{
\begin{equation}\label{eqn:RenyiEaverage}
\begin{split}
\mathcal{Z_{A}}=\sum_{\vec{\mathbf{m}}}\text{Tr}(\mathscr{S}^{k}_{n,A}(\ket{0}\bra{0})^{\otimes nk+1}({\hat {\bm K}}_{\vec m}^{\dagger}{\hat {\bm K}}_{\vec m})^{\otimes nk+1})\;\;\text{and}\;\;
\mathcal{Z}_{\varnothing}=\sum_{\vec{\mathbf{m}}}\text{Tr}((\ket{0}\bra{0})^{\otimes nk+1}({\hat {\bm K}}_{\vec m}^{\dagger}{\hat {\bm K}}_{\vec m})^{\otimes nk+1}). 
\end{split}
\end{equation}}
{Owing to the POVM condition
Eq.~\ref{LableEq-POVM-Condition},
$\lim_{k \rightarrow 0}\mathcal{Z}_{\varnothing}=1$,}
we can write the measurement averaged $n^{\text{th}}$ R\'enyi entropy as
\begin{equation}\label{eqn:RenyiEntropyAverage2}
\overline{S_{n,A}}=\frac{1}{1-n}\lim_{k\rightarrow 0}\frac{1}{k}\bigg(\frac{\mathcal{Z}_{A}}{\mathcal{Z}_{\varnothing}}-1\bigg)=\frac{1}{1-n}\bigg(\frac{\text{d}}{\text{d}k}\bigg|_{k=0}\frac{\mathcal{Z}_{A}}{\mathcal{Z}_{\varnothing}}\bigg){.}
\end{equation}
{Using Eq. \ref{Eq:MeasurementsSummedonLattice}, 
\ref{eqn:vaccumpathint},
\ref{eqn:ContinuumLimitofTwoLatticeOperators} and following the arguments in 
{the} derivation of Eq. \ref{eqn:NthMomentReplicaFieldTheory}, $\mathcal{Z}_{A}$ can be written as
}
\begin{eqnarray}
&&\mathcal{Z_{A}}=\sum_{\vec{\mathbf{m}}}\text{Tr}(\mathscr{S}^{k}_{n,A}(\ket{0}\bra{0})^{\otimes nk+1}({\hat {\bm K}}_{\vec m}^{\dagger}{\hat {\bm K}}_{\vec m})^{\otimes nk+1})\propto\nonumber\\
&&\propto\int \prod_{a=1}^{nk+1}D\phi^{(a)} \,e^{-\sum_{a=1}^{nk+1}S_{{*}}^{(a)}+\Delta\int dx \Phi(x)}\,\text{Tr}(\mathscr{S}^k_{n,A}\ket{\{\phi^{(a)}(x,0^{+})\}}\bra{\{\phi^{(a)}(x,0^{-})\}})\label{eqn:eerepl}
\end{eqnarray}
where,
\begin{eqnarray}
&&\Phi(x)=\sum_{\substack{a,b=1 \\ a\neq b}}^{R}
\mathcal{E}^{(a)}(x,0) \mathcal{E}^{(b)}(x,0) \text{ and }\ket{\{\phi^{(a)}(x,0^{\pm})\}}=\bigotimes_{a=1}^{R}\ket{\{\phi^{(a)}(x,0^{\pm})\}}
\end{eqnarray}
\end{widetext}
{Upon making use of the definition Eq.~\ref{Eq:REEinAQuantumTrajectory},
the factor $\text{Tr}\left (\mathscr{S}^k_{n,A}
\ket{\{\phi^{(a)}(x,0^{+})\}}\bra{\{\phi^{(a)}(x,0^-)\}}\right)$ in Eq.~\ref{eqn:eerepl} does the job gluing the $nk+1$ replicas into $k$ {$n$-sheeted} Riemann surfaces, as illustrated in Fig.~\ref{fig:EntanglementGluing}:}
{Each of these $k$ $n$-sheeted Riemann surfaces contain $n$ replicas which are glued in the spatial region $A$ along the $\tau=0$ equal-time slice, and there is one additional replica representing a plane that remains unglued.}
{Thus we conclude that the  
ratio $\mathcal{Z}_{A}/\mathcal{Z}_{\varnothing}$ of partition functions equals the correlation function of two 
twist fields describing these  $k$  $n$-sheeted Riemann surfaces. Thus we can express the measurement-averaged R\'enyi entropies from
Eq. \ref{eqn:RenyiEntropyAverage2} in terms of the twist fields as}
\begin{equation}\label{eqn:RenyiEntropyAverage3}
\overline{S_{n,A}}=\frac{1}{1-n}\frac{\text{d}}{\text{d}k}\bigg|_{k=0}\langle \prod_{j=1}^{k}\big(\Tau^{(j)}_{n}(u,0)(\Tau^{(j)}_{n})^{-1}(v,0)\big)\rangle_{\Delta_{*}},
\end{equation}
{where}
the superscript $j$ on the twist fields 
{denotes}
{the}
Riemann surface (out of $k$) {to which} the twist field corresponds{,} and the subscript $n$ 
{indicates} 
that we are dealing with twist fields for a $n$-sheeted Riemann surface. 
{$(\Tau^{(j)}_{n})^{-1}$ denotes the twist field conjugate (``inverse'') to 
$\Tau^{(j)}_{n}$.}
\vskip .8cm
{
\section{OPE coefficient of two Twist Fields into  \texorpdfstring{$\Phi(x)$}{Lg}
\label{appendix:OPEtwistsPhi}}
}

{In order to compute the scaling dimension of the twist field at the new fixed point to 1-loop order in the small parameter $\epsilon=$~$3/(m+1)$,}
we will need the {OPE}  coefficient 
{with which the perturbation} $\Phi(x)$ 
{(from Eqs.~\ref{LabelEqGeneralmReplicaFieldTheory}, 
\ref{eqn:GeneralMEnergy}) appears}
in the OPE of the twist fields.
This is equivalent to finding the following three point
{correlation function}
in the unperturbed replica theory ({with action in Eq. \ref{eqn:GeneralLGZAction}})
\begin{multline}\label{eqn:BigPhismallphiandTwists}
    \langle \prod_{j}^{k}\big(\Tau^{(j)}_{n}(u,0)(\Tau^{(j)}_{n})^{-1}(v,0)\big)\Phi(x)\rangle=\\\sum_{\substack{a,b=1 \\ a\neq b}}^{nk+1}\langle \prod_{j}^{k}\big(\Tau^{(j)}_{n}(u,0)(\Tau^{(j)}_{n})^{-1}(v,0)\big)\mathcal{E}^{(a)}(x,0)\mathcal{E}^{(b)}(x,0)\rangle.
\end{multline}
We know from Ref. \cite{calabrese2004,*calabrese2009entanglement} that
\begin{multline}\label{eqn:RiemannsheetstoReplica}
    \langle \mathcal{E}^{(a)}(x,0)\mathcal{E}^{(b)}(x,0) \rangle_{(\mathcal{R}_{n})^{k}+1}=\\\frac{\langle \prod_{j=1}^{k}\big(\Tau^{(j)}_{n}(u,0)(\Tau^{(j)}_{n})^{-1}(v,0)\big)\mathcal{E}^{(a)}(x,o)\mathcal{E}^{(b)}(x,0)\rangle}{\langle \prod_{j}^{k}\big(\Tau^{(j)}_{n}(u,0)(\Tau^{(j)}_{n})^{-1}(v,0)\big)\rangle}.
\end{multline}
The 
{LHS}
of the above equation calculates the correlator for the two 
specified 
fields
{$\mathcal{E}^{(a)}$ and $\mathcal{E}^{(b)}$}
in the geometry shown in Fig. \ref{fig:EntanglementGluing}, which involves $k$
{copies of a} $n$-sheeted Riemann 
{surface}
and 
{{\it one}  plane, denoted by the subscript $(\mathcal{R}_{n})^{k}+1$ on the correlator.}
{On the LHS of the above equation, the index $a$ is to be thought of as equal to a combined index $(i,\alpha)$. 
The index $i$ here indicates either a Riemann surface out of the $k$ 
{copies of the} $n$-sheeted Riemann 
{surface (when $i\in \{1, ..., k\}$)}
or it indicates the plane 
{(when $i=k+1$)}. When the index $i$ corresponds to a Riemann surface, the index $\alpha$ denotes the Riemann sheet of that $n$-sheeted Riemann surface on which the field is located~\footnote{When the index $i$ corresponds to the plane, there is no ambiguity of the Riemann sheet 
{to}
which the field belongs and $\alpha$ can be taken to be zero.}.}
On the other hand, {the} correlators on the 
{RHS}
of the above equation are evaluated in the  $nk+1$ independent replicas of the {$m^{\text{th}}$} minimal model and the 
{labels $a$ and $b$ indicate}
the {replica copy} of the theory.\par
Now to evaluate
{the correlator} $\langle \mathcal{E}^{(a)}(x,0)\mathcal{E}^{(b)}(x,0) \rangle_{(\mathcal{R}_{n})^{k}+1}$, we can use {a} conformal transformation to map each
{of the $k$ copies of the $n$-sheeted Riemann surface}
to 
{a plane}. 
In particular, we note that since $\mathcal{E}$ is a 
{(Virasoro) primary} field, its 
{expectation value} on the plane vanishes~\footnote{{
All primary fields of a CFT  are by convention subtracted so that their expectation values in the infinite plane vanish identically.}}.
So $\langle \mathcal{E}^{(a)}(x,0)\mathcal{E}^{(b)}(x,0) \rangle_{(\mathcal{R}_{n})^{k}+1}$ is zero unless both $\mathcal{E}^{(a)}(x,0)$ and $\mathcal{E}^{(b)}(x,0) $ lie on the same $n$-sheeted Riemann 
{surface}{, i.e. $a=(j,\alpha)$ and $b=(j,\beta)$ for the same Riemann surface $j$~\footnote{Here, the symbol $\beta$ should not be confused with the inverse-temperature.}
.} 
{The correlator $\langle \mathcal{E}^{(\alpha)}(x,0)\mathcal{E}^{(\beta)}(x,0) \rangle_{\mathcal{R}_{n}}$ }
for a {single} $n$-sheeted Riemann surface $\mathcal{R}_n$
can be calculated using the {following} conformal transformation 
\begin{equation}
    z=f(w)=\bigg(\frac{w-u}{w-v}\bigg)^{\frac{1}{n}}.
\end{equation}
In particular, {if $w$ corresponds to a point $(x,0)$ 
{on}
the $\alpha^{\text{th}}$ sheet in the $n$-sheeted Riemann surface,}
\begin{equation}
    z=f(\underbrace{\alpha,(x,0)}_{\substack{\text{(x,0) {on}  the} \\ \text{$\alpha^{\text{th}}$ sheet}}})=\bigg(\frac{x-u}{x-v}\bigg)^{\frac{1}{n}}e^{\frac{2\pi \alpha}{n}i}\text{with }\alpha\in {\{1,2,\dots,n\}}. 
\end{equation}
{Since} $\mathcal{E}$ is a 
{(Virasoro) primary}
field,
\begin{multline}\label{eqn:nRiemannSheetCorrelator0}
   \langle \mathcal{E}^{(\alpha)}(x,0)\mathcal{E}^{(\beta)}(x,0) \rangle_{\mathcal{R}_{n}}=\\ \bigg|\frac{\text{d}w_{1}}{\text{d}z_{1}}\bigg|^{-X_{\mathcal{E}}}\bigg|\frac{\text{d}w_{2}}{\text{d}z_{2}}\bigg|^{-X_{\mathcal{E}}} \langle \mathcal{E}^{(\alpha)}(z_{1})\mathcal{E}^{(\beta)}(z_{2}) \rangle_{\text{plane}},\\
\end{multline}
where $w_{1}$ 
{denotes position}
$(x,0)$ in the $\alpha^{\text{th}}$ \text{sheet} and $w_{2}$  
{denotes position}
$(x,0)$ in the $\beta^{\text{th}}$ \text{sheet}. Also,
\begin{equation}
\begin{split}
    \frac{\text{d}w_{1}}{\text{d}z_{1}}=n\frac{(x-v)(x-u)}{u-v}\bigg(\frac{x-v}{x-u}\bigg)^{\frac{1}{n}}e^{-\frac{2\pi \alpha}{n}i},\\
   \frac{\text{d}w_{2}}{\text{d}z_{2}}=n\frac{(x-v)(x-u)}{u-v}\bigg(\frac{x-v}{x-u}\bigg)^{\frac{1}{n}}e^{-\frac{2\pi \beta}{n}i},\\
   \text{and }|z_{1}-z_{2}|=2\bigg|\bigg(\frac{x-u}{x-v}\bigg)^{\frac{1}{n}}\sin\bigg(\frac{\pi(\alpha-\beta)}{n}\bigg)\bigg|.
\end{split}
\end{equation}
Then using Eq.~\ref{eqn:nRiemannSheetCorrelator0}
{as well as $\langle \mathcal{E}^{(\alpha)}(z_{1})\mathcal{E}^{(\beta)}(z_{2}) \rangle_{\text{plane}}=$~$1/ \left ( |z_1-z_2|^{2X_{\mathcal{E}}}\right)$, we obtain}
\begin{align}\label{eqn:nRiemannSheetCorrelator1}
  \langle \mathcal{E}^{(\alpha)}(x,0)&\mathcal{E}^{(\beta)} (x,0)\rangle_{\mathcal{R}_{n}}=\nonumber
  \\&= 
    \Bigg|\frac{(u-v)}{2n(x-v)(x-u)\sin\big(\frac{\pi(\alpha-\beta)}{n}\big)}\Bigg|^{2X_{\mathcal{E}}}\nonumber\\&=\langle \mathcal{E}^{(a)}(x,0)\mathcal{E}^{(b)}(x,0)\rangle_{(\mathcal{R}_{n})^k+1},
\end{align}
where in the last equality we recall that $a=(j,\alpha)$ and $b=(j,\beta)$.
{The last equality in the above equation follows
{because, as already mentioned above,} we are interested in the case when both the fields $\mathcal{E}^{(a)}$ and $\mathcal{E}^{(b)}$ lie on the same Riemann surface $\mathcal{R}_{n}$, since otherwise the correlator is zero.}
Finally{,} from Eq. \ref{eqn:BigPhismallphiandTwists}, \ref{eqn:RiemannsheetstoReplica} and 
{
$\langle\mathcal{T}_n(u,0)(\mathcal{T}_n)^{-1}(v,0)\rangle=
1/ {|u-v|^{2d_{n}}} $},
we {obtain
{for the desired three point function}}
\begin{align}
     \langle \prod_{j}^{k}\big(\Tau^{(j)}_{n}(u,0)&(\Tau^{(j)}_{n})^{-1}(v,0)\big)\Phi(x)\rangle=\nonumber\\&= \frac{C_{n,k}}{|u-v|^{2kd_{n}-2X_{\mathcal{E}}}|x-u|^{2X_{\mathcal{E}}}|x-v|^{2X_{\mathcal{E}}}}{,}\nonumber\\\\
     \text{with } C_{n,k}=k&\sum\limits_{\substack{\alpha,\beta=1 \\ \alpha \neq \beta}}^{n}\frac{1}{\big|2n\sin\big(\frac{\pi(\alpha-\beta)}{n}\big)\big|^{2X_{\mathcal{E}}}}{.}
\end{align}
Thus{,} 
 the required OPE coefficient is
\begin{equation}
    C_{n,k}=k\sum\limits_{\substack{\alpha,\beta=1 \\ \alpha\neq \beta}}^{n}\frac{1}{\big|2n\sin\big(\frac{\pi(\alpha-\beta)}{n}\big)\big|^{2X_{\mathcal{E}}}}=\frac{k}{2}\sum_{\alpha=1}^{n-1}\frac{1}{\big(\sin\big(\frac{\pi \alpha}{n})\big)^{2X_{\mathcal{E}}}}.
\end{equation}
Moreover{,} as 
{$2X_{\mathcal{E}}=
1-\epsilon
$,
with $\epsilon = 3/(m+1)$,} the {above} OPE coefficient can be expanded {in powers of $\epsilon$ as}
\begin{equation}\label{eqn:OPEcoeff0}
    C_{n,k}=\frac{k}{2}\sum_{\alpha=1}^{n-1}\frac{1}{\sin\big(\frac{\pi \alpha}{n})}+{\mathcal{O}(\epsilon).}
\end{equation}
\par 
{To obtain the von Neumann entanglement entropy}, we {also want to be able to}
analytically continue
{the $n$-dependence in the above expression to $n \to 1$.}
{Thus,} we want an expression for the {above} OPE coefficient 
{which is an analytic function of $n$ at $n=0$, and}
{which is valid for all real numbers $n$ and which reduces to Eq. \ref{eqn:OPEcoeff0} when $n$ is a natural number $(\geq 2)$.}
Following Ref. \cite{BlagouchineMoreau2023finite}, we can write 
{$1/\sin(\pi x)$}
{in the form}
(see also Ref. \cite{andrews_askey_roy_1999})
\begin{equation}
\begin{split} 
    \frac{1}{\sin(\pi x)}=\frac{1}{\pi}\int_{0}^{\infty}\text{d}t\;\frac{t^{x-1}}{1+t}\hspace{0.6cm}x\in(0,1) \ \ 
    {\text{implying}}
    \\
    \frac{1}{\sin(\frac{\pi \alpha}{n} )}=\frac{1}{\pi}\int_{0}^{\infty}\text{d}t\;\frac{t^{\frac{\alpha-n}{n}}}{1+t}=\frac{n}{\pi}\int_{0}^{\infty}\text{d}s\;\frac{s^{\alpha-1}}{1+s^n}.
\end{split} \qquad
\end{equation}
From Eq. \ref{eqn:OPEcoeff0}, the OPE coefficient then can be written as {$C_{n,k}=kI_{n}$, where $I_{n}$ is defined as}
\begin{equation}\label{eqn:OPEcoeff1}
    I_{n}{\stackrel{\text{def}}{=}}\frac{n}{2\pi}\int_{0}^{\infty}\text{d}s\;\frac{1-s^{n-1}}{(1-s)(1+s^n)}+
    {\mathcal{O}(\epsilon)}.
\end{equation}
{We make use of Eq.~\ref{eqn:OPEcoeff1} in Sect.
{\ref{sec:EntanglementEntropy}}
 below Eq.~\ref{eqn:In}.}
\vskip .8cm
{\section{Higher Loop Orders in {the}  RG and the Ising Critical Point\label{app:HigherLoopIsing}}}
We noted in Section \ref{Sec:IsingModel} that when $m=3$, i.e. in the case of 
{the}
Ising critical point, there is an additional subtlety associated with the RG analysis of {the}
replica action in Eq. \ref{LabelEqGeneralmReplicaFieldTheoryOddm}~--~\ref{eqn:GeneralLGZActionOddm}. 
This is due to the 
term in Eq. \ref{EqLinearReplicaEnergy}, which appears in 
{the}
OPE
in Eq. \ref{EqPhiPhiOPE} 
{for a generic number $R$ of replicas,}
and although it is irrelevant for minimal models $m\geq4$, it becomes marginal at $m=3$. 
We noted in Section \ref{Sec:IsingModel} 
{that, as shown in Eq.~\ref{EqPhiPhiApp} below,}
the coefficient of
{this}
marginal term in Eq. \ref{EqLinearReplicaEnergy} comes with a factor of $(R-1)$ in the OPE in Eq. \ref{EqPhiPhiOPE}. 
Thus, in {the}
 $R\rightarrow1$ limit, this term 
{is not generated by the RG to second  (1-loop) order  in the coupling constant $\Delta$ of the perturbation $\Phi$.}
In this appendix, we will show that the term in Eq. \ref{EqLinearReplicaEnergy} cannot be generated
{by the RG
to any (higher-loop) order in
the coupling $\Delta$ in the replica limit $R \to 1$, relevant for Born-rule measurements.}
{We will  provide two different arguments, (i) and (ii).}

{(i): In the first argument we use the fact that 
{all terms that could possibly}
be generated  under the RG at arbitrary order in the coupling
$\Delta$}
can be obtained by analyzing the {multiple} OPE 
\begin{equation}\label{EqPhiTimesN}
\underbrace{\Phi(x)\times\Phi(x)\times\cdots\times\Phi(x)}_{n\;\; \# \;\;\text{of} \;\;\Phi(x)}
\end{equation}
where $\Phi$ from Eq. \ref{EqIsingPhi} is {the perturbation} given by,
\begin{eqnarray}\label{EqIsingPhiApp}
    \Phi=\sum_{\substack{a,b=1 \\ a\neq b}}^{R}
\mathfrak{s}^{(a)} \mathfrak{s}^{(b)}.
\end{eqnarray}
{We use the well-known fact (see e.g. Refs.~\onlinecite{JLCardy_1986RGOPE, LUDWIG198797, cardy_1996, LUDWIGWIESE}) which states that the only operators that can be generated under the RG to any order are the operators that appear in the multiple OPE of the perturbation in Eq.~\ref{EqPhiTimesN}, and 
{in multiple OPEs} of the operators that appear 
in Eq.~\ref{EqPhiTimesN}. --  A brief review of this can be 
{found,  if desired, in App.~\ref{app:RGfromOPE}. --}
By analyzing these OPEs, we will show below that the marginal operator, Eq. \ref{EqLinearReplicaEnergy}, does not appear in any of these multiple OPEs in the limit $R \to 1$, and thus cannot be generated in this limit in any order.}\\
{We begin by discussing}
the possible operators 
that can occur in a multiple OPE of the operator $\Phi$ in Eq. \ref{EqPhiTimesN}.
{The {CFT describing the} Ising critical point has three primary fields: $I$, $\mathfrak{s}$ and $\mathfrak{e}$
{(identity, spin, and energy)}. The OPEs between these primary fields are 
given~\footnote{{
 In the above OPEs, we have 
 \textit{not} written 
 the explicit coefficients 
which accompany
 the fields in the OPE and which depend on position of the fields. 
(For Ising critical point, this is same as the fusion rules of the theory.)}.} 
by,
\begin{eqnarray}
    &&\mathfrak{s}\times \mathfrak{s}=I+\mathfrak{e},\label{EqIsingOPE}\label{EqOPESSE}\\
    &&\mathfrak{s}\times\mathfrak{e}=\mathfrak{s},\\
    &&\mathfrak{e}\times \mathfrak{e}=I,\\
    I\times \mathfrak{s}=\mathfrak{s},\;\;\; 
    &&I\times \mathfrak{e}=\mathfrak{e},\;\;\; I\times I=I\label{EqTrivialOPE}
\end{eqnarray}
For the purpose of the following discussion, we only care 
{about whether or not}
a field appears in the OPE of 
{two  given}
fields, and its exact coefficient  is immaterial. 
}
Let us now {first}  consider the OPE in Eq. \ref{EqPhiTimesN} for $n=2$. {Using the above OPE relations in each replica copy, we {obtain}}
\begin{eqnarray}\label{EqPhiPhiApp}
&&\Phi\times\Phi=4(R-2)\Phi+4a_1(R-1)\sum_{a=1}^{R}\mathfrak{e}^{(a)}+\nonumber\\ &&+a_2
{\sum_{\substack{a\neq b, a\neq c,\\b\neq c}}}\mathfrak{e}^{(a)}\mathfrak{s}^{(b)}\mathfrak{s}^{(c)}+a_3
{\sum_{a,b,c,d}^{\substack{\text{all indices are}\\\text{pairwise distinct}}}}
\mathfrak{s}^{(a)}\mathfrak{s}^{(b)}\mathfrak{s}^{(c)}\mathfrak{s}^{(d)}\nonumber\\
\end{eqnarray}
where $a_1,\;a_2$ and $a_3$ are $R$ independent numerical constants. 
We see that the marginal term $\sum_{a=1}^{R}\mathfrak{e}^{(a)}$ comes with a prefactor $(R-1)${, which} vanishes in {the}
$R\rightarrow1$ limit.
For $n=3$, the OPE in Eq. \ref{EqPhiTimesN} can be obtained by contracting 
{the RHSs} of Eqs. \ref{EqPhiPhiApp} and \ref{EqIsingPhiApp}. 
{In particular}, $\Phi$ on 
{the}
RHS of Eq. \ref{EqPhiPhiApp} can be contracted with $\Phi$ in Eq. \ref{EqIsingPhiApp}, and the 
{term} $\sum_{a=1}^{R}\mathfrak{e}^{(a)}$ will be again be produced with a $(R-1)$ {prefactor}. Moreover, the term 
${\sum_{\substack{a\neq b, a\neq c,\\b\neq c}}}\mathfrak{e}^{(a)}\mathfrak{s}^{(b)}\mathfrak{s}^{(c)}$
in Eq. \ref{EqPhiPhiApp} can contract with $\Phi=\sum_{a\neq b}\mathfrak{s}^{(a)}\mathfrak{s}^{(b)}$ to give,
\begin{equation}\label{EqPhiPhiEnergy}
    \bigg({\sum_{\substack{a\neq b, a\neq c,\\b\neq c}}}\mathfrak{e}^{(a)}\mathfrak{s}^{(b)}\mathfrak{s}^{(c)}\bigg)\times \Phi=
    \binom{R-1}{2}\sum_{a}\mathfrak{e}^{(a)}+\text{other terms}
\end{equation}
where 
again 
the marginal term $\sum_{a}\mathfrak{e}^{(a)}$ comes with a factor which vanishes in the {limit} 
{
$R\rightarrow1$ of interest. Finally, we note that {when contracted with $\Phi$}
the last term in Eq.~\ref{EqPhiPhiApp} {cannot produce the marginal term}~\footnote{{Note that the term $\mathfrak{s}^{(a_1)}\mathfrak{s}^{(a_2)}\mathfrak{s}^{(a_3)}\mathfrak{s}^{(a_4)}$ when contracted with $\mathfrak{s}^{(b_1)}\mathfrak{s}^{(b_2)}$ gives terms of the following form: $\mathfrak{s}^{(c_1)}\mathfrak{s}^{(c_2)}$, $\mathfrak{s}^{(c_1)}\mathfrak{s}^{(c_2)}\mathfrak{e}^{(c_3)}$ and $\mathfrak{s}^{(c_1)}\mathfrak{s}^{(c_2)}\mathfrak{e}^{(c_3)}\mathfrak{e}^{(c_4)}$. Thus, when contracted with $\Phi$, the last term on the RHS of the OPE in Eq. \ref{EqPhiPhiApp} cannot produce the marginal operator $\sum_a \mathfrak{e}^{(a)}$}}.}
Thus, one can conclude that as $R\rightarrow1$, the marginal term $\sum_{a}\mathfrak{e}^{(a)}$ does {also}  not occur in the OPE in Eq. \ref{EqPhiTimesN} for $n=3$.\par
We now present an induction argument for the absence of the 
{operator}
$\sum_{a}\mathfrak{e}^{(a)}$ in the OPE {of} Eq. \ref{EqPhiTimesN} for
{any 
$n$}
{number of operators $\Phi$, and for the set of operators appearing in this OPE.}
{To this end, let}
us assume that the 
{operator}
$\sum_{a}\mathfrak{e}^{(a)}$ 
{does not appear}
in the $R\rightarrow1$ limit in {the} OPE 
{of} 
{$n$ operators $\Phi$.}
{Let us also assume that the}
most general replica term that can occur in the OPE of
$n$
{operators}
$\Phi$ is
\begin{eqnarray}
\label{LabelEqGeneralTermOPEPhi}
{\sum_{\substack{a_1,a_2,\dots,a_{2k},\\b_1,b_2,\dots,b_k}}^{\substack{\text{all indices are}\\\text{pairwise distinct}}}}
     \mathfrak{s}^{(a_1)} \mathfrak{s}^{(a_2)}\cdots &&\mathfrak{s}^{(a_{2k})} \mathfrak{e}^{(b_1)}\mathfrak{e}^{(b_2)}\cdots\mathfrak{e}^{(b_l)}, \nonumber\\
      &&\text{where $2k+l\leq 2n$}.
     \qquad
\end{eqnarray}
{For
$n=2$ this corresponds to the terms appearing in 
Eq.~\ref{EqPhiPhiApp}, and this is our first step in the induction.}
Now to 
{obtain}
the OPE of
$(n+1)$ {operators}  $\Phi$ in Eq. \ref{EqPhiTimesN}, we have to contract all the terms 
{that appear}
in 
Eq.~\ref{LabelEqGeneralTermOPEPhi} above
with $\Phi=\sum_{a\neq b}\mathfrak{s}^{(a)} \mathfrak{s}^{(b)}$.
Since only the terms in the same replica copy can be contracted with each other,
out of {all the sub-terms shown in Eq. \ref{LabelEqGeneralTermOPEPhi} which appear in OPEs of 
{$n$ operators $\Phi$},} 
only the following terms can be contracted with {another} $\Phi$ to get the marginal $\sum_{a}\mathfrak{e}^{(a)}$ term:
\begin{eqnarray}
   \sum_{a_i\neq a_j}\mathfrak{s}^{(a_{i})} \mathfrak{s}^{(a_{j})}=\Phi,\\
    {\sum_{\substack{a_{i}\neq a_{j}, a_i\neq b_k,\\a_j\neq b_k}}}\mathfrak{s}^{(a_i)} \mathfrak{s}^{(a_j)}\mathfrak{e}^{(b_k)}.\label{LabelEqsse}
\end{eqnarray}
From Eq. \ref{EqPhiPhiApp} and Eq. \ref{EqPhiPhiEnergy} we see that both of these
{sub-terms}
when contracted with $\Phi$ produce the marginal $\sum_{a}\mathfrak{e}^{(a)}$, and 
{that the corresponding} coefficient vanishes in {the}
$R\rightarrow1$ limit in both 
cases.
{Moreover, all terms that can appear in the OPE of Eq.~\ref{LabelEqGeneralTermOPEPhi} with the perturbation $\Phi$ are again of the form of Eq.~\ref{LabelEqGeneralTermOPEPhi} with $n$
replaced by {$n+1$}. This completes the induction argument.}
{Thus, in summary, we have proven so far that \\
(a) the only operators that can appear in the multiple OPE of
$n$ 
operators $\Phi$ are the operators of the form appearing in Eq.~\ref{LabelEqGeneralTermOPEPhi}, and \\
(b) of those the marginal operator, having $k=0$ and $l=1$, appears with a combinatorical coefficient that vanishes in the limit $R\to 1$.}

{Finally, since the operators appearing in Eq.~\ref{LabelEqGeneralTermOPEPhi} can all be generated in the OPE in Eq.~\ref{EqPhiTimesN}, and thus could be generated by the RG
(with combinatorical coefficients that we have not determined), we would 
have finished demonstrating that the marginal operator cannot be generated in the RG to any order, if we could show that the marginal operator cannot appear in the limit $R\to 1$ in the OPE of an arbitrary number of operators of 
the type listed in Eq.~\ref{LabelEqGeneralTermOPEPhi}.  We will
now show that this is indeed the case. \\
First, we observe that it is sufficient to show that this is the case for only {\it two}
such operators, because by definition the operators appearing in Eq.~\ref{LabelEqGeneralTermOPEPhi} form a closed set of 
operators under the OPE~\footnote{I.e., they form a closed Operator Algebra.}. 
Namely, when we consider an arbitrary number of successive OPEs of operators of the form of
Eq.~\ref{LabelEqGeneralTermOPEPhi}, the marginal operator would not be generated  in this multiple OPE
if it was not generated in any of the individual successive OPEs in this limit (which involves only two operators).
On the other hand, we can show as follows that in
the OPE of {\it two} operators from Eq.~\ref{LabelEqGeneralTermOPEPhi}
the marginal operator can only appear with a combinatorical coefficient that vanishes in the limit $R \to 1$:\\
{Note that only the terms in the same replica copy can contract under
{the}
OPE. Moreover, the marginal operator $\mathfrak{e}$ is only produced either when two $\mathfrak{s}$ fields are fused (see Eq. \ref{EqOPESSE}) or when the $\mathfrak{e}$ field fuses with the identity (see Eq. \ref{EqTrivialOPE}). Therefore, the general term which appears in Eq. \ref{LabelEqGeneralTermOPEPhi} can produce the marginal operator $\sum_a \mathfrak{e}^{(a)}$ 
{only (i) when} it contracts with 
{itself (i.e., both operators have the same values of $k$ and $l$), i.e. with}
\begin{eqnarray}
\label{LabelEqGeneralTermOPEPhi2nd}
     {\sum_{\substack{a_1,a_2,\dots,a_{2k},\\b_1,b_2,\dots,b_k}}^{\substack{\text{all indices are}\\\text{pairwise distinct}}}}
     \mathfrak{s}^{(a_1)} \mathfrak{s}^{(a_2)}\cdots &&\mathfrak{s}^{(a_{2k})} \mathfrak{e}^{(b_1)}\mathfrak{e}^{(b_2)}\cdots\mathfrak{e}^{(b_l)},\nonumber\\
\end{eqnarray}
{or (ii)  when} it contracts with {another operator of the form in
Eq.~\ref{LabelEqGeneralTermOPEPhi} with the same value of $k$ but with $l$ replaced by $l+1$, namely with}
\begin{eqnarray}
\label{LabelEqGeneralTermOPEPhi3rd}
     {\sum_{\substack{a_1,a_2,\dots,a_{2k},\\b_1,b_2,\dots,b_k}}^{\substack{\text{all indices are}\\\text{pairwise distinct}}}}
     \mathfrak{s}^{(a_1)} \mathfrak{s}^{(a_2)}\cdots &&\mathfrak{s}^{(a_{2k})} \mathfrak{e}^{(b_1)}\mathfrak{e}^{(b_2)}\cdots\mathfrak{e}^{(b_l)}\mathfrak{e}^{(b_{l+1})}. \nonumber\\
\end{eqnarray}
Let us first consider the OPE of the term in Eq. \ref{LabelEqGeneralTermOPEPhi} itself, i.e. with the term in Eq. \ref{LabelEqGeneralTermOPEPhi2nd}.
Since we are interested in the coefficient of the marginal operator $\sum_{a}\mathfrak{e}^{(a)}$, we can consider two 
{identical} $\mathfrak{s}^{(a)}$ 
{fields, one in each of the two identical operators from 
Eq.~\ref{LabelEqGeneralTermOPEPhi}}
we are considering the OPE of, 
and contract 
{these}
two fields to produce the field $\mathfrak{e}^{(a)}$, while the rest of the fields
{in these two operators,}
which includes both $\mathfrak{s}^{(a_i)}$ and $\mathfrak{e}^{(b_{i})}$, should contract to produce the identity. 
Since $a_{i},b_{i}\neq a$, the number of choices for the replica indices of the remaining $\mathfrak{s}^{(a_i)}$ and $\mathfrak{e}^{(b_{i})}$ fields is given by $\binom{R-1}{2k+l-1}$. 
Thus, the marginal operator $\sum_{a}\mathfrak{e}^{(a)}$ appears with  a prefactor $\binom{R-1}{2k+l-1}$ in the OPE of the general term in Eq. \ref{LabelEqGeneralTermOPEPhi} with 
{itself, and this prefactor thus vanishes in the limit $R \to 1$.}
When $k=1$ and $l=0$, this statement is the same as $\sum_{a}\mathfrak{e}^{(a)}$ appearing with a prefactor of $(R-1)$ as shown in Eq. \ref{EqPhiPhiApp}.
Analogously, one 
{sees}
that in the OPE of Eq. \ref{LabelEqGeneralTermOPEPhi} with Eq. \ref{LabelEqGeneralTermOPEPhi3rd} 
the marginal operator $\sum_{a}\mathfrak{e}^{(a)}$ appears with  a prefactor 
{$\binom{R-1}{2k+l}$, thus also vanishing in the limit $R \to 1$.}
When $k=1$ and $l=0$ in Eqs. \ref{LabelEqGeneralTermOPEPhi} and \ref{LabelEqGeneralTermOPEPhi3rd}, this statement implies that the marginal operator  appears with a prefactor of $\binom{R-1}{2}$ in the OPE of Eq. \ref{EqIsingPhiApp} and Eq. \ref{LabelEqsse}, which was verified in Eq. \ref{EqPhiPhiEnergy}.
Thus, we see that whenever the marginal operator is produced under the OPE of two (same or different) general operators of  {the} type shown in Eq. \ref{LabelEqGeneralTermOPEPhi}, it always comes with a prefactor which vanishes in $R\rightarrow 1$ limit. This concludes our proof.}}
{Thus, to summarize our argument (i),  we conclude that 
in the limit $R\rightarrow1${,} the 
{operator}
$\sum_{a}\mathfrak{e}^{(a)}$ \textit{cannot} be generated {under the RG in any order in perturbation theory with $\Delta$.}} \\
{One can check that 
this result also holds 
if we include higher replica terms 
arising from higher cumulants  discussed in App. \ref{LabelAppB1HigherCumulantsCriticalIsing}, and the 
{proof of this statement}
proceeds analogous to the above discussion.}
{Finally, we note that
{besides 
$\Phi$, 
the exactly marginal operator and the higher replica terms discussed in App. \ref{LabelAppB1HigherCumulantsCriticalIsing}},
{all other 
terms 
that could be generated under the RG are of the form 
of those in Eq.~\ref{LabelEqGeneralTermOPEPhi}, involving a mixture of $\mathfrak{s}$ and $\mathfrak{e}$ (the term in Eq.~\ref{LabelEqsse} being the simplest example)}, and
are all irrelevant under {the} RG, as terms with support on the $\tau=0$ time-slice.}
\vskip .1cm
{(ii):} We will now give another argument 
which is perhaps more physical, for why 
{the operator}
$\sum_{a}\mathfrak{e}^{(a)}$ cannot be {generated} under 
{the} RG in the limit {$R \to 1$ of relevance to} 
Born-rule measurements. 
{The operator}
$\sum_{a}\mathfrak{e}^{(a)}$, if
{it were to be generated} 
at any order in RG, can be handled non-perturbatively by using the exact solution due to Bariev \cite{Bariev1979}, and  McCoy and Perk \cite{McCoyPerk1980}.
{Using the} exact 
{solution one sees that} when the Ising critical point {CFT} is perturbed with 
{the exactly marginal operator $\mathfrak{e}(x,\tau)$ supported on the one-dimensional time-slice}
and 
in the absence of any other perturbation, 
the power law exponent of
{the} $\mathfrak{s}(x,\tau)$ {two-point} correlation function {along the time-slice (defect line)} should change continuously with the 
{coupling}
strength of the
{marginal operator supported on the defect line.} 
In our replica field theory, in addition to
{a} possible 
perturbation $\sum_{a}\mathfrak{e}^{a}$ 
{generated under the} RG, we will also have the defect perturbation $\Phi$ {itself, from}  
Eq. \ref{EqIsingPhiApp}.
Ignoring
higher cumulants, which
{cannot change the low energy details (see App. \ref{LabelAppB1HigherCumulantsCriticalIsing})}, 
{the same}
replica field theory
{would also arise}
when we consider performing measurements with $\hat{\sigma}_{i}^{z}$ on the 
state 
\begin{equation}\label{EqDeformedStatePsi}
 \ket{\psi}=\exp{\{\kappa \sum_{i}\hat{\sigma}^{z}_{i}\hat{\sigma}^{z}_{i+1}\}}\ket{0}\;\;\;\;\;(\kappa\neq0)
\end{equation}
where $\ket{0}$ is the ground state at the Ising critical point and the operator $\hat{\sigma}^{z}_{i}\hat{\sigma}^{z}_{i+1}$ 
{represents}
the continuum field $\mathfrak{e}$ at the Ising critical point.
{If we insert in Eq. \ref{eqn:NthmomentsetupIsing} in place of the 
state $\ket{0}$ the state 
$\ket{\psi}$ from Eq. \ref{EqDeformedStatePsi}, we 
see that upon going}
{over} to the continuum 
{formulation,} we will 
{obtain}
the discussed replica theory with {{\it both},} the marginal $\sum_{a}\mathfrak{e}^{a}$ 
{as well as the} $\Phi(x)$ {interaction
added along the one-dimensional time-slice.} 
Thus, if the marginal $\sum_{a}\mathfrak{e}^{a}$ term 
{were to be} generated {under} {the}  RG in the replica field theory for $\hat{\sigma}_{i}^{z}$ measurements
{performed on the}
critical ground state $\ket{0}$, we get the same replica theory as that for $\hat{\sigma}_{i}^{z}$ measurements 
{performed on the} state $\ket{\psi}$.
This is a contradiction because Eq. \ref{eqn:bornruleavgN=1} tells us that the measurement averaged correlation function of
{the} $\hat{\sigma}_{i}^{z}$ operator, which 
{represents}
the field $\mathfrak{s}(x,\tau)$, should be the same as that in the unmeasured state, be it $\ket{0}$ or $\ket{\psi}$.
Since the $\hat{\sigma}_{i}^{z}$ correlation function has different power law behavior in states $\ket{0}$ and $\ket{\psi}$, they \textit{cannot} be described by the same replica field theory at any energy scale.
Thus, in the replica field theory for {the}  Ising critical ground state $\ket{0}$ under Born-rule $\hat{\sigma}_{i}^{z}$ measurements, the marginal $\sum_{a}\mathfrak{e}^{a}$ term cannot be generated at any order in
{the} RG.
\vskip .8cm
{\section{Brief Review - RG equations from the Operator Product Expansion (OPE)
\label{app:RGfromOPE}}}

In general one is interested in computing expectation values of ${\cal O}$, representing an operator or a product of operators in the perturbed theory such as in Eq. {\ref{LabelEqGeneralmReplicaFieldTheoryOddm}},
\begin{eqnarray}
\label{LabelEqExpectationO}
{\langle {\cal O}\rangle}_{\Delta_0}
={Z_{{*}}\over
Z_{\Delta_0}}
\ \langle
{\cal O} \ e^{+\Delta_0 \int dx \Phi(x)}
\rangle_*,
\end{eqnarray}
where expectation values $\langle \dots \rangle_*$ are taken in the unperturbed (i.e. critical) CFT  (in the present case the Ising CFT, compare e.g.
Eq.~\ref{LabelEqLGZActionIsing}).
Here $Z_*=$ $Z_{\Delta_0=0}$ is the partition function of the {unperturbed} CFT,  and $Z_{\Delta_0}$ is the fully interacting partition function
obtained from Eq.~\ref{LabelEqExpectationO} by letting ${\cal O} \to 1$.
The RG equations for all operators generated in perturbation theory to any order in $\Delta_0$ is obtained by expanding the exponential on the right hand side of Eq.~\ref{LabelEqExpectationO},
\begin{eqnarray}
\nonumber
&&\langle \dots \  1 \rangle_*+
\langle \dots 
\Bigl (\Delta_0 \int_{x_1}\Phi(x_1)
\Bigr )\rangle_* +
\\ \nonumber
&&+
\langle \dots \  
\Bigl ({\Delta_0^2\over 2!} \int_{x_1} \int_{x_2} \Phi(x_1) \Phi(x_2)
\Bigr )\rangle_*+ 
\\ \label{LabelEqExpansionOfExponentialOPE}
&&+
\langle \dots \  
\Bigl ({\Delta_0^3\over 3!} \int_{x_1} \int_{x_2}\int_{x_3} \Phi(x_1) \Phi(x_2)
\Phi(x_3)
\Bigr )\rangle_*+ 
\dots \qquad
\end{eqnarray}
{\footnote{{We note that in the above formula (and also in the subsequent formulae in this appendix) the symbol $\int_{x}$ is a short hand for $\int \frac{d^2x}{a^{1-X_{\mathcal{A}}}}$, i.e. in addition to the integral over coordinate $x$ of the integrand field $\Phi_{\mathcal{A}}(x)$, the measure of the integral is normalized with a factor of short distance cutoff $a$ raised to an appropriate power involving scaling dimension $X_{\mathcal{A}}$ of the field $\Phi_{\mathcal{A}}$ so that the corresponding coupling constant (like $\Delta_0$ when $\Phi_{\mathcal{A}}=\Phi$) is dimensionless.}}}.
The following discussion is independent of the operator(s) ${\cal O}$, indicated by the ellipses,  present in the expectation 
value~\footnote{{There is an analogous procedure to handle the RG equation of operators present in the expectation value, but this is not elaborated on here.}}.
In a general 
{term in Eq.~\ref{LabelEqExpansionOfExponentialOPE}}
we use the OPE which
expands the product of 
{$n$}
operators $\Phi$ into a complete set of operators $\Phi^{\cal A}$ located at, say, the position 
$x_{n}$ of the last operator,
\begin{eqnarray}
\nonumber
&&
\Phi(x_1) \dots  
\Phi(x_{n-1}) \Phi(x_{n})
=
\\ \label{LabelEqMultiplePhiOPE}
&&
=\sum_{{\cal A}}
\boldsymbol{C}_{\cal A}{\scriptstyle [(x_1-x_{n}), (x_2-x_{n}), ..., (x_{n-1}-x_{n})]}
\ \  \Phi^{{\cal A}}(x_{n}). \qquad
\end{eqnarray}
Most of the possible operators $\Phi^{{\cal A}}$ that appear
are irrelevant, and we will mostly be interested in relevant or marginal ones.
The integrals 
$\boldsymbol{I}_{n-1}$ over the $n-1$ 
relative coordinates appearing in the OPE coefficient $\boldsymbol{C}_{\cal A}$ are performed against a  suitable ``cutoff function''  which restricts the absolute values of all relative coordinates within the range between a short-distance cutoff $a$ and a long-distance cutoff $L$. (There are many options for the ``cutoff function'', and our discussion and result will not depend on this choice.)
Inserting these integrals into Eq.~\ref{LabelEqExpansionOfExponentialOPE}
the latter reads
\begin{eqnarray}
\nonumber
&&\langle \dots \  1 \rangle_*+
\langle \dots 
\Bigl (\Delta_0 \int_{x}\Phi(x)
\Bigr )\rangle_* +
\\ \nonumber
&&+
\langle \dots \  
\Bigl ({\Delta_0^2\over 2!} \sum_{\cal A} \boldsymbol{I}^{\cal A}_1\int_{x}   \Phi^{\cal A}(x)
\Bigr )\rangle_*+ 
\\ \nonumber
&&+
\langle \dots \  
\Bigl ({\Delta_0^3\over 3!} \sum_{\cal A} \boldsymbol{I}_2^{\cal A} \int_{x} 
\Phi^{\cal A}(x)
\Bigr )\rangle_*+ 
\dots =
\\ \nonumber
&&
{=
\langle \dots \Big ( 1+
\Delta \int_x \Phi(x) +\sum_{\Phi^{\cal A}\not = \Phi} \lambda_{\cal A}
\int_x\Phi^{\cal A}(x) + ...\Big )
\rangle_*}
\\ \label{LabelPartitionFunctionPhi}
&&=
\langle \dots
e^{\Delta \int_x \Phi(x) +\sum_{\Phi^{\cal A}\not = \Phi} \lambda_{\cal A}
\int_x\Phi^{\cal A}(x)}
\rangle_*
\end{eqnarray}
where 
we have re-exponentiated in the last line of Eq.~\ref{LabelPartitionFunctionPhi} (using standard logic) and we defined
\begin{eqnarray}
\label{LabelEqDeltaRenormalized}
&&
\Delta\left(\Delta_0, {a \over L}
\right)=
\Delta_0 \left[1
+ {\Delta_0\over 2!} \boldsymbol{I}_1({a\over L})
+ {\Delta_0^2\over 3!} \boldsymbol{I}_2({a\over L}) + 
\dots \right ],
\qquad
\\ \label{LabelLambdaARenormalized}
&&{\rm and}
\\ \nonumber
&&
\lambda^{\bf {\cal A}}\left(\Delta_0, {a \over L}
\right)
=
 \left[ {\Delta_0^2 \over 2!} \boldsymbol{I}^{{\cal A}}_1({a\over L})
+ {\Delta_0^3\over 3!} \boldsymbol{I}^{{\cal A}}_2({a\over L}) + \dots
\right ], 
\quad
\\ \nonumber
&& 
{\rm for} \ 
\  \Phi^{{\cal A}} \not = \Phi.
\end{eqnarray}
Here we used the abbreviation
\begin{eqnarray}
\nonumber
 \boldsymbol{I}_k({a\over L}) :=
\boldsymbol{I}^{\cal A}_k({a\over L}), \ \ {\rm when} \ \ \Phi^{\cal A}=\Phi, 
\  \ k=1, 2, ...
 \end{eqnarray}
The dependence of the RG equations on $\Delta$, at any order, is then obtained
for both of the couplings $\Delta$ and $\lambda^{\cal A}$ in the standard manner:
The RG equation for $\Delta$ reads 
 \begin{eqnarray}
 \nonumber
 &&
 {d \Delta(\ell)\over d\ell}=
 y_{\Delta} \cdot \Delta(\ell)  
 +(a{\partial \over \partial a})_{|\Delta_0} \ \Delta\left(\Delta_0, {a \over L}
\right)=
\\  \label{LabelRGDeltaAllOrders}
&&
=y_{\Delta} \cdot \Delta(\ell)  
 + b_2  \ \Delta^2(\ell)
 + b_3 \ \Delta^3(\ell) + \dots, 
 \end{eqnarray}
 where $\ell=\ln(L/a)$ and $\Delta(\ell=0)=\Delta_0$ is kept fixed,
 while  Eq.~\ref{LabelEqDeltaRenormalized} is used to re-express the right hand side order-by-order in terms of 
 $\Delta\left(\Delta_0, {a \over L}
\right)=$
 $\Delta(\ell)$. Here, $y_{\Delta}$  is the RG eigenvalue of the coupling $\Delta$ in the unperturbed CFT, i.e.
 $y_{\Delta}=$
 $1- X_{\Delta}=$ $3/(m+1)$;  compare Eq.~\ref{EqScalingDimensionofEnergyNEW},
 but now with $m=odd$. The terms of up {to} order $\Delta^2(\ell)$ (1-loop order) are those listed in Eq.~\ref{eqn:1loopRGdisorderstrength}.\\
 The RG equations for $\lambda^{\cal A}$ read
 \begin{eqnarray}
 \nonumber
&& {d \lambda^{\cal A}(\ell)\over d\ell}=
 y_{\cal A} \cdot  \lambda^{\cal A}(\ell)  
 +(a{\partial \over \partial a})_{|\Delta_0} \ \lambda^{\cal A}\left(\Delta_0, {a \over L}
\right) =
\\  \label{LabelRGlambdaAAllOrders} 
&&
=
 y_{\cal A} \cdot  \lambda^{\cal A}(\ell) 
+ b^{\cal A}_2  \ \Delta^2(\ell)
 + b^{\cal A}_3 \  \Delta^3(\ell) + \dots,
\\ \nonumber
&&
{\rm for} \ \
\Phi^{\cal A} \not = \Phi,
 \end{eqnarray}
 where $\lambda^{\cal A}(\ell=0)=0$, and
again Eq.~\ref{LabelEqDeltaRenormalized} is used to re-express the right hand side order-by-order in terms of 
 $\Delta\left(\Delta_0, {a \over L}
\right)=$
 $\Delta(\ell)$.
 Here,
 $y_{\cal A}$ are the RG eigenvalues of the couplings $\lambda^{\cal A}$ in the unperturbed CFT.
 
At this stage of the discussion only powers of $\Delta(\ell)$ appear on the right hand side of both RG equations Eqs.~\ref{LabelRGDeltaAllOrders},~\ref{LabelRGlambdaAAllOrders}, while we see from the latter equation that in general non-vanishing couplings $\lambda^{\cal A}$ of operators $\Phi^{\cal A} \not = \Phi$
are generated. These will then appear in the argument of the exponential of the last line of
Eq.~\ref{LabelPartitionFunctionPhi}. The key point then is the following: {
Upon the RG coarse-graining process
these thereby generated couplings will generate additional terms in the RG equations. These  
additional terms
can easily be understood and incorporated into our existing discussion due to the fact that the set of all thereby generated operators $\Phi^{\cal A}$ with couplings $\lambda^{\cal A}$ form the set of operators
listed in Eq.~\ref{LabelEqGeneralTermOPEPhi} which is  closed under the  Operator Product
Expansion (OPE).} This means that all the terms generated by the RG from the couplings
$\lambda^{\cal A}$ of operators $\Phi^{\cal A} \not = \Phi$ can be understood by simply generalizing Eq.~\ref{LabelEqMultiplePhiOPE} to multiple OPEs of the operators appearing in Eq.~\ref{LabelEqGeneralTermOPEPhi}, namely to
\begin{align}
&
\Phi^{{\cal A}_1}(x_1)
\dots  
\Phi^{{\cal A}_{n-1}}(x_{n-1}) \Phi^{{\cal A}_n}(x_{n})
=\nonumber
\\
&
=\sum_{{\cal A}}
\boldsymbol{C}_{\cal A}^{{\cal A}_1,  ..., {\cal A}_{n-1}, {\cal A}_n}
{\scriptstyle [(x_1-x_{n}), (x_2-x_{n}), ..., (x_{n-1}-x_{n})]}
\ \Phi^{{\cal A}}(x_{n}).\label{LabelEqMultiplePhiOPEPHIlambda}
\end{align}
Employing the same logic that led to the   RG equations Eqs.~\ref{LabelRGDeltaAllOrders},~\ref{LabelRGlambdaAAllOrders} now leads to the same RG equations but with arbitrary powers of the coupling constants $\lambda^{\cal A}$  appearing on the right hand side of these equations.
\par
{The key result of this analysis is that the only couplings $\lambda^{\cal A}$ that can be generated under the RG are those of operators $\Phi^{\cal A}$ that can appear on the right hand side of the multiple OPE in Eq.~\ref{LabelEqMultiplePhiOPEPHIlambda}. Such  contributions would be represented by a term  of order $\lambda^{{\cal A}_1} \dots \lambda^{{\cal A}_{n-1}} \  \lambda^{{\cal A}_n}$ 
on the right hand side of the RG  equation for $\lambda^{\cal A}$.
However we show in argument (i) 
of App.~\ref{app:HigherLoopIsing} 
that all  the OPE coefficients  in 
Eq.~\ref{LabelEqMultiplePhiOPEPHIlambda},
involving on the left hand side operators
appearing in Eq.~\ref{LabelEqGeneralTermOPEPhi},
vanish in the limit $R \to 1$, when  $\Phi^{\cal A}$ and $\lambda^{\cal A}$ appearing on the right hand side corresponds to the exactly marginal operator
in 
Eq.~\ref{EqLinearReplicaEnergy}
and its corresponding coupling constant. This then implies that the exactly marginal operator cannot be generated under the RG to any order in perturbation theory in the coupling $\Delta$ in the limit $R \to 1$.}
\vskip .8cm
{\section{Irrelevance of Higher Cumulants from Avoided Level Crossings\label{app:AvoidedLevelCrossingsHigherCumulants}}}

The  ``higher-moment operators'' in Eq.~\ref{EqHigherCumulantsContinuum}
and
in Eq.~\ref{HigherCumulantsSimplifiedIsing}  whose scaling dimensions 
at the $\Delta_* \not =0$ fixed point are of interest
in App.~\ref {LabelAppB1HigherCumulantsTricriticalIsing} and App.~\ref{LabelAppB1HigherCumulantsCriticalIsing}, respectively, are conformal boundary operators: In the standard manner, these operators with support on the
one-dimensional $\tau=0$ time-slice  in space-time can be viewed upon folding  the space-time
along 
this time-slice 
\cite{FendleyLudwigSaleur1995},
\cite{LeclairLudwig}
as operators with support on the boundary, the real axis, of the tensor product of two identical copies of the  (non-random but replicated) bulk CFT, located in the upper half complex plane. The fixed point $\Delta_* \not =0$
describes a scale- and conformally invariant boundary condition  $B_{\Delta_*}$ on the  two copies of the bulk CFT in the upper half plane, which determines all the universal properties of interest to us in this paper.
After conformal mapping from the upper half complex plane to
the interior of an infinitely long strip of finite width $L$ with identical boundary conditions $B_{\Delta_*}$ on both sides, the spectrum of the Hamiltonian ${\hat H}_{\Delta_*}$, generating translations along the strip, is universally related  by finite-size scaling 
\cite{CARDY1986BoundaryContent},\cite{AffleckLudwig-Kondo1991}
and the operator-state correspondence to the scaling dimensions of the set of all operators with support on the boundary $B_{\Delta_*}$, of interest to us here.

Here we {will discuss, 
for each value of $m$ or equivalently of $\epsilon=3/(m+1)$,}
the evolution (``spectral flow'') of a particular set of energy levels of the Hamiltonian ${\hat H}_{\Delta(\ell)}$ where the coupling constant 
$\Delta=\Delta(\ell)$
{(here $\ell=\ln(L/a)$)}, flows under the RG from the RG-unstable zero coupling fixed point
$\Delta=0$ in the ultraviolet to the RG-stable finite-coupling
fixed point 
{$\Delta_* = \Delta_*({\epsilon})$ in the infrared.}
{For each value of $m$ 
({or equivalently}
$\epsilon$) this}
describes an RG flow between two conformally invariant boundary conditions.
In the intermediate regime of length scales $\ell$ away from the two fixed points, the corresponding boundary condition $B_\Delta(\ell)$ will not be scale- nor conformally invariant, and the entire spectrum of the corresponding
Hamiltonian ${\hat H}_{\Delta(\ell)}$  will undergo an evolution, i.e. a spectral flow with the length scale $\ell$. Because the operator coupling to $\Delta$ is invariant under the group $S_R$ of permutations of the $R$ replicas, we can classify all eigenstates of ${\hat H}_{\Delta(\ell)}$ according to irreducible representations of the permutation group $S_R$.
\par
{For each value of $m$ we} consider the spectrum of this Hamiltonian in each symmetry sector separately. Since the operators  in Eq.~\ref{EqHigherCumulantsContinuum}
and
in Eq.~\ref{HigherCumulantsSimplifiedIsing}  of interest to us are singlets under permutations, we restrict attention to the  $S_R$-singlet sector of the spectrum of ${\hat H}_{\Delta(\ell)}$. We know that this Hamiltonian is, due to the large conformal symmetry,  integrable at the ultraviolet ($\Delta=0$)  as well as the infrared fixed point ($\Delta_*\not =0$). However, at all intermediate scales $\ell$  away from the two fixed points this Hamiltonian is not expected to be integrable since the operator $\Phi$ coupling to $\Delta$ 
(and thus setting the boundary condition $B_{\Delta(\ell)}$ which determines the spectrum of ${\hat H}_{\Delta(\ell)}$) is not expected to conserve a macroscropic number of the conformal conservation laws present at the two fixed points.
Given that the Hamiltonian ${\hat H}_{\Delta(\ell)}$ is not integrable at intermediate scales, the evolution of its spectrum as a function of scale $\ell=\ln(L/a)$ in the $S_R$-singlet sector is expected to exhibit {\it avoided level crossings}. 
{Now for each value of $m$ (or $\epsilon$),}
as discussed in App.~\ref {LabelAppB1HigherCumulantsTricriticalIsing} and App.~\ref{LabelAppB1HigherCumulantsCriticalIsing}, 
the scaling dimensions of the operators in  Eq.~\ref{EqHigherCumulantsContinuum}
and
in Eq.~\ref{HigherCumulantsSimplifiedIsing} at the ultraviolet fixed point $\Delta=0$
{are equal to 
$2k \times X_\epsilon= 2k \times X_{\varphi_{1,2}}$
when $m=$ 
{\it even}, and 
$2k \times X_{\epsilon}= 2k \times X_{\varphi_{1,2}}$ 
when 
$m=$ {\it odd}, respectively, and thus are strictly ordered in both cases. Here 
{$X_\epsilon=X_{\varphi_{1,2}}$}
is given by  Eq.~\ref{EqScalingDimensionofEnergyNEW} for both cases,  $m=$ {\it even} and  $m=$ {\it odd} {(see footnote \cite{Note26})}. 
Note that in the limit $m \to \infty$ ($\epsilon \to 0$), these dimensions become
$2k \times (1/2) =k$ since 
{$X_\epsilon=X_{\varphi_{1,2}}$} 
$\to 1/2$.}
{In either case, all} these operators with $k>1$  thus 
{have, for any value of $m$ (even or odd),} scaling dimensions larger than the {respective} perturbation $\Phi$ {(}which corresponds to $k=1${)} 
at the ultraviolet fixed point $\Delta=0$.
{For each case, i.e. for any {\it even} and for any {\it odd} value of $m$ corresponding to 
Eq.~\ref{EqHigherCumulantsContinuum}
and
Eq.~\ref{HigherCumulantsSimplifiedIsing} respectively, we expect,}
given the avoided level 
crossings,
as we increase the scale $\ell=\ln(L/a)$ to  run the RG via finite-size scaling 
from
the ultraviolet to the infrared fixed point 
{$\Delta_*(\epsilon)$,}
that the relative ordering of these scaling dimensions for different values of $k$ is preserved. In particular, we expect the scaling dimensions of all these operators with $k>1$ to remain larger than the scaling dimension of the operator which has the  smallest scaling dimension at the ultraviolet fixed point $\Delta=0$, which is the one with $k=1$, corresponding to the perturbation $\Phi$. But since we know that the perturbation $\Phi$ must be irrelevant (i.e. must have scaling dimension $>1$) at the infrared fixed point (as given by the slope of the corresponding RG beta function for $\Delta$), we conclude that
the scaling dimensions of all the operators with $k>1$ will also  need to 
{be}
certainly larger than unity, due to avoided level crossings. 
This provides an argument supporting the irrelevance {at the infrared fixed point} of 
the operators {in} Eq.~\ref{EqHigherCumulantsContinuum}
and
in Eq.~\ref{HigherCumulantsSimplifiedIsing} 
arising from all higher cumulants, both in the  tricritical Ising (App.~\ref {LabelAppB1HigherCumulantsTricriticalIsing})  
as well as in the Ising (App.~\ref{LabelAppB1HigherCumulantsCriticalIsing}) case.\par
We close by noting that avoided level crossings of scaling dimensions of bulk
operators in RG flows between two 
(bulk) 2D RG fixed points, arising from perturbations breaking the integrability  of the ultraviolet CFT, have been observed
explicitly via the Truncated Conformal Space approach
\cite{YurovZamolodchikov1991}, 
{e.g. see Ref. \cite{KonikPalmaiTakacsTsvelik2015} Sect. IV.C, Figs. 7, 8.}
Flows of boundary scaling dimensions in RG flows between two different boundary fixed points of the same bulk CFT have also been studied using the Truncated Conformal Space approach, see e.g. 
Ref. \cite{FeveratiGrahamPearceTothWatts2006};
the latter particular investigation is of less direct relevance for us since in this study only an integrable boundary perturbation is discussed, but it demonstrates the ability to study RG flows between boundary fixed points effectively within the Truncated Conformal Space approach. Finally, all spectra numerically obtained from 
K. G. Wilson's numerical renormalization group approach {to} the Kondo- and other quantum impurity problems
({see e.g. Ref.}
\onlinecite{Wilson1975,PangCox1991})
precisely observe \cite{AffleckLudwig-Kondo1991}
related spectra of boundary RG flows between different fixed point boundary conditions on a fixed bulk CFT, {exhibiting avoided crossings in a given symmetry sector.}
\bibliography{DefectBib}
\end{document}